\documentclass[preprint2,times]{aastex701}

\usepackage{savesym}
\savesymbol{tablenum}

\usepackage{graphicx}%
\usepackage{multirow}%
\usepackage{amsmath,amssymb,amsfonts}%
\usepackage{amsthm}%
\usepackage{mathrsfs}%
\usepackage[title]{appendix}%
\usepackage{xcolor}%
\usepackage{xspace}

\usepackage{chemformula}
\usepackage{hyperref}
\usepackage[separate-uncertainty=true]{siunitx}
\usepackage{float}
\usepackage{lipsum}

\usepackage[T1]{fontenc}

\newcommand{\dd}{\mathop{}\!\mathrm{d}}

\newcommand\chili{\textsc{CHILI}\xspace}

\DeclareSIUnit\bar{bar}
\DeclareSIUnit\AU{AU}
\DeclareSIUnit\dex{dex}
\DeclareSIUnit\erg{erg}
\DeclareSIUnit\day{day}
\DeclareSIUnit\year{yr}
\newcommand{\WPMS}{\watt\per\meter\squared}

\shorttitle{CHILI Part I --- Earth and Venus Evolution}

\begin{document}

\title{Coupled atmospHere Interior modeL Intercomparison (\chili). I.\\ Evolutionary Modelling -- Primordial Magma Oceans of Earth and Venus}

\author[orcid=0000-0002-8368-4641]{Harrison Nicholls}
\email[show]{harrison.nicholls@ast.cam.ac.uk}
\affiliation{Institute of Astronomy, University of Cambridge, UK}

\author[orcid=0000-0001-6878-4866]{Joshua Krissansen-Totton}
\email{joshkt@uw.edu}
\affiliation{University of Washington, USA}

\author[orcid=0000-0002-3286-7683]{Tim Lichtenberg}
\email{tim.lichtenberg@rug.nl}
\affiliation{Kapteyn Astronomical Institute, University of Groningen, The Netherlands}

\author[orcid=0000-0003-2915-5025]{Laura Schaefer}
\email{lkschaef@stanford.edu}
\affiliation{Stanford University, USA}

\author{Keiko Hamano}
\email{keiko.i.hamano@gmail.com}
\affiliation{National Astronomical Observatory of Japan, Japan}

\author[orcid=0000-0001-8804-120X]{Maxime Maurice}
\email{maxime.maurice@uca.fr}
\affiliation{Université Paris-Saclay, CNRS, GEOPS, 91405, Orsay, France}

\author[orcid=0000-0002-3740-9235]{Henri Samuel}
\email{samuel@ipgp.fr}
\affiliation{Universit\'e Paris Cit\'e, Institut de physique du globe de Paris, CNRS, Paris, France}

\author[orcid=0009-0002-0821-062X]{Alexandra Papesh}
\email{agp99@uw.edu}
\affiliation{University of Washington, USA}

\author[orcid=0000-0003-2679-813X]{Carlos Ortiz-Quintana}
\email{}
\affiliation{University of Central Florida, USA}

\author[orcid=0000-0002-9032-8530]{Junellie Perez}
\email{jgonza70@jhu.edu}
\affiliation{Johns Hopkins University, USA}

\author[orcid=0000-0002-0747-8862]{Yamila Miguel}
\email{ymiguel@strw.leidenuniv.nl}
\affiliation{SRON Netherlands Research Organisation Netherlands, Niels Bohrweg 4, 2333CA Leiden, The Netherlands}
\affiliation{Leiden Observatory, University of Leiden, Einsteinweg 55, 2333CA Leiden, The Netherlands}

\author[orcid=0000-0001-8832-5288]{Denis E. Sergeev}
\email{denis.sergeev@bristol.ac.uk}
\affiliation{School of Physics, University of Bristol,
HH Wills Physics Laboratory, Tyndall Avenue, Bristol BS8 1TL, United Kingdom}


\author[orcid=0000-0001-9284-0143]{Philipp Baumeister}
\email{philipp.baumeister@fu-berlin.de}
\affiliation{Institut f\"ur Geologische Wissenschaften, Freie Universit\"{a}t Berlin, Germany}

\author[orcid=0000-0003-3652-3690]{Spandan Dash}
\email{}
\affiliation{Universit\'e Paris Cit\'e, Institut de physique du globe de Paris, CNRS, Paris, France}

\author[orcid=0009-0002-9902-731X]{Leoni Janssen}
\email{ljanssen@strw.leidenuniv.nl}
\affiliation{SRON Netherlands Research Organisation Netherlands, Niels Bohrweg 4, 2333CA Leiden, The Netherlands}
\affiliation{Leiden Observatory, University of Leiden, Einsteinweg 55, 2333CA Leiden, The Netherlands}

\author[orcid=0009-0008-5219-8772]{Jonathan W. Keathley}
\email{jonathan.keathley@ucf.edu}
\affiliation{University of Central Florida, USA}

\author[orcid=0009-0003-4258-4366]{Alexandre de Larminat}
\email{delarminat@ipgp.fr}
\affiliation{Universit\'e Paris Cit\'e, Institut de physique du globe de Paris, CNRS, Paris, France}

\author[orcid=0000-0002-1924-641X]{Emmanuel Marcq}
\email{emmanuel.marcq@latmos.ipsl.fr}
\affiliation{Laboratoire Atmospheres Milieux Observations Spatiales, France}

\author[orcid=0000-0001-8817-1653]{Lena Noack}
\email{lena.noack@fu-berlin.de}
\affiliation{Institut f\"ur Geologische Wissenschaften, Freie Universit\"{a}t Berlin, Germany}

\author[orcid=0009-0005-5234-6673]{Hugo Pelissard}
\email{hugo.pelissard@u-bordeaux.fr}
\affiliation{University of Bordeaux, France}

\author[orcid=0009-0009-6098-296X]{Bo Peng}
\email{bpengeps@stanford.edu}
\affiliation{Dept. of Earth \& Planetary Sciences, Doerr School of Sustainability, Stanford University}

\author[orcid=0009-0009-5036-3049]{Emma Postolec}
\email{e.n.postolec@rug.nl}
\affiliation{Kapteyn Astronomical Institute, University of Groningen, The Netherlands}

\author[orcid=0000-0001-7553-8444]{Ramses Ramirez}
\email{Ramses.Ramirez@ucf.edu}
\affiliation{University of Central Florida, USA}

\author[orcid=0009-0008-7799-7976]{Mariana Sastre}
\email{m.c.villamil.sastre@rug.nl}
\affiliation{Kapteyn Astronomical Institute, University of Groningen, The Netherlands}

\author[orcid=0000-0001-6974-6714]{Andrea Zorzi}
\email{zorzi@stanford.edu}
\affiliation{Stanford University, USA}


\begin{abstract}
Earth and Venus represent two evolutionary outcomes arising from initially molten `magma ocean' periods, followed by lifetimes of chemical and geophysical divergence. Their physics is common to all rocky planets and is accessible to simulations that adopt coupled interior-atmosphere modelling approaches. Our understanding of planet histories and interpretation of current states is dependent on this modelling, yet existing codes vary in their approximations. 
Here, we present the first results from the Coupled atmospHere Interior modeL Intercomparison (\chili) project; benchmarking planetary evolution codes in the context of Earth and Venus to identify key model sensitivities. Our 
`nominal' Earth models predict magma ocean solidification timescales within 4~Myr of thermal evolution, and are consistent with empirical constraints on Earth's early history. Venus scenarios exhibit more diverse behaviours where prolonged magma ocean stages can be conditionally sustained for 50\,Myr. Cooling timescales correlate with initial hydrogen and carbon budgets, but model-specific treatments of volatile partitioning and vertical energy transport introduce substantial inter-model variance. Different parametrisation of mantle geodynamics, convection, melting curves, rheological properties, and radiative transfer give rise to divergent evolutionary behaviours. Discrepancies in atmospheres generated by magma ocean outgassing underscore these differences, although C-H-O compositions with surface pressures exceeding 100\,bar are favoured. 
This intercomparison identifies critical sensitivities in volatile partitioning, escape processes, mantle viscosity, and melting curves. Validating these treatments is essential for enabling deep insight  into the early histories of the Solar System's terrestrial planets, and for drawing meaningful interpretations from ongoing observational exoplanet campaigns.
\end{abstract}



\section{Introduction} 
\label{sec:intro}

Rigorous intercomparison of the planetary evolution models and their physics is critical for interpreting current observations of exoplanets. The Coupled atmospHere Interior modeL Intercomparison (\chili) project \citep{chili_protocol_2025} identifies key uncertainties and limitations in these models, as part of the established CUISINES model intercomparison framework \citep{Sohl24_cuisines}. Here, we present Part\,I of \chili by simulating Earth and Venus scenarios and comparing outcomes from a range of magma ocean evolution models.

All rocky planets are born into hot states with global `magma oceans' \citep{ Schaefer2018RSPTA, Abe1985, canup_origin_2001, cameron_the_1976}. Yet, Earth and Venus' present climates differ substantially from each other, despite these two rocky planets orbiting the same star and being of similar sizes \citep{lodders_planetary_1998, genda_origin_2016, javoy_where_2005}. How did Earth become habitable and Venus did not \citep{Hamano2024arXiv}? Why does Earth have plate tectonics in contrast to Venus' different tectonic setting \citep{solomatov_stagnant_1996,gillmann_atmosphere_2014,Gulcher2020NatGe,Byrne2021PNAS}? Does physics dictate these to be common outcomes of early planetary evolution, or two niche scenarios? Earth and Venus' differential climates and geodynamics represent billions of years of evolution, shaped by the interplay of fluid dynamic, chemical, and physical processes \citep{abe_thermal_1997, Lichtenberg2023}. 

Isotopic measurements of zircon minerals suggest that the Earth's mantle solidified before 4.2\,Gyr ago \citep{cavosie_magmatic_2005, sole_hadean_2025}. Given these measurements, Earth's last global magma ocean period then dates to between $\approx$100~and~400\,Myr after its formation \citep{dalrymple_age_2001}. Giant impact events -- such as those suggested having formed the Moon -- would have re-melted Earth's mantle during the period between its formation and its final solidification \citep{canup_origin_2001, borg_moon_2014, barboni_early_2017}, potentially allowing multiple sequential magma ocean stages to modulate Earth's structure and volatile budget \citep{Abe1985, citron_the_2018, canup_origin_2001}. Xenon isotopes provide further evidence for efficient, magma ocean mediated volatile degassing within Earth's first 100\,Myr \citep{rzeplinski_hadean_2022}. When Earth first solidified, and whether solidification occurred once or episodically, is dependent on the processes regulating its early cooling \citep{Schaefer2017ApJ, boukare_solidificat_2025, Salvador2023_SSR, chao_lava_2021}. 

The formation of a solid crust remains a necessary condition for condensing water oceans and geochemical weathering \citep{foley_the_2015, pierrehumbert_the_2002}, potentially active between magma ocean epochs \citep{Mojzsis2001Natur,dasgupta_ingassing_2013}. So, careful comparison between computer modelling and empirical constraints on Earth's earliest stages contextualise the conditions for the origin of life and our current climate \citep{Elkins-Tanton2012, zahnle_earths_2010}. 

Venus' age is comparable to Earth's, but its early thermal and geodynamical history are less well constrained \citep{halliday_the_2023, lammer_formation_2020}. Direct insights into Venus' geothermal history are precluded due to the absence of samples and global resurfacing by active volcanic processes within the last $\sim750$--$150$\,Ma  \citep{marchi_long_2023,strom_the_1994, ivanov_the_2013}. We expect that Venus' early magma ocean modulated escape of its initial volatiles $\sim10$~to~$\sim100$~Myr after the planet formed \citep{Hamano2013}. Indications of a water-poor Venusian interior \citep{constantinou_venus_2025} and depletion of radiogenic $^{40}\mathrm{Ar}$ in its atmosphere \citep{orourke_venus_2015, kaula_venus_1999} potentially suggest efficient early magma ocean degassing. Magma ocean solidification was followed by several Gyr of uncertain atmosphere-interior interactions \citep{Rolf2022SSRv,warren_narrow_2023,Ghail2024SSRv, gillmann_the_2022}. After magma ocean solidification, Venus may have sustained an uninhabitable post-runaway greenhouse state throughout its lifetime, or it could have temporarily hosted a water ocean \citep{Turbet2021, warren_narrow_2023, way_venus_2016, krissansen2021venus}, a view which is supported by the possible felsic compositions of highland tessera terrain \citep{gilmore_venus_2017}. Understanding the lifetime history of our sister planet is strongly dependent on modelled predictions to interpret limited in-situ measurements.


Planetary evolutionary models -- informed by known physics -- can predict the feasible historical pathways potentially undertaken by a given planet, based on its measured size, irradiation, and composition \citep{ElkinsTanton2008, Lichtenberg2025Science,krissansen2022understanding}. Numerical simulation methods have remained the leading approach for quantifying planetary lifetime histories since \cite{Abe1985}. More recently, computer modelling developments have appealed to exoplanet environments starkly different to the Solar System; e.g. the peas-in-a-pod TRAPPIST-1 planets \citep{Agol2021, krissansen2022predictions}, or permanent `lava worlds' \citep{Herath2024MNRAS, shorttle_distinguish_2024, zilinskas2025,seidler_impact_2024}. Improvements in the accuracy of climate calculations \citep{Molliere2019AA,villanueva_modeling_2024,selsis_cool_2023,nicholls_agni_2025} and volatile partitioning behaviours \citep{Schaefer2017ApJ,Suer2023FrEaS,Bower2025ApJ} enable the modelling of these extrasolar scenarios. Consequently, multiple distinct simulation codes have emerged \citep{chili_protocol_2025}. The exchange of energy and mass between these planets' interiors and atmospheres represents a strong and sensitive control on their thermal histories and melting states \citep{Schaefer2018RSPTA, Lichtenberg2023, Lichtenberg2025Science}. Coupled  model codes aim to treat this  problem of interconnected interior-atmospheric evolution, and have already been successfully applied to a range of (exo)planetary scenarios. For example, \citet{Krissansen-Totton_2024} used PACMAN to show that early, chemically reducing primordial envelopes can give rise to clement, water-rich environments on rocky worlds. Meanwhile, \citet{nicholls_volatile_2026} used PROTEUS to show that super-Earth compositions measured by JWST reflect unexpected volatile-rich formation scenarios. The VPLanet interior-climate evolution code (not part of \chili) is able reproduce `observable' metrics and habitability predictions of pre-industrial Earth \citep{gilbert_a_2026}.

The current assortment of magma ocean evolution models differ inhomogeneously by their adopted approximations and simplifications \citep{chili_protocol_2025}. Our expectations for the melting state and atmospheric compositions of specific planets rest largely on simulations, so model sensitivities and uncertainties necessarily influence interpretations of our measurements and observations \citep{Lichtenberg2025Science, schaefer2016}. Ultimately, the inner Solar System planets will always represent our only `ground truth' for benchmarking and comparing complex simulation codes, and the sensitive interplay of  physical, chemical, and dynamical which they aim to resolve.

The \chili protocol describes a controlled framework for intercomparison of coupled interior-atmosphere magma ocean evolution models, and presented initial results from `static' and evolutionary calculations. \citet{chili_protocol_2025} highlighted important quantitative differences between calculations of Earth's magma ocean solidification timescales (from $10^5$ to $10^9$\,yr) and resultant atmospheric compositions (with \ch{H2O} partial pressures from 100 to 700\,bar). Simulated pathways to Earth solidification qualitatively diverged between models, either following protracted semi-molten periods of radiative equilibrium or directly solidifying. However, initial comparisons of static and evolutionary models' atmospheric structures showed good agreement. 

This paper (Part I) presents the first results from \chili by considering Earth and Venus case studies. A second \chili paper (Part II) simulates the evolution of highly irradiated exoplanets, orbiting the M-type star TRAPPIST-1, as an alternative basis for model intercomparison. A third \chili paper (Part III) considers `static' non-evolving atmosphere simulations to focus specifically on the physics and chemistry probed by telescope remote-sensing techniques. Part III depends on atmospheric compositions and interior states informed from Part I and Part II. These comparisons will guide future model developments. The eventual resolution of these epistemological uncertainties on early planetary evolution will provide rich insight into Earth's deep past, while enabling meaningful interpretation of exoplanet observations.

Section \ref{sec:methods} reviews the key physics resolved by the time-evolving \chili models and common parametrisations of these processes, and provides specific details about how the models differ. Section \ref{sec:results} presents simulated evolution pathways for Earth and Venus, testing sensitivity to volatile budgets, and diagnoses the physics and approximations driving models' different predicted pathways. Section \ref{sec:discuss} discusses inter-model commonalities and deviations, with implications for interpreting the early histories of Earth and Venus, highlighting the processes and approximations deserving of future development and research. Section \ref{sec:conclude} itemises our conclusions.

\section{Methods} 
\label{sec:methods}

\subsection{Intercomparison of `Nominal' and gridded scenarios}
\label{sec:methods_intercomp}

This part of \chili uses Earth and Venus as case studies for comparing simulated interior-atmosphere evolution pathways. We consider `nominal' scenarios for both planets, initialised with $4.7\times10^{20}$\,kg of hydrogen (equivalent to three Earth oceans\footnote{Here, `oceans' of hydrogen content is a unit describing an amount of hydrogen atoms, equivalent to the mass of hydrogen atoms within Earth's surface water ocean reservoir. One ocean corresponds to \SI{1.4e21}{\kilo\gram} of \ch{H2O} or \SI{1.6e20}{\kilo\gram} of hydrogen atoms \citep{clark_carbon_1982}.}) and $2.73\times10^{20}$\,kg of carbon in the mantle-atmosphere system. If outgassed entirely as \ch{CO2}, this carbon inventory corresponds to 192.4\,bar of surface pressure. These budgets represent the initial conditions on the hydrogen and carbon contents which are free to cycle between the modelled mantle and atmosphere reservoirs according to their solubility and chemistry \citep{abe_thermal_1997}.  Nominal Venus is identical to Nominal Earth except its mass, radius and orbital distance are adjusted to match Venus' present day values \citep{lodders_planetary_1998}. We refer to cases with these nominal hydrogen and carbon budgets as `Nmnl', below. 

Magma ocean cooling timescales are sensitive to planetary volatile budgets, particularly that of hydrogen, because thick atmospheres induce strong blanketing and greenhouse effects which delay mantle solidification \citep{abe_thermal_1997,nicholls_redox_2024, Nikolaou2019, Hamano2015}. Yet, multiple physical processes and circumstantial factors collectively determine the volatiles obtained by a planet during its formation \citep{ikoma_constraints_2006,Krijt2023ASPC5341031K}. For example, a planet could form from volatile-richer or volatile-poorer material, depending on its location exterior or interior to the water ice line or soot line \citep{Drazkowska2023ASPC534717D,Bergin2026arXiv260210308B}, while being subject to desiccation and mixing of multiple protoplanetary disk sources \citep{lichtenberg_a_2019, ikoma_constraints_2006, sossi_physicochem_2025}. Giant impacts, later-stage delivery, and volatile redistribution would then  modulate the amount of volatile elements distributed across early mantles and atmospheres \citep{Lupu2014ApJ, roche_atmospheric_2025, nakajima_scaling_2021, Kegerreis2020ApJ901L31K, Lock2024PSJ528L}. As a result, there are large empirical uncertainties on Earth's initial \citep{wang_the_2018} and present volatile budget: its combined mantle and atmosphere water content is estimated between $\sim2$ and $\sim10$ oceans \citep{peslier_water_2017}. There is substantial uncertainty in the additional amounts of species of light elements incorporated in Earth's metallic iron core \citep{Hirose2013,Shahar2026}. To respect the uncertainty and potential range of outcomes arising from these processes, we additionally consider a range of evolution scenarios,  across three hydrogen inventories (\mbox{[Hlow, Hmid, Hhigh]}$ = $[1.60, 7.80, 16.00] $\times~10^{20}$ kg\,H) corresponding to 1, 5, and 10 oceans of \ch{H2O} \citep{peslier_water_2017, javoy_where_2005}. We simultaneously consider three carbon inventories (\mbox{[Clow, Cmid, Chigh]}$ = $[1.36, 2.73, 5.44] $\times~10^{20}$ kg C) to form a grid of nine volatile-budget combinations.  Our exploration of various initial volatile budgets reflects our uncertainty on the amount of \ch{H2O} and \ch{H2} expected to be exogenously delivered \citep{ormel_how_2021, Krijt2023ASPC5341031K} to young planets, and endogenously formed within their interiors by redox chemistry  \citep{ikoma_constraints_2006, genda_origin_2016, kite_water_2021, miozzi_experiments_2025}. 

For a given planet case, we anticipate a range of evolutionary behaviours to arise from the grid of nine volatile budget combinations considered, since the \chili models incorporate various physical parametrisations of atmospheric opacity, chemistry, and partitioning \citep{chili_protocol_2025}. This grid brackets the configuration adopted for our `nominal' Earth and Venus scenarios. Some \chili models do not simulate carbon, so up to ten different compositional compositions are considered for each planet.

We initialise simulations at a stellar age of 50\,Myr \citep{rubie_tungsten_2025, dalrymple_age_2001}. The models adopt different initial temperatures and entropies, while ensuring that these correspond to a fully molten state. This initial condition ensures that the planet's core has fully differentiated from the mantle, the Sun has entered its main sequence phase, and the solar nebula has dispersed \citep{halliday_the_2023, schoenberg_new_2002, Baraffe2015}. Additional accretion of material onto these young planets is not simulated by any of the \chili models, but would be minimal after 50\,Myr \citep{Dauphas2017Natur541521D,halliday_the_2023}. To curtail model complexity, all of these simulations neglect clouds or aerosols and their radiative effects. Instead, they enforce a constant 10\% Bond albedo scale factor applied to incoming stellar radiation \citep{pluriel_modeling_2019, cmiel2025}. Various treatments of atmospheric escape are applied, so a 30\% energy-limited photoevaporative escape efficiency is adopted where applicable. 

\subsection{Common parametrisations of key physics}
\label{sec:methods_physics}

To contextualise and enable an understanding of how the \chili models arrive upon their predicted evolutionary pathways for interior-atmosphere evolution, we outline commonly adopted parametrisations of key physical processes. 

The rheological properties of mantle-building materials are sensitive to their melting state, mineralogy, elemental composition, temperature-pressure conditions, and their incorporation of light elements \citep{petford_rheology_2003, karki_viscosity_2010, russell_models_2022, di_rheology_2026}. Importantly, since these materials comprise multiple components and mineral phases, two `rheological transitions' can occur as a parcel of material transitions from a liquid to solid state during magma ocean solidification  \citep{costa_rheology_2009, petford_rheology_2003}. Crystals initially form within a largely-molten fluid, while with continued cooling, melt instead becomes trapped within a largely-solidified crystalline structure . A transition in rheological properties (e.g. viscosity) occurs between 20--60\% melt fraction: firstly, a solid-liquid transition near $\Phi\approx50\%$, secondly a connectivity transition near $\Phi\approx7\%$ \citep{costa_rheology_2009, scott_the_2006}. These known sensitivities limit our understanding of Earth's deep interior, yet require some representation in these codes, since the rate at which mantle convection transports heat energy is related to its rheology \citep{Schaefer2018RSPTA, Miyazaki2019}. We may treat this microphysics with a single rheological transition around a critical melt fraction $\phi_c$. All models except LINCS and PROTEUS adopt a $\phi_c=0.4$ as the melt fraction representative of this transition \citep{solomatov_ToG_2015}. Correspondingly, the \chili models adopt various piecewise analytical relationships between mantle viscosity and melt fraction or temperature $T$.

Dynamic viscosity $\eta$ represents the resistance to deformation within a fluid, related to its kinematic viscosity and density as $\eta=\rho\nu$. A first-order approximation for the viscosity of (semi-)molten mantle building materials expresses $\eta$ as a function of its temperature $T$ through a generalised form of the Vogel–Fulcher–Tammann (VFT) equation,
\begin{equation}
    \eta_l = \eta_{0l} \exp \Big[A_l+\frac{B_l}{T-C_l}\Big],
    \label{eq:vft}
\end{equation}
where the constants are obtained empirically \citep{Nikolaou2019, ikeda_understandi_2013}. The constant $A_l$ may be non-zero to represent dissolved \ch{H2O} acting to decrease $\eta$ by disrupting crystal matrix polymerisation \citep{costa_rheology_2009,turcotte2002geodynamics,sossi_solubility_2023}. This equation can be corrected for the effect of solid cumulates suspended within the fluid through a melt fraction term \citep{roscoe1952viscosity}, applicable as $\Phi\rightarrow\Phi_c$,
\begin{equation}
    \eta = \eta_l \Big(1-\frac{1-\Phi}{1-\Phi_c}\Big)^{-2.5}
    \label{eq:vft_corr}
\end{equation}
although this correction represents a poor approximation as $\Phi\rightarrow0$ \citep{kobayashi_viscosity_2000, karki_viscosity_2010}. Models may transition to (or towards) an Arrhenius law at smaller melt fractions, below $\Phi_c$, where the viscosity is controlled by that of the solid matrix, which is weakened by residual melt \citep{Lebrun2013} as
\begin{equation}
    \eta = \eta_s \exp \left(-\alpha_n \Phi\right)
    \label{eq:arrh_melt}
\end{equation}
where $\alpha_n=26$ for diffusive creep. The viscosity of the solid matrix $\eta_s$ follows an Arrhenius law written for diffusive creep as
\begin{equation}
    \eta_s = \frac{\mu}{2A_s}\Big(\frac{h}{b}\Big)^{2.5} \exp \Big[\frac{E_s+PV}{R_g T}\Big]
    \label{eq:arrh}
\end{equation}
where $P$ is the pressure, $R_g$ is the gas constant, $E_s$ is an effective activation energy,$\mu$ is the shear modulus, $A_s$ is a pre-exponential factor, $h$ is the size of crystal grains, and $b$ is the Burgers vector length describing the crystal lattice structure. Equation~\ref{eq:arrh} applies in the low melt-fraction limit where the crystal matrix becomes well connected \citep{karato_rheology_1993, Lebrun2013, schaefer2016}. Variations upon these approaches are adopted by the \chili models, including where the activation volume $V$ may be set to zero \citep{marsh_on_1981}.

In the context of planetary magma oceans, convection is a multidimensional, often turbulent, energy transport process which acts across a range of spatial scales \citep{turner_convection_1986, siggia_high_1994}. Convection-resolving simulations are computationally prohibitive, so it must be parametrised within interior-atmosphere evolution models \citep{meier_mantle_2026, gillmann_atmosphere_2014}. In the boundary layer theory (BLT) of convection, rising parcels of fluid are assumed to cool adiabatically, so mantle temperature structures would follow an adiabat defined by a single potential temperature $T_\mathrm{pot}$ -- equal to the `true' temperature at a reference top-of-mantle pressure. That is, a mantle temperature profile can be derived by integration of
\begin{equation}
    \dd T/\dd P = \alpha T /( \rho c_p) 
    \label{eq:mantle_ad}
\end{equation} 
from the top-of-mantle downwards, with increasing pressure. $\alpha$ is the thermal expansion coefficient, $c_p$ the specific heat capacity, and $\rho$ is the density \citep{turcotte2002geodynamics}.

Under this fully-convective assumption, the heat flux arising from the mantle scales as a function of the temperature contrast and the Rayleigh number, Ra -- which characterises the relative importance of  buoyancy-driven  convective motion compared to the resistive effects of thermal and momentum diffusion \citep{turcotte2002geodynamics}. Ra can be used to indirectly quantify the mantle's transition from an initially turbulent convective regime towards laminar convective flow \citep{turner_convection_1986}. The relationship may be stated as
\begin{equation}
    F_c \propto \frac{\rho c_p \kappa_t}{L} (T_\mathrm{pot}-T_\mathrm{surf})\Big(\frac{\mathrm{Ra}}{\mathrm{Ra_{crit}}}\Big)^n
    \label{eq:fconv}
\end{equation}
in the soft turbulence limit, where the empirical exponent $n$ is often taken to be $1/3$ \citep{turcotte2002geodynamics, abe_thermal_1997, ElkinsTanton2008}. $L$ is the length scale over which the temperature gradient $T_\mathrm{pot}-T_\mathrm{surf}$ occurs (e.g., the mantle thickness) and $\kappa_t$ the thermal diffusivity. Additional terms and constants would be adopted in different flow regimes (e.g., $2/7$ in the hard turbulence limit) depending on the domain's aspect ratio: here, the ratio of mantle depth to planet radius \citep{siggia_high_1994, ahlers_heat_2009}. BLT parametrisations assume a relationship between the convective heat flux and the Rayleigh number, which is    proportional to the ratio of the mantle viscosity and temperature contrast to the mantle viscosity, 
\begin{equation}
    \mathrm{Ra}= \frac{\rho \alpha g}{\kappa_t \eta  }(T_\mathrm{pot}-T_\mathrm{surf}) L^3,
    \label{eq:rayleigh}
\end{equation}
where $g$ is the gravity. In this soft-turbulence regime, Ra is related to the Nusselt number (the dimensionless heat flux) through the empirical relationship $\mathrm{Nu}=0.089 \mathrm{Ra}^{1/3}$  \citep{solomatov_ToG_2015}.

An important consequence of BLT is the requirement of using scalar values for these dynamical quantities to represent the dynamics of the whole mantle; in reality, the mantle's properties would vary spatially subject to local temperature-pressure-composition conditions \citep{bower_numerical_2018}. We compare models' predicted values for representative `bulk' melt fractions $\Phi$ and viscosities $\eta$ in Section~\ref{sec:results_interior}. 

Following \citet{abe_thermal_1997}, it is commonplace to parametrise the melt fraction $\phi$ of mantle-building materials as a function of their temperature $T$ or specific entropy,
\begin{equation}
    \phi = \frac{T-T_s}{T_l-T_s},
    \label{eq:meltfrac}
\end{equation}
where $T_s$ and $T_l$ are the solidus and liquidus temperatures, for some pressure and assumed composition. Equation \ref{eq:meltfrac} can be volumetrically integrated across a region of the mantle bounded by the planetary radius, $R_p$, and a bottom boundary, $R_l$, to obtain the representative bulk mantle melt fraction,
\begin{equation}
    \Phi = \frac{3}{R_p^3-R_l^3}\int_{R_l}^{R_p} \phi(r) ~ r^2 dr,
    \label{eq:meltfrac_avg}
\end{equation}
such that BLT-adopting models obtain representative mantle viscosities using $\Phi$ within their preferred viscosity-melt fraction relationship. The \chili models adopt different choices for $R_l$, which represent different interpretations about the extent of the reservoir of molten mantle into which volatiles partition. Following the CUISINES approach \citep{Sohl24_cuisines}, models are not modified for the purposes of this intercomparison, so each model reports $\Phi$ according to its own `original' definition (Section~\ref{sec:methods_models}). These are the same melt fractions that they adopt in their calculations of rheology and volatile solubility. Comparisons at a given $\Phi$ (e.g. 5\%) should be interpreted with these differences in mind. %

All mantles are assumed to solidify from the core-mantle boundary (CMB) upwards as $T_\mathrm{pot}$ decreases and the corresponding adiabat intersects the liquidus and solidus curves at sequentially decreasing pressure levels.  As shown in \citet{walbecq_the_2025}, using bulk melt fraction $\Phi$ to evaluate whole-mantle rheological properties is approximately equivalent to taking a volume-average harmonic mean of these quantities across the magma ocean region \citep{Nikolaou2019}, although this equivalence is not exact in general, since viscosity is a strongly non-linear function of local melt fraction and temperature, and the harmonic mean is dominated by the least viscous layers.

It has been hypothesised that a conductive boundary layer (CBL) may form at the surface between magma ocean interiors and their overlying atmospheres \citep{ElkinsTanton2008, Lebrun2013, Salvador2023_SSR}. This theorised layer could allow $\sim100$\,K temperature discontinuities between the upper-mantle and surface brightness temperature, thereby acting to depress radiative cooling rates, subject to its thermal conductivity $k_b$ and thickness $d_b$ \citep{ElkinsTanton2008, Lebrun2013}. However, this thin layer is not readily spatially-resolved by models. Several \chili models use Fourier's law of thermal conduction to relate the mantle temperature $T_m$ to the surface brightness temperature $T_\mathrm{surf}$ through its conductive heat flux,
\begin{equation}
    F_b=k_b (T_m-T_\mathrm{surf})/d_b.
    \label{eq:cbl}
\end{equation}
Solutions for this equation are obtained by ensuring that $F_b$ equals the mantle's convective heat flux and the net radiative energy loss from the atmosphere (the outgoing longwave radiation, OLR, minus the absorbed stellar radiation, ASR;  $F_b=F_c=\mathrm{OLR}-\mathrm{ASR}$) from which $T_\mathrm{surf}$ is determined; models adopt various choices for $T_m$ and $d_b$ described in subsections below. The mantle temperature $T_m$ may be taken as equal to its potential temperature $T_\mathrm{pot}$.

All \chili models simulate the solubility of volatiles into these planetary magma oceans using solubility relationships between volatile's dissolved concentrations and their partial surface pressures \citep{abe_thermal_1997, Schaefer2018RSPTA, zahnle_evolution_1988}. This common approach treats ingassing and outgassing as a single equilibrium process \citep{ElkinsTanton2008}. In reality, volatile degassing occurs throughout a region below the atmosphere-interior interface, determined by a sufficiently low pressure at for  volatiles dissolved within ascending parcels to exsolve, whilst ingassing occurs by diffusion at the atmosphere-interior interface itself \citep{sparks_bubble_1978, gaillard_bubble_2021}. Approaching these processes robustly requires microphysical treatments; \citet{walbecq_the_2025} calibrated a numerical implementation within the PlanAtMO code using 3D fluid models and empirical measurements.

Treatment of how the potential incorporation of volatiles within largely-solidified deep mantle regions is more varied; two separate processes may be considered. Firstly,  through the presence of volatiles stored within remnant melt pockets entrained within solidified mantle regions as the turbulent magma ocean crystallises \citep{laurent_earths_2020, schiano_primitive_2003,hier_the_2017}. Secondly, through the direct equilibrium partitioning of volatiles between molten and the mineral phases in direct contact within a `mushy', partially solidified mantle region. In the latter case, \ch{OH^-} are considered to be directly incorporated within the mineral lattice \citep{hauri_partitionin_2006,bell_water_1992} rather than stored by dissolution into the molten regions. Equilibrium partitioning of volatiles is  parametrised with partitioning coefficients \citep{Lebrun2013, ElkinsTanton2008}. 

Mantle oxidation state is expected to evolve due to multiple physical effects which can compound and negate each other \citep{ Schaefer2017ApJ,lichtenberg_redox_2021, hirschmann2022magma}. We may proxy the mantle oxidation state using thermochemical redox reactions; e.g., the iron-w\"ustite buffer,
\begin{equation}
    \ch{2 Fe + O2 $\Leftrightarrow$ 2 FeO}.
    \label{eq:iw}
\end{equation}
This reaction is important and often adopted as a proxy because (i) iron atoms can take on different oxidation states, (ii) iron is abundant in planets due to its pile-up during stellar nucleosynthesis \citep{lodders_planetary_1998, kippenhahn2012stellar}, and (iii) oxygen is the most abundant element in rocky planets \citep{wang_the_2018, sossi_physicochem_2025}. For these reasons, the fugacity of oxygen in the upper-mantle ($f$\ch{O2}) can be used as a proxy for the mantle redox state, by its quantification relative to the $f$\ch{O2} value of the iron-w\"ustite buffer \citep{frost_chapter_2018}. This reaction is temperature dependent, so iron redox thermochemistry and differential melt-solid partitioning of iron (e.g. between ferric and ferrous states) may sequentially oxidise cooling magma oceans \citep{ kress_the_1991, Schaefer2024JGRE}. However, this temperature effect is potentially modulated by atmospheric escape of oxygen atoms, mantle dynamics, and core formation, which dynamically change the availability of iron and oxygen atoms in the mantle \citep{wade_core_2005, wordsworth2018redox, kasting_mantle_1993}. Critically, $f$\ch{O2} also controls chemical speciation in these models' outgassed atmospheric compositions because it is also treated as the partial pressure of \ch{O2} molecules; e.g. via \ch{2 H2 + O2 $\Leftrightarrow$ 2 H2O} and \ch{2 CO + O2 $\Leftrightarrow$ 2 CO2}. To enable a controlled comparison at simulation end points, all models' initial conditions are calibrated to achieve a final oxygen fugacity of IW+4\footnote{i.e., atmospheric oxygen fugacity $f$\ch{O2} being $10^4$ times larger than the $f\mathrm{O^{IW}_2}$ of the iron-w\"ustite buffer, \ch{2 Fe + O2 $\Leftrightarrow$ 2 FeO}, evaluated at the same temperature.}, which is representative of Earth's upper mantle at the present day \citep{rollinson_archaean_2017, frost_the_2008}.  None of the \chili models explicitly account for chemical interactions or partitioning of volatiles between the metallic (iron) core and planetary mantles -- the models treat the core as chemically inert, with a fixed 55\% interior radius fraction informed by Earth's interior structure \citep{lodders_planetary_1998}. 

\subsection{Description of the participating models}
\label{sec:methods_models}

The \chili models' configuration files and the resultant simulation data, in addition to intercomparison analysis scripts and plots, are freely available on \chili's  GitHub repository:\url{https://github.com/projectcuisines/chili/}. This GitHub repository is archived on Zenodo: \url{https://zenodo.org/records/20680020}  \citep{github_zenodo}. 

The seven evolutionary models compared in this paper are detailed in the following subsections Table~\ref{tab:models}. This suite differs from the \chili protocol by: (a) considering only explicit time-tracking codes, since static models are left for \chili Part III, (b) including NEONGOOEY as a successor to GOOEY, and (c) including MOAI, LINCS, and PlanAtMO as additional codes. 

\begin{figure*}[htb]
    \centering
    \includegraphics[width=0.97\linewidth]{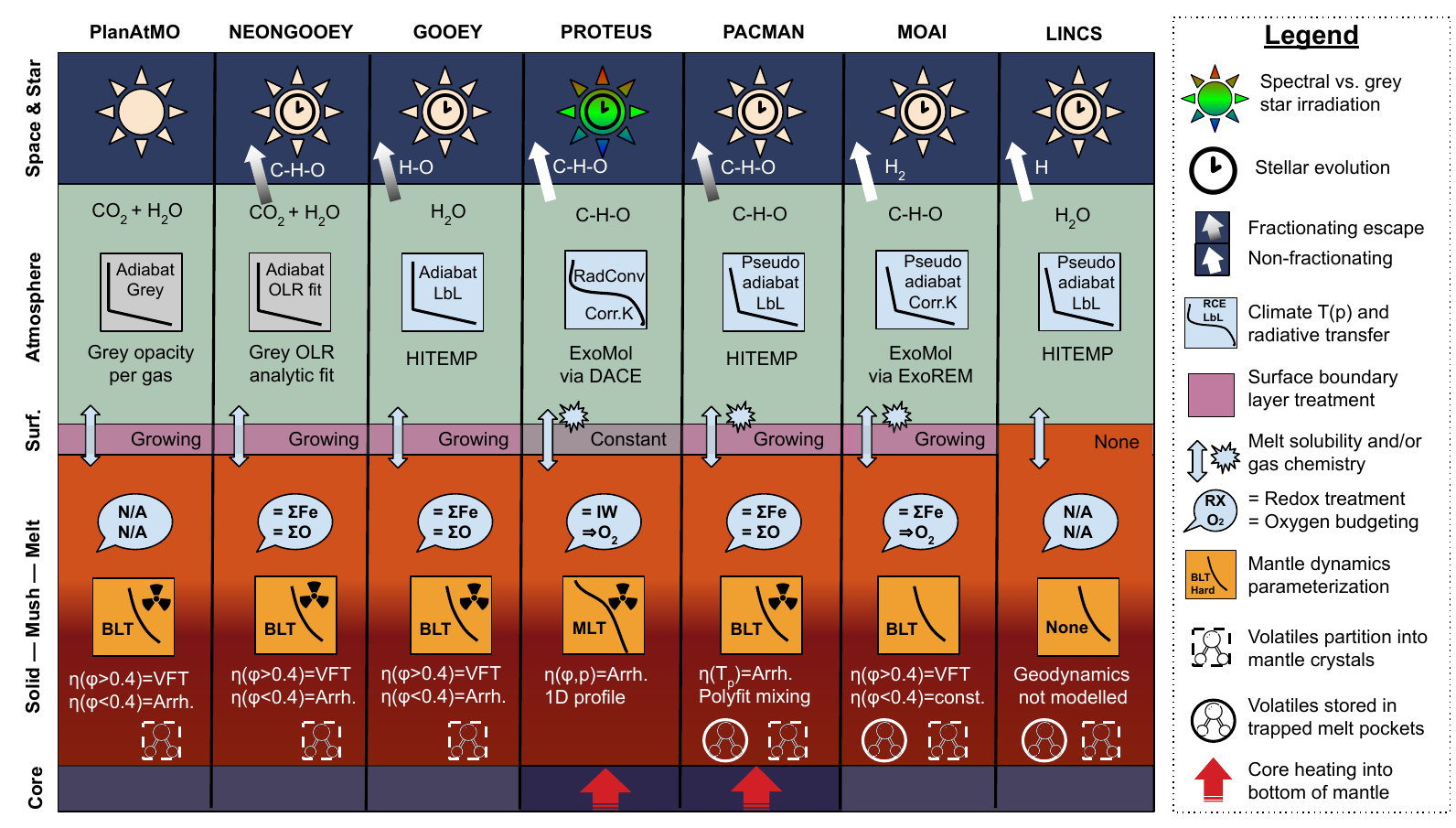}
    \caption{Cartoon of the time-resolving evolutionary models presented in this paper, highlighting key differences in their parametrisations. Various atmosphere profiles are adopted (blue and grey boxes). Radiative transfer is then applied through grey, correlated-k, or line-by-line (LbL) approaches. Redox treatments vary: some codes track multiple redox states of Fe ($\Sigma\mathrm{Fe}$), while others prescribe it (IW), related to whether they ensure conservation of oxygen ($\Sigma\mathrm{O_2}$), while others do not parametrise redox at all. Viscosity $\eta$ is commonly treated as a function of melt fraction $\Phi$ using analytic scalings.}
    \label{fig:models}
\end{figure*}

The coupled problem of magma ocean interior-atmospheric evolution across planetary time-scales and spatial-scales involves sensitive physical processes. The parametrisation of these processes within computer models -- which represents simplification upon reality -- is set by our choices balancing model complexity versus simplification, numerical expenditure, and empirical uncertainties. As a result of these challenges, the \chili codes adopt different physical treatments and are inhomogeneously more or less `complete' in their modelling, which enables diagnosis of the physics their predicted planetary evolution pathways. The cartoon in Figure~\ref{fig:models} visualises the interior-atmospheric-space domains parametrised by these models and symbolically indicates how the active physics within each domain is treated per-model. Similarly, Table~\ref{tab:models} references each codes' adopted melting solidus/liquidus curve, viscosity parametrisation, and implied mantle composition/mineralogy. 

\begin{table*}[htb!]
    \centering
    \caption{Models' adopted soldius and liquidus melting curves, viscosity treatments, and the implied mantle mineralogies/compositions used in this paper. }
    \begin{tabular}{ l l l l l l}
    Model & Solidus & Liquidus & Viscosity & Mineralogy/composition \\
    \hline
    PlanAtMO\footnote{\citet{walbecq_the_2025} -- \url{https://marcq.page.latmos.ipsl.fr/radconv1d.html}}  & \citet{Nikolaou2019} & \citet{Nikolaou2019}  & VFT to Arrhenius & Perovskite, lherzolite \\
    
    NEONGOOEY\footnote{C.~Ortiz \& R.~Ramirez -- \url{https://stars.library.ucf.edu/planetary-habitability-atmospheric-models/2/}} & Linearised \citet{hirschmann_mantle_2000} & Solidus + 600\,K & VFT to Arrhenius & Dry peridotite \\
    
    GOOEY\footnote{\cite{schaefer2016} -- \url{https://purl.stanford.edu/rk050tc3031}} & Linearised \citet{hirschmann_mantle_2000}  & Solidus + 600\,K & VFT to Arrhenius & Dry peridotite \\
    
    PROTEUS\footnote{\cite{lichtenberg2021} -- \url{https://proteus-framework.org}} & \citet{Andrault2011} & \citet{Hamano2013} & Arrhenius per layer & Dry enstatite \\
     
    PACMAN\footnote{\cite{Krissansen-Totton_2024} -- \url{https://github.com/joshuakt/Primary\_to\_Secondary}} & \citet{hirschmann_mantle_2000} + $P_\mathrm{surf}$ & Solidus + 600\,K & Arrhenius-polyfit &  Dry peridotite \\
    
    MOAI\footnote{\cite{maurice_redox_2023} -- \url{https://framagit.org/MaximeMaurice/MOAI}} & \citet{fiquet_melting_2010} &  \citet{fiquet_melting_2010} & VFT to constant & Dry peridotite \\
    
    LINCS\footnote{\cite{Hamano2015} -- Code not available online} & \citet{Hamano2013} & \citet{Hamano2013} & --- & Dry peridotite KLB-1 \\

    \label{tab:models}
    \end{tabular}
\end{table*}

\subsubsection{PlanAtMO}

PlanAtMO is a parametrised model for simulating the coupled interior-atmosphere evolution, cooling, and chemistry of early planetary mantles and climate states described in \citet{walbecq_the_2025}. PlanAtMO can optionally simulate non-equilibrium volatile outgassing \citep{walbecq_the_2025}, although this functionality is disabled for \chili's intercomparison. Therefore, similarly to the other \chili models, here PlanAtMO assumes that saturated volatiles within ascending magma ocean fluid parcels efficiently form bubbles that degas at the surface; the \ch{CO2} and \ch{H2O} are distributed between the atmosphere and the molten mantle regions in equilibrium using Henry's law solubility relationships (\citet{pan_the_1991} and \citet{holloway_volatiles_1994}, respectively). The model does not calculate mantle redox conditions and volatile trapping within melt pockets is neglected, however, the two volatiles are sequentially partitioned into solidified perovskite and lherzolite mineral phases as the mantle solidifies, using separate partition coefficients \citep{Lebrun2013}. Lherzolite forms at lower pressures and takes up volatiles into its crystal structure more efficiently than perovskite phases \citep{Schaefer2024JGRE}. 

The atmosphere is composed of \ch{CO2} and \ch{H2O}, in a 1\,bar background of \ch{N2}, and is assumed to be in radiative-convective equilibrium using the RADCONV1D atmosphere model \citep{Marcq2017}, where the atmosphere temperature profile follows a moist \ch{H2O} pseudoadiabat up to a 200\,K isothermal stratosphere. With this profile, grey radiative transfer calculations are performed to obtain the outgoing radiation flux, using opacities for \ch{CO2} and \ch{H2O} which are held constant, although these species' abundances evolve with time. Escape processes are neglected by PlanAtMO and stellar evolution is disabled. 

The thermal evolution of the planetary interiors is parametrised with BLT, following previous works, so PlanAtMO assumes that the magma ocean and the solidified mantle are adiabatic \citep{Lebrun2013,Salvador2017,Nikolaou2019}.  However, in contrast with these aforementioned studies, PlanAtMO considers that when a shallow magma ocean and underlying a mushy mantle coexist,  the heat loss at the surface of the planet only contributes to the cooling of the magma ocean and the mushy mantle regions do not cool below the rheological threshold.  \citet{walbecq_the_2025} found that this assumption is better justified by the considerable reduction of convective vigour in solid-state-like mushy mantles compared to liquid like magma ocean. This also results in considerably faster  magma ocean solidification compared to redistributing the heat loss at the surface to the cooling of the entire planet. PlanAtMO defines $R_l$ in Equation~\ref{eq:meltfrac_avg} equal to the core-mantle boundary radius. Since PlanAtMO neglects melt trapping, $\Phi$ does not include a trapped-melt contribution. The thermal contribution of the core is neglected, while latent heat during mantle solidification is accounted for. Mantle radiogenic heating follows the heat-producing element abundances given in \citet{Nikolaou2019}. The surface conductive boundary layer thickness dynamically evolves depending on the thermal state, subject to Fourier's law. The surface temperature is obtained by equating the magma ocean's convective heat flux (Equation~\ref{eq:fconv}), the conductive flux across this layer (Equation~\ref{eq:cbl}), and the net atmospheric energy loss \citep{Lebrun2013}. The mantle adiabat and its thermodynamic properties follow \citet{Nikolaou2019}. %

Viscosity is treated as a function of evolving potential temperature, surface pressure, and bulk melt fraction. For the local melt fractions $\phi>0.4$ within the convecting magma ocean probed by PlanAtMO in this work, the VFT equation (Equation~\ref{eq:vft}) is used with $\eta_{l0}=2.4~10^{-4}$~Pa~s, $B_l=4600$~K and $C_l=1000$~K \citep{karki_viscosity_2010} with the correction of Equation~\ref{eq:vft_corr}. PlanAtMO considers no  matrix compaction or melt percolation processes in partially molten regions, whose melt fractions are above 40\%. 

The solidus and liquidus follow piecewise experimental melting curves depending on the pressure \citep{Nikolaou2019, walbecq_the_2025}. PlanAtMO simulations terminate when the melt fraction at the surface temperature-pressure conditions correspond to a melt fraction of 40\%, through Equation~\ref{eq:meltfrac}.

\subsubsection{GOOEY}

GOOEY model was presented in the \chili protocol paper and is described by \citet{schaefer2016}. This model simulates the outgassing of pure-\ch{H2O} atmospheres, neglecting the contribution of carbon-bearing species. \ch{H2O} dissolves into the magma ocean through a Henry's law fit to \citet{papale_modeling_1997}; valid for $T>1000$\,K and $P<1$\,GPa. In addition to solubility of \ch{H2O} into the magma ocean, GOOEY parametrises its equilibrium partitioning at the interface between the magma ocean and solidified mantle regions using a constant equilibrium partitioning coefficient $k_\mathrm{H2O}=0.01$. Fractionating energy-limited escape of O-H outflow is modelled using binary diffusion coefficients to determine O and H mass fluxes \citep{zahnle_mass_1986}. Oxygen molecules implicitly formed by this fractionation act to oxidise the magma ocean containing FeO droplets, thus, GOOEY tracks the planet's evolving oxygen budget and its effect on the redox state of iron within the mantle \citep{kress_the_1991}.

The atmospheric temperature profile is assumed to follow a dry adiabat from the surface, up to an isothermal stratosphere at the radiative skin temperature of the planet \citep{pierrehumbert_principles_2010}. Based on these atmospheric profiles, GOOEY interpolates pre-tabulated values for the outgoing longwave radiative flux from the atmosphere, which are derived from line-by-line radiative transfer calculations using HITEMP2010 opacities for line-absorption \citep{Rothman2010} and MT\_CKD opacities for continuum absorption \citep{clough_line_1989}.  The bolometric stellar irradiation follows \citet{Baraffe2015} evolution tracks; the X-ray and ultraviolet component is assumed equal to 0.1\% of the total stellar luminosity and factors into the atmospheric escape model. 

Mantle convection is parametrised in GOOEY with boundary layer theory \citep{lay_coremantle_2008}. The mantle cools through a conductive surface boundary layer into the bottom of the atmosphere, whose thickness $d_b$ evolves with time according to Equation~\ref{eq:cbl}. The surface temperature is evolved in time alongside the potential temperature, subject to the balance between this conductive flux and the net atmospheric energy loss \citep{schaefer2016}. The mantle is uniformly heated by radioactive decay and the latent heat released by mantle solidification. The mantle adiabat is calculated assuming a constant specific heat capacity of 1200\,J\,kg$^{-1}$\,K$^{-1}$, a thermal expansivity of $2\times10^{-5}$\,K$^{-1}$, and a uniform mantle density. %
The GOOEY liquidus is 600\,K hotter than the solidus temperature, which is evaluated dynamically as the surface overburden pressure evolves, evaluated using a linearisation of the \citet{hirschmann_mantle_2000} dry peridotite solidus.  The bulk melt fraction $\Phi$ is calculated through Equation~\ref{eq:meltfrac_avg} -- which is evaluated using Equation~\ref{eq:meltfrac} with $T_\mathrm{pot}$ against the solidus and liquidus temperatures -- having a radial extent considering a dynamic $R_l$ defined by the base of the magma ocean, where the local melt fraction exceeds zero. GOOEY does not model trapped melt. The magma ocean's dynamic viscosity follows the VFT law (Equation~\ref{eq:vft}) with the constants of \citet{karki_viscosity_2010}, evaluated at $T_\mathrm{pot}$ and corrected for suspended crystals through Equation~\ref{eq:vft_corr} using the melt fraction implied by $T_\mathrm{pot}$ at the surface. When this melt fraction falls below $\phi_c$, it follows an Arrhenius law (Equation~\ref{eq:arrh}) \citep{Abe1985}. GOOEY simulations terminate when the surface cools to the solidus temperature (1420\,K at zero pressure).  

\subsubsection{NEONGOOEY}

NEONGOOEY is a FORTRAN successor to GOOEY which additionally accounts for magma ocean outgassing of \ch{CO2} and an initial 10\,bar primordial \ch{H2} envelope. The NEONGOOEY rewrite achieves this by modifying the GOOEY atmospheric escape framework, following the elemental fractionation treatment described by \cite{zahnle1990} and \cite{odert2016}, based on binary diffusion within the thermosphere. This updated model also adjusts the calculated outgoing longwave radiation fluxes by using an analytical OLR calculation -- which depends on the gaseous composition and surface temperature -- following the grey gas approach described by \cite{carone2025}. 

The mantle heat flux, bulk melt fraction, melting curves, viscosity parametrisation, and termination criteria are all implemented in NEONGOOEY identically to GOOEY. We present simulation results from both NEONGOOEY and GOOEY in this work to test the effects of their modelling differences. GOOEY is retained from the \chili protocol paper, although its known radiative transfer error (Section~\ref{sec:results_radtrans}) has been corrected in NEONGOOEY; differences arising from this error should not be interpreted as physical uncertainty. The NEONGOOEY simulations terminate when the surface temperature reaches the solidus temperature.

\subsubsection{PROTEUS}

PROTEUS is a modular framework for simulating planetary evolution \citep{lichtenberg2021, nicholls_redox_2024} which participated in the initial intercomparison presented within the \chili protocol paper \citep{chili_protocol_2025}. Two important differences between PROTEUS and the other \chili models are that its interior-atmosphere states are modelled with 1D-resolved approaches, and that it models interior evolution in terms of specific entropy rather than temperature. 

Here, PROTEUS simulates the partitioning of C-H-O volatiles between the atmosphere and magma ocean regions, while neglecting their distribution into the deep solidified mantle. Importantly, PROTEUS prescribes the surface oxygen fugacity relative to a temperature-dependent implementation of the iron-w\"ustite reaction \citep{oneill_the_2002}, rather than explicitly tracking oxygen atoms, so oxygen atoms are not conserved throughout its simulated evolution. Atmospheric escape is parametrised as a non-fractionating process, so the escaping mass fluxes of C-H-O inventories are proportional to their outgassed mass-mixing ratios. The hydrodynamic energy-limited escape rate is determined subject to a spectroscopically evolving stellar irradiation spectrum \citep{gueymard_proposed_2002, johnstone2021active, Baraffe2015}. 

At each time-step of the PROTEUS evolutionary calculations, it obtains an energy-conserving solution for the atmosphere climate state using AGNI: a  radiative-convective-chemical atmosphere model \citep{nicholls_convective_2025, nicholls_agni_2025}. Rather than prescribing a pseudoadiabtic profile, AGNI self-consistently accounts for the potential formation of non-convective atmosphere layers which modulate the outgoing radiation fluxes \citep{selsis_cool_2023, cmiel2025}. It parametrises atmospheric convection  using canonical dry mixing-length theory \citep{joyce_a_2023}. Radiation fluxes are obtained using a 48-band correlated-$k$ two-stream scheme \citep{manners_a_2024}. The absorption coefficients of gas species are obtained from the DACE database cross-sections, which are derived from ExoMol and HITEMP linelists \citep{grimm_helios_2021,Rothman2010, tennyson_exomol_2012}. Continuum absorption is treated with MT\_CKD for \ch{H2O} \citep{Mlawer2012} and with the HITRAN continuum tables otherwise \citep{Gordon2022}. Atmosphere profiles are presented in Appendix~\ref{app:climate}.

PROTEUS time-steps 1D mantle dynamics using the SPIDER code \citep{bower_numerical_2018}. This model adopts a mixing-length convection scheme, rather than BLT, so regions of the mantle where convection becomes sluggish are permitted to become sub-adiabatic \citep{abe_thermal_1997, turner_convection_1986}. SPIDER simultaneously accounts for mantle heat transport fluxes from gravitational settling of crystals and melt-solid phase separation, latent heating from solidification -- these flux terms represent the energetic effects of early crystal settling through molten regions, and drainage of melt through largely-solidified regions at later (cooler) conditions.  It also includes thermal conduction, radiogenic heating, and  core cooling \citep{bower_numerical_2018}. The melt fraction $\phi$ is calculated locally at each mantle layer using the lever rule in an entropy formalism, rather than as a function of temperature. Rheological properties of the partially-molten `mush' phase and solid phase are evaluated separately by SPIDER at each mantle layer, and then combined volume-additively using a harmonic mean weighted by the local melt mass-fraction \citep{bower_numerical_2018}. Viscosities are obtained for each layer and phase using an Arrhenius law (Equation~\ref{eq:arrh}) with a pressure-dependent activation volume \citep{rozel_continental_2017}, and corrected to an Earth-like Mg/Si ratio \citep{wolf_eos_2018}. SPIDER's bridgmanite melting curves are derived from a hybrid of empirical measurements and molecular dynamics simulations \citep{stixrude_thermodynam_2009, mosenfelder_the_2009, Andrault2011, Hamano2013}. The thermodynamic properties of the melt phase are obtained from the MgSiO$_3$ equation of state of \citet{wolf_eos_2018}. Alongside its mixing-length treatment of mantle dynamics which returns 1D mantle temperature profiles, PROTEUS also treats the surface conductive boundary layer differently, by taking a constant thickness $d_b=1$\,cm and conductivity $k_b=2$\,W\,m$^{-1}$\,K$^{-1}$, and setting the mantle temperature $T_m$ equal to the topmost node of the mantle temperature profile. At each coupling time-step, AGNI solves for the $T_\mathrm{surf}$ at which the conductive flux through this layer (Equation~\ref{eq:cbl}) equals the net upward energy flux at the base of the atmosphere. SPIDER then uses this flux as its upper boundary condition. The bulk melt mass-fraction $\Phi_m$ is defined across the whole mantle by setting $R_l$ equal to the core mantle boundary, and includes all melt present in each mantle layer, since PROTEUS does not model a separate trapped-melt reservoir. %
In this study, PROTEUS simulations terminate when $\Phi_m \leq 5\mathrm{\,wt\%}$. The melt mass-fraction is converted to a volumetric-fraction for intercomparison with the other CHILI models.

\subsubsection{PACMAN}

PACMAN is an open source planetary evolution model which also featured in the \chili protocol paper. This code simulates outgassed atmospheric compositions comprising CO$_2$, CO, CH$_4$, H$_2$, H$_2$O, and O$_2$; accommodating arbitrary C-H-O volatile compositions and graphite condensation \citep{Krissansen-Totton_2024, krissansen2022predictions}. Equilibrium atmosphere-melt dissolution, solid-melt partitioning, and dynamic melt trapping via compaction within the solidified mantle \citep{hier_the_2017, Sim2024JGRE}, together distribute C-H-O inventories between the atmosphere, molten interior, and solid phases of the planet. Dissolved volatiles include \ch{CO2}, \ch{H2}, and \ch{H2O}: assumed to be well mixed in the magma ocean, with a mantle composition based on Bulk Silicate Earth \citep{iacono_new_2012, hirschmann_solubility_2012}. PACMAN treats atmospheric escape as being diffusion-limited or energy-limited, depending on the atmospheric composition and shortwave irradiation exposure, and includes the relative fractionation of C-H-O atoms.  The iron redox state and oxygen fugacities of the solid and molten mantle regions co-evolve according to thermal and compositional effects \citep{kress_the_1991}; calibrated here with an initial FeO mole fraction of 0.06 to satisfy IW+4 from the \chili protocol. 

Outgoing longwave radiation fluxes from the pseudoadiabatic atmosphere are computed using a correlated-$k$ method \citep{Molliere2019AA}. In addition to line lists for H$_2$O, CO$_2$, and CO \citep{Rothman2010}, we include continuum opacities for H$_2$O–H$_2$O, CO$_2$–CO$_2$, H$_2$–H$_2$, and CO-CO collisionally induced absorption  \citep{Molliere2019AA}.

Convective heat fluxes from the mantle are treated using boundary layer theory in the soft turbulence regime. The mantle adiabat is therefore assumed to extend from the surface boundary layer to the core-mantle boundary, and is linear in radius with a constant specific heat capacity of 1200\,J\,kg$^{-1}$\,K$^{-1}$ and thermal expansivity of $2\times10^{-5}$\,K$^{-1}$. The time-evolution of mantle potential temperature $T_\mathrm{pot}$ is determined by equations representing the magma ocean and solid-state convection phases (a formulation of Equation~\ref{eq:fconv}): governed by the balance of heat loss from the surface (modulated by atmospheric opacity), latent heat of fusion during magma ocean solidification, an imposed metallic core heat flux, and radiogenic heat production in the mantle (\citet{krissansen-totton2021}~Supplemental\,Materials\,A.1). At each step, $T_\mathrm{surf}$ is solved such that the convective heat flux equals the net atmospheric energy loss. Treatment of the surface CBL follows \citet{Lebrun2013} and \citet{schaefer2016}, where the boundary layer thickness evolves in time (Equation~\ref{eq:cbl}). The PACMAN solidus curve is a smooth function between a linear fit to the solidus for low pressure dry peridotite, and a linear fit to the solidus for the high-pressure lower mantle \citep{hirschmann_mantle_2000, schaefer2016}, evaluated at pressures which account for overburden of surface volatiles. Following \citet{schaefer2016}, the liquidus is 600\,K warmer than the solidus at all pressures. 

Mantle viscosity $\eta$ is calculated as a function of $T_\mathrm{pot}$, where pure-solid and pure-melt-phase viscosities are evaluated at the solidus and liquidus temperatures using an Arrhenius law. These end-member viscosities are combined using a polynomial fit within the mixed-phase region, constructed to be continuous at these boundaries. This $\eta$ parameterisation follows a variation of  \citet{Lebrun2013,Salvador2017,schaefer2016} modified to ensure that Earth’s modern mantle potential temperature does not result in a magma ocean or mush-state mantle. The bulk melt fraction $\Phi$ reported for PACMAN is the whole-mantle volume average of local melt fractions given by Equation~\ref{eq:meltfrac} along the mantle adiabat, with $R_l$ in Equation~\ref{eq:meltfrac_avg} at the core-mantle boundary. Since $\phi$ depends only on temperature, melt trapped within the solidified mantle is not included in $\Phi$. %
As a result, a critical melt fraction is implicitly derived from the assumed melting curves and the temperature-dependent viscosity parametrization (which is chosen to recover behaviour of \citet{costa_rheology_2009}). PACMAN simulations terminate when the surface temperature reaches a 1473\,K solidus temperature.

\subsubsection{MOAI}

Similarly to the other evolutionary codes, the MOAI model employs a two domain interior-atmosphere approach. Assuming that the part of the magma ocean where the melt fraction is above the rheological transition to be well mixed and equilibrated with the atmosphere at the surface P-T conditions, outgassed compositions are computed by solving thermochemical equilibrium of 12 species including \ch{CO}, \ch{CO2}, \ch{CH3}, \ch{HCN}, \ch{O2}, \ch{H2}, and \ch{H2O}. MOAI simultaneously models how these species dissolve into the magma ocean using the solubility laws provided in Table\,A.1 of \citet{maurice_MOAI_2024}. A chemical and solubility solution is obtained by requiring conservation of H, C, N and S, and  the buffering of \ch{O2} fugacity by  equilibrium between FeO and Fe$_2$O$_3$ using the parametrization from \citet{hirschmann2022magma}. MOAI neglects compaction of melt pockets which may occur during complete crystallisation, so the solidified deep mantle stores a volatile content corresponding to the content of the trapped melt, in addition to that accommodated in the solid crystals at equilibrium.  The redox evolution is tracked by accounting for differential solid-melt compatibility of Fe$^{3+}$ and Fe$^{2+}$ states into mineral phases \citep{Schaefer2024JGRE}. A minimal bulk silicate planet concentration of 1\,ppm is set for nitrogen and sulfur abundances. Atmospheric escape of \ch{H2} is treated standalone; oxygen and carbon are retained within the planet.

MOAI constructs hydrostatic atmospheric temperature profiles following either a dry adiabat or a moist \ch{H2O} pseudo-adiabat, and then performs radiative transfer using the exo\_k code \citep{leconte_spectral_2021} to obtain the outgoing radiation fluxes from correlated-$k$ tables for \ch{H2O}, \ch{CO}, \ch{CO2} and \ch{CH4} from the ExoREM model. Rayleigh scattering is enabled in MOAI's radiative transfer calculations, unlike for the other \chili models. 

The model treats the magma ocean dynamics with BLT, assuming that it comprises a vigorously convecting liquid layer the `soft' turbulence regime \citep{solomatov_ToG_2015}, where the temperature profile follows an adiabat. Regions of the mantle which solidify are assumed to retain a local melt fraction $\phi$ corresponding to the rheological transition ($40\%$), which no longer equilibrates with the atmosphere and freezes \textit{in situ} upon further cooling. The bulk melt fraction $\Phi$ is defined through Equation~\ref{eq:meltfrac_avg} with $R_l$ at the core-mantle boundary. Retained melt contributes to $\Phi$ as $\Phi$ decreases below 40\% to reach the 1\% termination criterion. The mantle's density is evaluated using a linear fit of the liquid \ch{MgSiO3}, a constant heat capacity of 1000\,J\,kg$^{-1}$\,K$^{-1}$, and a thermal expansivity following the VFT relation \citep{abe_thermal_1997}. The effect of phase change on the density, heat capacity and thermal expansivity are neglected. The MOAI melting curves are obtained from \citet{fiquet_melting_2010} and do not evolve with time. The mantle mineralogy is calculated dynamically according to the evolving temperature-pressure conditions \citep{Miyazaki2019}.

The viscosity of the magma ocean follows a piecewise temperature- and $\Phi$-dependent VFT relation, following \citet{Nikolaou2019}. When the local melt fraction $\phi<40\%$, viscosity is held constant to a solid phase end-member value of $10^{21}$ Pa\,s. The characteristic viscosity governing the convective heat flux of the magma ocean (Figure~\ref{fig:interior}d) is that of the convecting region above the rheological front, so it remains far below this value. The surface temperature is related to the mantle potential temperature through a thin conductive  boundary layer, with $T_\mathrm{surf}$ solved by equating the convective heat flux to the net radiative energy loss, following \citet{Lebrun2013}. %
Heat from the metallic core and radioactive decay are neglected, but latent heat release from mantle phase change is included. The termination criterion for MOAI is when the whole-mantle melt fraction $\Phi\leq1\%$.

\subsubsection{LINCS}

The LINCS (Linked INterior-atmosphere Co-evolution System) model simulates the coupled evolution of the rocky interior and the atmosphere during magma ocean solidification, based on the numerical framework developed by \citeauthor{Hamano2013} (\citeyear{Hamano2013}, \citeyear{Hamano2015}). The partitioning of \ch{H2O} between the atmosphere and the magma ocean is assumed to be in solubility equilibrium at the planetary surface. The water solubility is determined by a polynomial function of pressure fitted to data calculated using the model of \citep{papale_modeling_1997} with a primary N-MORB composition \citep{Presnall&Hoover1987} at 2000~K. Atmospheric escape of hydrogen is treated within the energy-limited regime with heating efficiency of 10\%. To align with the \chili protocol, LINCS was modified to compute the absorbed stellar radiation based on the specified Bond albedo.

The atmosphere consists solely of \ch{H2O} and follows a dry or moist pseudo-adiabat with no carbon-bearing species incorporated into the simulations. The outgoing thermal radiation is determined via line-by-line calculations under Earth's gravity, using HITEMP 2010 \citep{Rothman2010} for line opacity and MT\_CKD2.5 \citep{Mlawer2012} for continuum opacity. These radiation data are tabulated as a function of surface temperature and pressure, and then interpolated at each time step.  For the Venusian cases, the outgoing thermal radiation data are corrected to account for the difference in gravitational acceleration. 

The LINCS version used in this comparison project does not model mantle geodynamics; the heat flux from the interior is the difference between the outgoing planetary radiation and the absorbed stellar radiation, plus the latent heat of fusion from mantle solidification. LINCS does not require a viscosity parametrisation; the effects of magma and mantle viscosity are implicitly incorporated into the heat capacity of the magma ocean based on the assumed thermal structure. Similarly to models adopting BLT, LINCS approximates the magma ocean regions with an adiabatic temperature profile characterized by a single potential temperature, while the underlying solidified mantle temperature is assumed to remain constant at the solidus. This treatment effectively corresponds to a choice of critical melt fraction $\Phi_c=0$ and an $R_l$ at the core-mantle boundary. The heat capacity of the entire mantle was derived as a function of potential temperature, calculated from the enthalpy change relative to its incremental variation. The adiabatic mantle profile is calculated based on the parametrisation by \citet{abe_thermal_1997}. The mass melt fraction in each layer assumed to depend linearly on temperature between the liquidus and solidus. The solidus and liquidus temperatures follow \citet{Hamano2013}, which assumes a bulk mantle composition of peridotite KLB-1. The liquidus curve up to 30~GPa is fitted to experimental data grouped by liquidus phase. Above 30~GPa, the liquidus curve was obtained as a regression curve using the melting experimental data of peridotite KLB-1, where the melting data of periclase and perovskite were first fitted using the Lindemann law and their activity coefficients were treated as parameters. The solidus curve below 22.5~GPa adopts the compilation by \citet{Herzberg2000}, and a polynomial fit is used above 22.5~GPa. All the experimental data and specific references used to derive these liquidus and solidus curves are detailed in the Supplementary Information of \citet{Hamano2013}. 

Upon solidification, LINCS considers a constant 1\% of interstitial melt trapping and melt-solid partitioning of water. Percolation and compaction in the mixed-phase regions are neglected. LINCS simulations terminate when the surface temperature reaches the 1,370\,K solidus temperature -- corresponding to a bulk mantle melt fraction of 0. The trapped interstitial melt is therefore not included in $\Phi$.

\section{Results} 
\label{sec:results}

\subsection{Nominal Earth and Venus cases}
\label{sec:results_earthvenus}
\begin{figure}
    \centering
    \includegraphics[width=\linewidth]{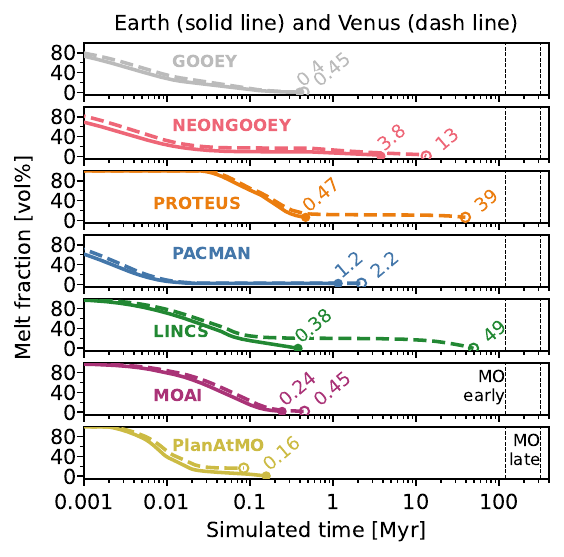}%
    \caption{
    Cooling and solidification of Nominal Earth and Venus cases (solid and dashed lines). Each panel plots melt fraction versus simulated time, calculated by each model (panels). Simulation end points are indicated by circular markers. Time is measured relative to some fully-molten state. Taking $t=0$ as Earth's final global re-melting event, e.g. the Moon-forming impact, vertical black lines correspond to two estimates for Earth's age when its mantle finally solidified \citep{sole_hadean_2025, cavosie_magmatic_2005}. 
    }
    \label{fig:earth_venus}
\end{figure}

Figure~\ref{fig:earth_venus} compares nominal Earth and Venus evolution tracks of volumetric mantle melt fraction, calculated by each model (panels). The time $x$-axis represents the simulated time from an initially molten state, such as following a particular giant impact event \citep{abe_thermal_1997}. All of these Nominal Earth cases qualitatively agree by cooling and terminating (solid scatter points) before two empirical estimates of maximum magma ocean solidification time from surviving Hadean rocks and zircons (vertical dashed lines; \cite{sole_hadean_2025}), with NEONGOOEY taking the longest time to solidify (3.8\,Myr). 

Venus cases (dashed lines) typically have longer solidification timescales than Earth; its enhanced irradiation flux offsets thermal emission to inhibit planetary cooling. However, the Earth-Venus evolutionary divergences vary between the models (solid vs dashed lines).  Simulations of Nominal Venus, except by MOAI and GOOEY, enter into states of global radiative equilibrium  in which stellar radiation absorption balances thermal emission, and the melt fraction becomes approximately constant in time \citep{van_onset_2025, nicholls_redox_2024}. We may classify these as `Type II' planet behaviours in contrast to the `Type I'  scenarios which continually cool to the point of solidification \citep{Hamano2013}. During these periods of radiative equilibrium, mantle melt fraction is approximately constant and the simulated Venusian primordial magma oceans are extended for up to 49\,Myr (dashed green line). Type II scenarios eventually solidify after long-term atmospheric escape processes sufficiently strip the atmospheric greenhouse.  LINCS and PROTEUS (green and orange lines) find markedly different evolutionary pathways between their Nominal Venus and Nominal Earth scenarios: predicting  Type I behaviour with early solidification for Earth, and Type II behaviour with late solidification for Venus. Magma oceans facilitate efficient interior-atmosphere exchange of volatiles, so Type II scenarios with extended magma ocean periods could offer greater exposure of chemical (e.g. H content) and isotopic signatures (e.g. D/H) to atmospheric escape processes \citep{grinspoon_implication_1993}. The modelling approaches which give rise to these different outcomes are identified in later sections.

\subsection{Earth evolution following post-formation outcomes}
\label{sec:results_grid}

\begin{figure}
    \centering
    \includegraphics[width=0.9\linewidth]{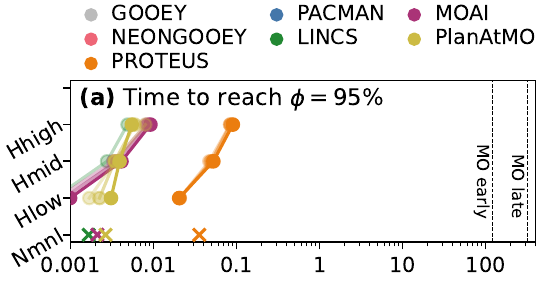}%
    \vspace*{-1mm}
    \includegraphics[width=0.9\linewidth]{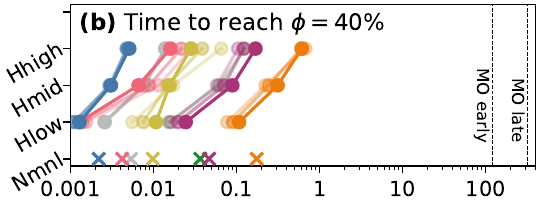}%
    \vspace*{-1mm}
    \includegraphics[width=0.9\linewidth]{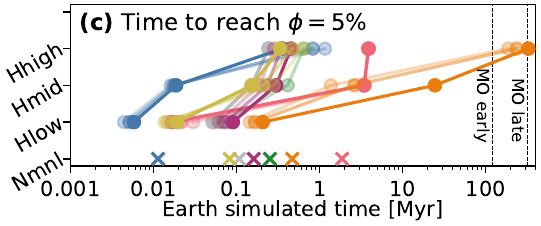}%
    \caption{
    Time taken for Earth scenarios to reach melt fractions of \textbf{(a)} 95\%, \textbf{(b)} 40\%, and \textbf{(c)} 5\%. Cooling timescales are sensitive to planetary inventories of hydrogen (y-axes) and carbon (marker opacity), for a given model (marker colour). Connected scatter points span three hydrogen inventories, for a given carbon inventory. Time is measured relative to some fully-molten state. Taking $t=0$ as Earth's final global re-melting event, e.g. the Moon-forming impact, vertical black lines correspond to two estimates of maximum magma ocean solidification time from surviving Hadean rocks and zircons \citep{sole_hadean_2025, cavosie_magmatic_2005}. 
    }
    \label{fig:cooling_time}
\end{figure}

Our comparisons between Nominal Earth and Venus show a range of divergent thermal evolution pathways for these two planets, depending on the model applied to simulate them. Since C-H-N-O-S volatile budgets are expected to vary between planets \citep{Krijt2023ASPC5341031K,Drazkowska2023ASPC534717D}, and chemistry is suggested to exert a leading-order control over planetary cooling timescales \citep{nicholls_redox_2024}, we investigated how these  models respond to our grid of initial H and C inventories. Figure~\ref{fig:cooling_time} plots the simulated time taken for our grid of Earth scenarios to cool to three distinct melt fraction thresholds: 95\% (\textbf{a}, mostly molten), 40\% (\textbf{b}, rheological transition), and 5\% (\textbf{c}, mostly solid). Circular scatter points show different hydrogen and carbon inventories, across our grid of parameters, compared to the Nominal Earth scenario (crosses).

For a given model (scatter colour), thermal evolution timescales correlate with the hydrogen inventory ($y$-axis): Hhigh cases take longer to reach all melt fraction thresholds (panels) compared to Hmid and Hlow scenarios. This known correlation between hydrogen content and cooling rate has been well established in the literature since additional volatiles (i) introduce a stronger atmospheric blanketing effect \citep{nicholls_redox_2024, Hamano2015}, and (ii) are sustained against escape processes for a longer time \citep{Hamano2013, schaefer2016}. Variations in carbon inventory (marker opacity) have a secondary effect to the hydrogen inventory, which is also an established trend \citep{nicholls_redox_2024, Miyazaki2019}, and explained by \ch{H2O} and high-pressure \ch{H2} having a stronger greenhouse potential than \ch{CO2} and \ch{CO} \citep{lichtenberg2021, Wordsworth2013, ramirez_warming_2014}.

The simulated time required for Earth to reach a given melt fraction (Figure~\ref{fig:cooling_time} panels) is more sensitive to the model applied (scatter colour) than to the volatile inventory considered (scatter opacity). An important corollary being that any inferred thermal histories (e.g. from the estimated age of ancient mineral samples) from initial volatile budgets is currently challenging, because it is more strongly dependent on model selection than on the inferred parameters.  

Non-linear cooling sequences are common across the models (Figure~\ref{fig:cooling_time}). Most models initially cool rapidly, with PROTEUS being the slowest to reach 95\% melt fraction (panel a), although this still occurs within 0.1\,Myr of simulated time. PROTEUS' departure from the other models, in this manner, is explained by its mantle dynamics treatment (Section~\ref{sec:results_interior}) and mantle melting curves (Appendix~\ref{app:melting_curves}). The time taken to reach 40\% melt fraction is also rapid on geologic timescales: less than 0.6\,Myr across all cases (panel b). However, the thermal evolution stalls at these partially-molten states near the rheological transition between 30--40\% melt fraction, across all models \citep{costa_rheology_2009, Abe1985}. This commonality arises because partially molten `mushy' magma ocean layers have increased viscosities with lower convective heat fluxes \citep{Lebrun2013, schaefer2016}.

In the following sections, we use diagnostic variables derived from Nominal Earth and Venus simulations to explain these different cooling timescales, which arise from the key physics that shapes these early stages of planetary thermo-compositional evolution.

\subsection{Volatile thermochemistry and partitioning}
\label{sec:results_chem}
\begin{figure}
    \centering
    \includegraphics[width=\linewidth]{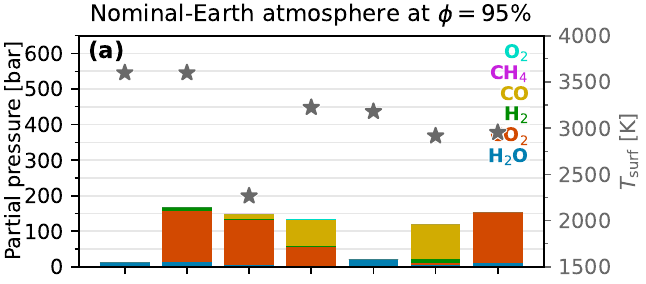}%
    \vspace*{1mm}
    \includegraphics[width=\linewidth]{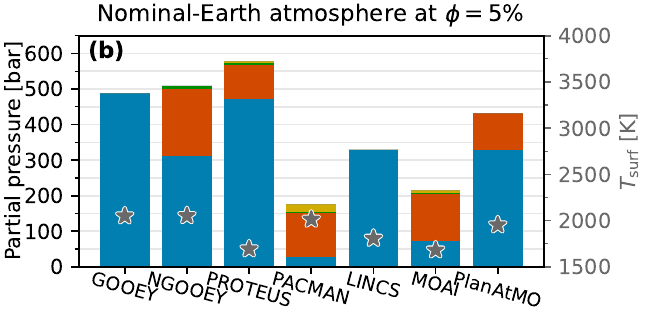}%
    \caption{
    Atmospheric compositions for Nominal Earth mantle melt fractions of \textbf{(a)} 95\% and \textbf{(b)} 5\%. For each model (x-axis), gas partial pressures [bar] are shown by stacked bar charts (colours), alongside surface temperature (grey stars). Models do not simulate the same set of gas species; e.g. by neglecting \ch{CO2}.
    }
    \label{fig:chem_earth}
\end{figure}

Cooling timescales and the emergence of Type I/II behaviours are sensitive to atmospheric mass and composition, because energy released from the planet's interior is largely blanketed by infrared-opaque atmospheric species \citep{nicholls_redox_2024, lichtenberg2021, zahnle_earths_2010}. The diversity in inter-model cooling timescales shown in Figures~\ref{fig:earth_venus}~and~\ref{fig:cooling_time} potentially reflects their range of calculated atmospheric compositions. To test this, Figure~\ref{fig:chem_earth} plots outgassed atmospheric compositions (bars) and surface temperatures (stars) for simulations of Nominal Earth during its early molten state (panel a) and nearly solidified state (panel b). 

Most models predict early \ch{CO}-\ch{CO2} dominated atmospheres (yellow and orange bars, panel a) because early molten silicate readily dissolves hydrogen bearing \ch{OH-} ions, while outgassing C-rich atmospheres \citep{sossi_solubility_2023, ElkinsTanton2008, bower_retention_2022}.  The total surface pressures of C-rich models (NEONGOOEY, PROTEUS, PACMAN, MOAI, PlanAtMO) show good quantitative agreement, with variation in total surface pressure from $\sim120$ to $150$\,bar.  For Nominal Earth at volumetric melt fraction $\Phi=95\%$ (Figure~\ref{fig:chem_earth}a), PROTEUS has the lowest surface temperature $T_\mathrm{surf}= 2266$\,K and produces an almost-entirely \ch{CO2} atmosphere (88\%), while MOAI and PACMAN favour \ch{CO}-dominated compositions. These three different \ch{CO2}/\ch{CO} ratios are partially explained by the models' different surface temperatures (grey stars) at 95\% mantle melt fraction; \ch{2 CO + O_2 $\Leftrightarrow$ 2 CO2} is an exothermic reaction which favours \ch{CO} at higher temperatures \citep{hirschmann_magma_2012}. 

\begin{figure}
    \centering
    \includegraphics[width=0.9\linewidth]{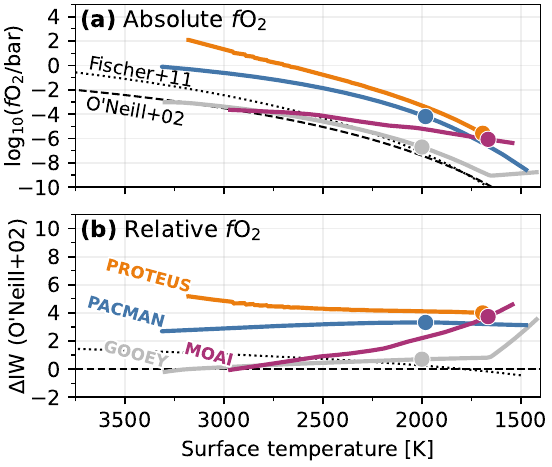}%
    \caption{
    Oxygen fugacity $f$\ch{O2} from each model's Nominal Venus simulation, plotted as a function of the surface temperature $T_\mathrm{surf}$, compared to the iron-w\"ustite (IW) buffer in absolute \textbf{(a)} and relative \textbf{(b)} terms. Line colour indicates the model, with circular markers at 5\% mantle melt fraction. Oxygen fugacity is a proxy for upper-mantle redox state. Dashed and dotted lines show empirically calibrated parametrisations of $f$O$_2^\mathrm{IW}(T)$ \citep{fischer_equation_2011, oneill_the_2002}. Some models' outgassing and $f$\ch{O2} calculations adopt different temperatures to their calculated $T_\mathrm{surf}$ (see Section~\ref{sec:methods_models}).
    }
    \label{fig:fo2}
\end{figure}

Differences in modelled carbon speciation are also driven by the upper-mantle redox state which sets the \ch{O2} partial pressure (Figure~\ref{fig:fo2}). PACMAN, GOOEY, NEONGOOEY, and MOAI track the concentration of iron-bearing phases at two different redox states (i.e. ferric $\mathrm{Fe^{3+}}$ and ferrous $\mathrm{Fe^{2+}}$ iron). These four models ensure conservation of oxygen atoms in the planet, simultaneously accounting for generation of \ch{O2} in the atmosphere caused by preferential escape of hydrogen \citep{wordsworth_ABIOTICO_2014,kasting_mantle_1993,katyal_effect_2020}, and temperature-dependent conversion between FeO and \ch{Fe2O3} in the magma ocean \citep{kress_the_1991}. MOAI and PACMAN also track the relative partitioning of ferric and ferrous iron between the solidified and molten mantle regions, although the mantle is not explicitly 1D-resolved \citep{sorbadere_the_2018}. All other \chili models prescribe oxygen fugacity relative to the iron-w\"ustite redox buffer \citep{oneill_the_2002}. 

Since PROTEUS pins $f$\ch{O2} relative to the iron-w\"ustite buffer, leading to $26\times$ the $p$\ch{O2} calculated by PACMAN (at 2750\,K, where the mantle is entirely molten), thereby favouring \ch{CO2} over \ch{CO} in Figure~\ref{fig:chem_earth}a. MOAI predicts a substantially more reduced initial magma ocean state (Figure~\ref{fig:fo2}), leading to its predicted $p$\ch{O2} being only $\sim 1/420$ that of PACMAN's, explaining MOAI's prediction of a \ch{CO}-dominated outgassed atmospheric composition (Figure~\ref{fig:chem_earth}a) despite its lower surface temperature than PACMAN. Together, these differences in carbon thermochemistry highlight the intimate relationship between mantle redox state and atmospheric composition.

GOOEY and LINCS outgas thin steam atmospheres (12~and~21~bar, respectively) because they neglect carbon chemistry. NEONGOOEY does not model \ch{CO} but does predict 87\% \ch{CO2}-dominated compositions comparable with PROTEUS and PlanAtMO. Since the Nominal Earth atmospheres in GOOEY and LINCS are thinner, they are expected to less effectively `blanket' the planet's interior, resulting in correspondingly larger outgoing radiation fluxes and shorter solidification times (Figure~\ref{fig:earth_venus}). However, the effect of including the opacity of carbon-bearing species on the planet's thermal evolution remains a second order effect, compared to the other modelling differences which dominate at these early evolutionary stages (Section~\ref{sec:results_interior}). The outgoing radiation flux from GOOEY is  comparable with PACMAN and PROTEUS, despite these latter two codes also simulating carbon chemistry and the additional opacities of \ch{CO2} plus \ch{CO} (Figure~\ref{fig:olr}b). This further demonstrates that inter-model differences yield larger changes to cooling timescale than variations in initial whole-planet carbon budgets (Figure~\ref{fig:cooling_time}).

\begin{figure}
    \centering
    \includegraphics[width=\linewidth]{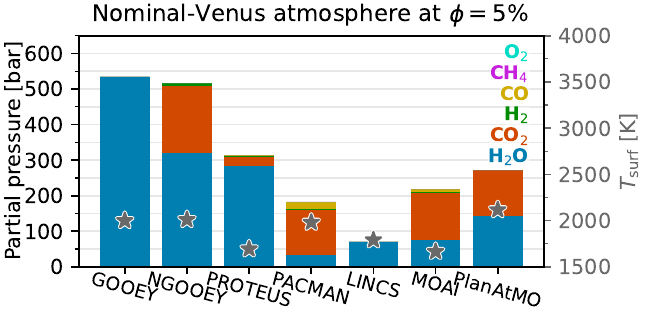}%
    \caption{
    Atmospheric compositions for Nominal Venus at 5\% mantle melt fraction. Analogous to Nominal Earth case in Figure~\ref{fig:chem_earth}b by adopting the same axis limits. 
    }
    \label{fig:chem_venus}
\end{figure}

Atmospheres overlying nearly-solidified mantles (Figure~\ref{fig:chem_earth}b) are generally found to be dominated by steam, but show substantially increased diversity in composition and total pressure, compared to initial fully-molten states. Bulk carbon inventories are largely outgassed at 95\% melt fraction, so magma ocean solidification towards 5\% melt fraction injects hydrogen atoms into initially C-rich atmospheres \citep{bower_retention_2022, nicholls_redox_2024, Nikolaou2019}. At the same time, cooler surface temperatures favour thermochemical conversion of \ch{CO} into \ch{CO2}, resulting in \ch{H2O}+\ch{CO2} dominated mixtures at later stages (Figures~\ref{fig:chem_earth}b~and~\ref{fig:chem_venus}). 

Using our simulations of Nominal Earth atmospheric compositions, shown in Figure~\ref{fig:chem_earth}, we are able to identify three classes of behaviours demonstrated by these models' evolving atmospheric compositions:
\begin{itemize}
    \item (\ch{CO2}-to-\ch{H2O}). NEONGOOEY, PROTEUS, and PlanAtMO predict early \ch{CO2}-rich atmospheres, with total pressures of approximately $\sim150$~bar, that become enriched \ch{H2O} which dominates the composition at later magma ocean stages.
    
    \item (\ch{CO}-to-\ch{CO2}). PACMAN and MOAI predict early $\sim150$~bar CO-rich compositions, that transition to \ch{CO2}-dominated with only minor \ch{H2O} degassing.
    
    \item (Growing-\ch{H2O}). GOOEY and LINCS do not model carbon chemistry, so predict very thin \ch{H2O} atmospheres at early stages, which thicken during magma ocean solidification due to `catastrophic' degassing of otherwise-soluble \ch{H2O}.
\end{itemize}
These three behaviour classes help interpret the outputs from these models' climate calculations in Section~\ref{sec:results_radtrans}.

\begin{figure}
    \centering
    \includegraphics[width=0.85\linewidth]{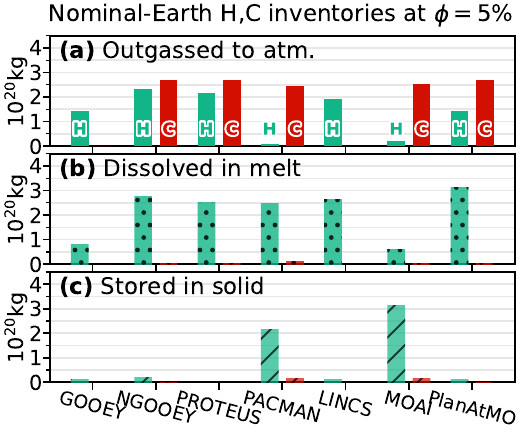}%
    \caption{
    Hydrogen and carbon mass budgets (green and red bars) distributed between the three reservoirs of Nominal Earth at 5\% mantle melt fraction. Panels show masses of H and C \textbf{(a)} outgassed into the atmosphere, (\textbf{b}, dotted) dissolved in the remnant magma ocean, and (\textbf{c}, hatched) stored in solidified mantle regions by various mechanisms depending on the model. LINCS and GOOEY do not simulate carbon. Plots in Appendix~\ref{app:overview} visualise the evolution of volatile partitioning as a function of time.
    }
    \label{fig:inv_earth}
\end{figure}

At $\Phi=5\%$ for Nominal Earth, PACMAN predicts the lowest surface pressure (176\,bar, 71\% \ch{CO2}) while PROTEUS predicts the largest (578\,bar, 82\% \ch{H2O}). This range primarily results from these models' differential treatment of  hydrogen partitioning within the planet's interior \citep{carone2025}. To demonstrate this, Figure~\ref{fig:inv_earth} plots the absolute mass inventories of hydrogen (green) and carbon (orange) distributed by each model between the atmosphere (panel a), molten magma ocean (panel b), and solid mantle regions (panel c). The PACMAN Nominal Earth simulation stores $\approx50\%$ of its hydrogen inventory in the solidified mantle, through its parametrised dynamic interstitial melt trapping \citep{laurent_earths_2020, schiano_primitive_2003}, which is theorised to be an efficient mechanism for deep sequestration of hydrogen \citep{Sim2024JGRE, kent_melt_2008, hier_the_2017}.  Similarly, MOAI's Nominal Earth simulation incorporates 79\% of its remaining hydrogen into the deep mantle via the interstitial melt pockets, with minimal (6\%) atmospheric storage after substantial mantle solidification and atmospheric escape (Figure~\ref{fig:escape}). These differences are an important source of model disagreement because PACMAN's and MOAI's predicted deep hydrogen storage leaves a hydrogen-poor atmosphere (Figure~\ref{fig:chem_earth}b), so H atoms are more readily retained against escape processes. In contrast, PROTEUS entirely neglects volatile storage within the solid phase of the mantle, so PROTEUS predicts massive hydrogen-enhanced atmospheres containing \ch{H2O} partial pressures exceeding 450\,bar (Figure~\ref{fig:chem_earth}b), alongside trace amounts of \ch{H2}, at these later stages of magma ocean evolution.

Whether within a global magma ocean or within pockets of trapped melt, hydrogen tracked as \ch{H2O} can efficiently dissolve into molten silicate as \ch{OH^-} cations, under oxidising conditions \citep{sossi_solubility_2023, hirschmann_magma_2012, sim_volatile_2024}. Rather than considering trapped pockets of melt within the solidified mantle regions, several \chili models instead treat the equilibrium redistribution of hydrogen between magma oceans and the mineral phases of underlying solidified mantle regions, where \ch{OH^-} are considered to be directly incorporated within the mineral lattice \citep{hauri_partitionin_2006,bell_water_1992}. This process is tracked through \ch{H2O} using parametrised equilibrium partitioning coefficients \citep{Lebrun2013, ElkinsTanton2008}. GOOEY, NEONGOOEY, and LINCS adopt the same coefficient value $k_\mathrm{H_2O}=10^{-2}\mathrm{\,kg/kg}$, which leads to small amounts of \ch{H2O} storage in the solid mantle phases (Figure~\ref{fig:inv_earth}c). The partition coefficient simulated by PlanAtMO depends on the pressure at which solids form, which determines their mineral phase \citep{Schaefer2024JGRE, walbecq_the_2025}, generally resulting in $k_\mathrm{H_2O}$ values less than $10^{-3}$. These models show good agreement on the amount of \ch{H2O} sequestered by equilibrium partitioning (Figure~\ref{fig:inv_earth}b,c). 

MOAI's predicted \SI{3e20}{\kilo\gram} of solid-mantle hydrogen storage is consistent with a lower estimate of \SI{3.4e20}{\kilo\gram} of hydrogen atoms (stored as water) within Earth's present mantle \citep{peslier_water_2017}. The other models retain substantial hydrogen inventories, comparable to Earth's present water content, within their remnant magma ocean regions (Figure~\ref{fig:inv_earth}b). Whether these hydrogen reservoirs would be finally outgassed during magma ocean crystallisation, or trapped within the mantle, is not resolved here \citep{Mojzsis2001Natur, foley_the_2015, oneill_conditions_2007}.

Earth and Venus are similarly sized, and their Nominal scenarios are initialised with the same carbon and hydrogen budgets (Section~\ref{sec:methods}). Thus, modelled differences in their thermal and compositional histories are reflective of their differing bolometric and high-energy instellation fluxes. Figure~\ref{fig:chem_venus} plots the outgassed atmospheric composition and surface temperature of Nominal Venus at $\Phi=5\%$. This scenario is directly analogous to the Nominal Earth case in Figure~\ref{fig:chem_earth}b and broadly exhibits quantitatively similar compositions, for a given model. Comparison of Figures~\ref{fig:chem_earth}b~and~\ref{fig:chem_venus}, which represent late magma ocean stages ($\Phi=5\%$), shows that total surface pressures on Nominal Venus are smaller than for Nominal Earth, despite the simulations being initialised with the same carbon and hydrogen budgets. The lower surface pressure on Venus arises because the planet is exposed to larger X-ray and ultraviolet radiation fluxes that drive atmospheric escape \citep{zahnle_mass_1986, hunten_mass_1987, owen_atmospheric_2019}. Atmospheric blanketing, greenhouse effects, and absorption of stellar radiation jointly act to delay or prevent magma ocean solidification \citep{Hamano2015, krissansen2021venus, Hamano2024arXiv}. Venus' exposure to a larger bolometric stellar flux than Earth means that a thinner atmosphere -- which has correspondingly weaker blanketing and greenhouse effects -- is necessary for the planet to solidify, all else equal (Figure~\ref{fig:chem_venus}). This is enabled by long-term photoevaporative escape \citep{Hamano2013, owen_atmospheric_2019}.

An exception to these Earth-Venus pressure differences is MOAI, which only considers the escape of \ch{H2}, of which only trace amounts are formed, so MOAI predicts similar surface pressures between Venus and Earth cases (218~and~215~bar, respectively). PACMAN also does not follow this trend of atmospheric pressure difference between Earth and Venus. The PACMAN model finds a surface pressure of 182~bar for Venus compared to 176~bar for Earth at the same melt fraction. This can be attributed to Venus' smaller mantle volume, and the treatment of melt trapping of volatiles and atmospheric escape in the PACMAN model. Competition between atmospheric radiative blanketing and heating by stellar radiation is consistent with later solidification under the Type II scenario where escape regulates the cooling timescale. Type II Venus is exhibited by all models except MOAI and PACMAN, for which both its Nominal Venus and Nominal Earth scenarios directly cool to solidification  (Figure~\ref{fig:earth_venus}).

Hydrodynamic outflow of escaping gas, driven by the absorption of stellar radiation, is expected to preferentially remove lighter atoms under most irradiation regimes \citep{hunten_mass_1987, zahnle_mass_1986, owen_atmospheric_2019}. This would increase the O/H and C/H mixing ratio of the atmosphere \citep{cherubim2025oxidation, krissansen-totton2021}, and potentially drive mantle oxidation by hydrogen loss \citep{wordsworth_ABIOTICO_2014, katyal_effect_2020}. GOOEY and PACMAN incorporate these effects but show close chemical correspondence between Nominal Venus and Nominal Earth at 5\% melt fraction. For example, at $\Phi=5\%$, PACMAN calculates atmospheric C/H mass ratios of 31.25 and 25.46 for Nominal Earth and Venus respectively -- indicative of systematic carbon enhancement on Earth (Figure~\ref{fig:inv_earth}), opposite to our expectations from fractionation by atmospheric escape. So, while fractionation by escape is known to have imprinted measurable isotopic signatures on Earth and Venus' present-day atmospheres, our modelling suggests that it played only a minor role in setting the volatile element ratios  of their atmospheres \citep{zahnle_mass_1986, grinspoon_implication_1993}. Instead, differences in bulk atmospheric C/H ratio primarily arise through melt trapping within the solid phase of the mantle; faster solidification more efficiently traps pockets of melt in the solid phase as the magma ocean solidifies upwards from the core-mantle boundary, and the volatile elements have different solubilities. PACMAN estimates that 54.6\% and 57.6\% of Nominal Venus and Nominal Earth final hydrogen inventories is distributed into trapped deep-mantle melt pockets, respectively. In comparison, the final carbon inventory of Nominal Earth is 91.4\% outgassed and 6.8\% trapped (Figures~\ref{fig:inv_earth}~and~\ref{fig:overview_pacman}). The relatively more efficient trapping of hydrogen within the deep interior -- increasingly effective under faster solidifying scenarios -- leads to higher atmospheric C/H ratios, despite hydrodynamic escape acting to drive a decrease in C/H ratio by thermospheric fractionation \citep{carone2025}.

As reviewed in Section~\ref{sec:methods}, the redox conditions within and upon planetary interiors  is expected to evolve throughout their lifetimes due to a range of processes \citep{ Schaefer2017ApJ,lichtenberg_redox_2021, hirschmann2022magma}. Thus, the \chili models take different approaches to modelling Earth and Venus' historical redox conditions. Figure~\ref{fig:fo2} plots the mantle $f$\ch{O2} calculated by four simulations of Nominal Venus -- used as a proxy for the surficial redox environment. In absolute terms, the simulated $f$\ch{O2} decreases over time as temperature decreases (Figure~\ref{fig:fo2}a). However, it is more appropriate to compare $f$\ch{O2} in relative terms \citep{frost_chapter_2018}. Figure~\ref{fig:fo2}b shows four models reproducing the \chili protocol's requirement that endpoint $f$\ch{O2} values tend towards IW+4. However, the models show different temperature-dependent behaviours. PROTEUS (orange line) prescribes $f$\ch{O2} relative to IW \citep{oneill_the_2002}, without accounting for O/H fractionation by atmospheric escape or ferric-ferrous iron partitioning during magma ocean crystallisation. PROTEUS' calculated whole-planet oxygen budgets are therefore not conserved and respond directly to thermochemistry modelled at the degassing conditions: constructed to hold $f$\ch{O2} near IW+4 across all temperatures (and corresponding planet ages). PROTEUS' deviation from IW+4 at $T_\mathrm{surf}>3000$\,K (Figure~\ref{fig:fo2}b) arises from departure between the surface temperature $T_\mathrm{surf}$ and degassing temperature $T_\mathrm{magma}$. In comparison, PACMAN does track oxygen mass-conservation while also accounting for hydrogen-oxygen fractionation in the escaping hydrodynamic outflow. Despite these additional physics and Venus' enhanced exposure to X-ray and ultraviolet radiation compared to Earth, PACMAN $f$\ch{O2} values remain near IW+4 across all temperatures: the blue line in Figure~\ref{fig:fo2}b shows only slight oxidation from IW+2.8 to IW+3.2 during simulated cooling of the surface temperature from 3250~to~2500~K. Hydrodynamic fractionation of O/H in the PACMAN case is minimal and therefore does not drive mantle oxidation; consistent with PACMAN's predictions of minimal atmospheric loss (Figure~\ref{fig:escape}),  and with its predicted preferential sequestration of hydrogen into the solid-phase of the mantle (Figure~\ref{fig:inv_earth}c), and production of carbon-rich atmospheres (Figure~\ref{fig:chem_venus}).

The oxygen fugacity calculated by GOOEY (grey line in Figure~\ref{fig:fo2}) only shifts from IW+0 to IW+4 during the final stages of the simulated cooling of Nominal Venus, at melt fractions less than 5\%. This late-stage oxidation is also not driven by hydrodynamic fractionation of O/H -- since GOOEY predicts minimal hydrogen loss (Figure~\ref{fig:escape}) -- and is caused entirely by the concentration of ferric iron in the remnant magma ocean layers. 

In comparison to GOOEY and PACMAN, MOAI shows a more continuous trend of f\ch{O2} as a function of temperature (and time) -- plotted in purple in Figure~\ref{fig:fo2}. MOAI's calculated $f$\ch{O2} also span the widest range: from initially IW+0 to IW+4 at 5\% mantle melt fraction. This behaviour arises, firstly, from MOAI's prediction of continuous and substantial hydrogen loss (as \ch{H2}), since its simulations retain their entire oxygen budget, which drives monotonic oxidation of the planet's mantle \citep{kasting_mantle_1993}. MOAI also simulates differential ferric-ferrous iron partitioning into the solidified mantle regions, causing oxidation of the magma ocean under these bottom-up fractional crystallisation scenarios \citep{maurice_redox_2023}.

Overall, these differences in volatile chemistry and partitioning highlight that: (a) interiors and atmospheres of Earth- and Venus-like scenarios are predicted to oxidise over time, under bottom-up fractional crystallisation scenarios; (b) processes which sequester hydrogen into deep planetary interiors strongly modulate atmospheric compositions, controlling whether carbon- or water-dominated atmospheres are produced at magma ocean solidification; (c) mantle redox evolution pathways regulate outgassed atmospheric \ch{CO2}/\ch{CO} ratios.

\subsection{Atmospheric escape}
\label{sec:results_escape}

Atmospheric escape incorporates multiple physical, chemical, and hydrodynamic processes, which resolve towards different escape `regimes' depending on a planet's size and irradiation exposure \citep{owen_atmospheric_2019}. No specific protocol is established for atmospheric escape; the \chili models take differing approaches to parametrising atmospheric escape, although they generally adopt the energy-limited approximation in which upper-atmosphere heating by absorption of X-ray and ultraviolet radiation launches a hydrodynamic outflow of gas \citep{hunten_mass_1987, zahnle_mass_1986, owen_atmospheric_2019}. PROTEUS treats the outflow as intrinsically non-fractionating, while MOAI and LINCS assume only \ch{H2} can escape, PACMAN accounts for C-H-O fractionation by thermospheric diffusion, and GOOEY/NEONGOOEY both treat O-H fractionation. PlanAtMO does not simulate atmospheric escape.

\begin{figure}
    \centering
    \includegraphics[width=0.95\linewidth]{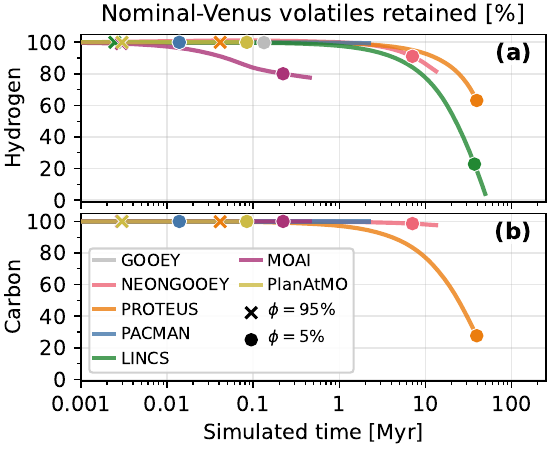}%
    \caption{
    Relative amounts of the initial inventories of hydrogen (\textbf{a}, top) and carbon (\textbf{b}, bottom) retained by the Nominal Venus scenario, as a function of simulation time. Atoms are lost from the planet due to hydrodynamic escape. Line colour indicates the model, with scatter points at 95\% and 5\%  melt fraction (crosses and circles, respectively).
    }
    \label{fig:escape}
\end{figure}

Figure~\ref{fig:escape} plots the hydrogen and carbon inventories of the Nominal Venus scenario, over time, which both decrease due to the cumulative effects of atmospheric escape processes. The circular markers show the amount of initial H and C retained by the planet by the time its mantle has reached a simulated melt fraction of 5\% -- corresponding to the outgassed atmospheric compositions shown in Figure~\ref{fig:chem_venus}. LINCS predicts that Venus retained only 22.9\% of its post-accretion hydrogen budget at the point of magma ocean solidification (green line), in contrast to GOOEY, which predicts 99.8\% hydrogen retention, despite these models both considering pure-\ch{H2O} atmospheres and equilibrium partitioning between solid-mantle, magma ocean, and atmosphere. Their different predictions arise from the prolonged magma ocean in the LINCS case, where Venus exhibits type II behaviour that exposes the planet's hydrogen content to atmospheric escape for 40\,Myr, while GOOEY solidifies in a much shorter timescale of 0.45\,Myr without entering a state of global radiative equilibrium. PROTEUS predicts a similar magma ocean cooling timescale to LINCS, while finding that a larger fraction (63.2\%) of the initial hydrogen budget is retained. This difference between LINCS and PROTEUS arises because PROTEUS calculates evolving bulk mass-loss escape rates, which are then partitioned into carbon and hydrogen mass fluxes, so PROTEUS' substantial outgassed carbon reservoirs offset the amount of hydrogen lost. The massive loss of carbon in the PROTEUS case corresponds with its prediction of a steam-dominated (i.e. oxygen and hydrogen) atmospheric composition, shown in Figure~\ref{fig:chem_venus}. 

MOAI finds that 20\% of the initial hydrogen is removed by escape during the course of simulated Nominal Venus magma ocean evolution, which corresponds with gradual mantle oxidation in Figure~\ref{fig:fo2}. PACMAN and GOOEY predict negligible amounts of hydrogen escape in the Nominal Venus case, and correspondingly predict only minor mantle oxidation until the mantle melt fraction reaches 5\%, at which point other processes drive substantial changes in $f$\ch{O2} (Section \ref{sec:results_chem}). Together, these three models' behaviours suggest that substantial escape of hydrogen does correlate with oxidation of rocky planet interiors, although it  acts simultaneously with interior mantle processes that can drive a final, rapid oxidation (Figure~\ref{fig:fo2}). 

Substantially different predictions of hydrogen and carbon losses between these models, shown by Figure~\ref{fig:escape}, is an important point of model disagreement. Venus' and Earths' surficial volatile inventories after magma ocean solidification likely played a key role in controlling their early climate states \citep{constantinou_venus_2026, sossi_redox_2020}. However, these differences are not unexpected, because accurate treatment of atmospheric escape processes requires vertically resolved hydrodynamic photochemical modelling of these planets' upper-atmospheres \citep{schulik_aiolos_2023, brain_atmospheric_2016, owen_atmospheric_2019}.   

\subsection{Radiative transfer and climate modelling}
\label{sec:results_radtrans}

The rate of energy radiated to space by a planet determines its cooling rate. This net energy flux is the balance of outgoing thermal radiation, incoming stellar radiation, and scattering processes \citep{pierrehumbert_principles_2010}. Thus, we must consider how the \chili models determine these flux terms in order to understand their predicted planetary cooling timescales \citep{lichtenberg2021, Lebrun2013}. Outgoing longwave radiation (OLR) quantifies bolometric radiation emitted by the planet to space, which is determined by thermal emission and attenuation by the atmosphere. The OLR is effectively controlled by the radiating temperature of the planet's photosphere, where the optical depth of longwave radiation $\tau_\mathrm{LW}=1$, the location of which is determined by the atmospheric composition and temperature profiles \citep{pierrehumbert_principles_2010, stamnes2017radiative}. Aerosol scattering effects are not modelled, per the \chili protocol, so a positive difference between the OLR and absorbed stellar radiation (ASR) represents net energy loss.

\begin{figure*}
    \centering
    \includegraphics[width=0.92\linewidth]{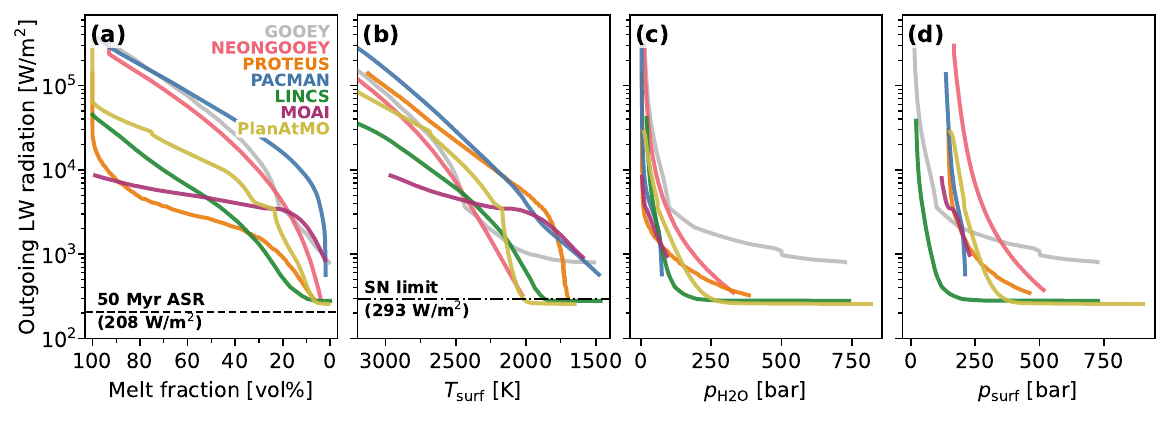}%
    \caption{
    Outgoing longwave radiation flux from Nominal Earth, plotted as a function of melt fraction \textbf{(a)}, surface temperature \textbf{(b)},\ch{H2O} partial pressure \textbf{(c)}, and total surface pressure \textbf{(d)}. Dashed black line quantifies absorbed stellar radiation: $\mathrm{ASR} = \frac{L_\odot(50\mathrm{\,Myr})}{4\pi (1 \mathrm{\,AU})^2}\times(1-0.1)\times 0.375\cos(48.19^\circ)$, with 10\% Bond albedo, 0.375 geometric factor, and $48.19^\circ$ Zenith angle \citep{Baraffe2015, cronin_on_2014}. Dash-dot black line indicates the Simpson-Nakajima steam atmosphere runaway limit on OLR \citep{nakajima_olr_1992}.
    }
    \label{fig:olr}
\end{figure*}

Figure~\ref{fig:olr} plots outgoing longwave radiation fluxes from Nominal Earth simulations, against the modelled melt fraction (\textbf{a}, left) and surface temperature (\textbf{b}, right).  The OLR is only indirectly related to mantle melt fraction through its relationships with surface temperature and outgassed volatile inventories, which are mediated by the vertical atmospheric temperature structure and composition. The ASR for young Earth is indicated in panel\,a for a circular 1\,AU orbit around a 50\,Myr Sun \citep{Baraffe2015}. All models' OLR curves show similar qualitative behaviours across all four panels, because thermal emission scales super-linearly with temperature. For a greybody emitter, we should expect OLR to behave as a log-linearly decreasing line in panel\,b, following the Stefan-Boltzmann radiation law $\varepsilon\sigma T^4$ as the planets cool from an initially hot state.

The OLR initially exceeds \SI{e5}{\WPMS} for all of these Nominal Earth simulations -- except MOAI -- with initial surface temperatures above 2500\,K that correspond to deep magma oceans. High OLR values are achieved, even for these optically-thick atmospheres extending far above the magma ocean surface, because temperatures at the $\tau=1$ atmospheric photosphere layer are large (Figure~\ref{fig:climate_profiles}).  At initially high temperatures, PROTEUS finds an $\mathrm{OLR}>\SI{e5}{\WPMS}$ because its AGNI atmosphere sub-module determines that near-surface atmosphere regions are convectively stable. In these regions, energy is transported by radiative diffusion at a shallower lapse rate than would be supported by the corresponding adiabat \citep{selsis_cool_2023, cmiel2025}, which has a steeper temperature gradient, and therefore produces increased photospheric temperatures and larger OLR values are predicted, subject to compositional differences \citep{nicholls_convective_2025, nicholls_massradius_2026}.

In general, higher partial surface pressures of \ch{H2O} and \ch{CO2} correspond with lower OLR fluxes (Figure~\ref{fig:olr} panels b and c): a `blanketing' effect, whereby the outgassing of thick, volatile-atmospheres decreases planetary cooling rates by absorbing thermal radiation from the deep atmosphere and magma ocean surfaces, thereby extending magma ocean lifetimes. Atmospheric blanketing of magma oceans has been established by previous studies \citep{nicholls_redox_2024, Hamano2015, ElkinsTanton2008} and justifies the trend of prolonged magma oceans under hydrogen-enhanced Earth scenarios demonstrate by all \chili models (Figure~\ref{fig:cooling_time}). The simulations agree that OLR decreases as the planet cools and the atmosphere thickens (Figure~\ref{fig:olr}b). Compositional evolution causes deviation from a simple log-linear relationship between OLR and surface temperature -- especially notable for MOAI and PROTEUS (orange line in Figure~\ref{fig:olr}b).

Specific compositions also determine the OLR through their control over the photospheric location \citep{pierrehumbert_principles_2010}, so differences in atmospheric outgassing are reflected in these models' OLR calculations. For example, PACMAN's outgassing code finds \ch{CO}-richer compositions compared to PROTEUS (Figure~\ref{fig:chem_earth}). The smaller absorption cross-section of \ch{CO} compared to \ch{CO2} \citep{Rothman2010} places the PACMAN's modelled photosphere at deeper pressures with higher temperatures, thereby yielding higher OLR values than PROTEUS for a given surface temperature, despite PACMAN enforcing pseudoadiabatic temperature profiles. 

MOAI calculates a substantially smaller initial OLR relative to the other models, when compared at a given surface temperature (Figure~\ref{fig:olr}b). This lower OLR arises because MOAI sets the atmosphere $T(p)$ structure with a \ch{H2O} pseudoadiabat and cool 200\,K stratosphere, which results in a cooler photospheric temperature and reduced thermal emission.  MOAI also predicts a highly-reduced early magma ocean which outgases \ch{CO}-dominated atmospheres (Figure~\ref{fig:fo2}).

NEONGOOEY, GOOEY, LINCS, and PlanAtMO find that OLR becomes approximately constant as a function of surface temperature, for surface temperatures less than 2000\,K (Figure~\ref{fig:olr}b). In the same regime, LINCS and PlanAtMO also predict that OLR is constant with increasing \ch{H2O} partial and total surface pressures beyond $\sim300\mathrm{\,bar}$. In these cases, the OLR approximates the Simpson-Nakajima runaway greenhouse limit (dash-dot black line) which arises from their modelling of saturated  pure-steam upper atmosphere layers that have temperature structures independent of the surface conditions and compositions \citep{nakajima_olr_1992, ingersoll_the_1969}. 

The OLR calculated by GOOEY asymptotically tends towards \SI{797}{\WPMS} since GOOEY imposes an isothermal stratosphere, held constant at the radiative skin temperature; once the deep adiabatic region has cooled sufficiently, the $\tau_\mathrm{LW}=1$ photosphere layer exists within the constant-temperature stratospheric region, that determines the OLR. MOAI does not resolve the Simpson-Nakajima limit at cooler temperatures (Figure~\ref{fig:olr}b), despite also following a pseudoadiabatic temperature structure, because its radiative transfer scheme does not incorporate the effects of \ch{H2O} collision-induced absorption and (Figure~\ref{fig:olr}c) it predicts smaller \ch{H2O} partial pressures \citep{lichtenberg2021, zahnle_evolution_1988}. PACMAN maintains a comparatively high OLR, above the Simpson-Nakajima limit, due to the low atmospheric water content (Figure~\ref{fig:chem_earth}).

Our understanding of the models' different approaches to parametrising planetary climates and radiative transfer can explain the calculated relationships between OLR and surface temperature in Figure~\ref{fig:olr}b. The models' different cooling behaviours reflect that both atmospheric composition and its temperature structure play a role in the cooling of magma ocean surfaces. 

Models which exhibit comparable behaviour in OLR-$T_\mathrm{surf}$ space (Figure~\ref{fig:olr}) are not necessarily consistent with the three types of atmospheric compositional evolution identified in Section~\ref{sec:results_chem}: \ch{CO2}-to-\ch{H2O}, \ch{CO}-to-\ch{CO}, and growing-\ch{H2O}. Similarity between Figure~\ref{fig:olr} panels c~and~d, showing \ch{H2O} partial pressure and total surface pressure respectively, is consistent with expectations that \ch{H2O} opacity is a strong control over the outgoing radiation fluxes.

OLR predictions from NEONGOOEY and PlanAtMO radiative transfer calculations, as a function of surface temperature, are in good agreement (Figure~\ref{fig:olr}b, yellow and pink lines). This corresponds to their similar predictions of atmospheric compositional evolution, which exhibit a \ch{CO2}-to-\ch{H2O}  transition (Figure~\ref{fig:chem_earth}). However, PROTEUS exhibits the same \ch{CO2}-to-\ch{H2O} compositional transition but predicts a substantially larger OLR (orange line) than both NEONGOOEY and PlanAtMO, for a given surface temperature. Furthermore, comparing these three models' OLR values at a $\sim40$\% melt fraction (Figure~\ref{fig:olr}a) --- rather than as a function of temperature --- instead ranks the outgoing radiation flux from PlanAtMO as much lower than both NEONGOOEY and PROTEUS (yellow, pink, and orange lines). PROTEUS' atmosphere model solves for radiative-convective equilibrium without condensation, which differs from NEONGOOEY and PlanAtMO's atmosphere profiles that are defined by \ch{H2O}-saturation curves, so PROTEUS derives cooler upper-atmospheric temperatures with correspondingly lower thermal emission fluxes (Figure~\ref{fig:climate_profiles}).

PlanAtMO (\ch{CO2}-to-\ch{H2O}), NEONGOOEY (\ch{CO2}-to-\ch{H2O}), and LINCS (Growing-\ch{H2O}) reproduce the Simpson-Nakajima OLR limit under cooler conditions at later stages of the modelled evolution \citep{nakajima_olr_1992}. This corresponds with their predicted \ch{H2O}-dominated compositions at later stages where upper-atmosphere temperature structures are set by \ch{H2O} saturation. GOOEY simulates purely-\ch{H2O} atmospheric compositions but is unable to reproduce the Simpson-Nakajima limit due to an error in its radiative transfer calculations which is corrected by its NEONGOOEY successor (see Section~\ref{sec:methods_models}).

Using Figure~\ref{fig:chem_earth}, PACMAN and MOAI were classified as predicting similar \ch{CO}-to-\ch{CO2} atmospheric composition transitions. However, PACMAN calculates the highest OLR and MOAI the lowest OLR at high surface temperatures (Figure~\ref{fig:olr}b, blue and purple lines). These models eventually converge in OLR-$\mathrm{T_{surf}}$ space at somewhat cooler conditions $<2000$\,K since both codes adopt cool isothermal stratospheres that determine the photospheric brightness temperature (and thus the OLR) at later stages.

Models that predict similar compositions (e.g. PACMAN and MOAI) are able to produce substantially different OLR values, for a given surface temperature (Figure~\ref{fig:olr}b). In parallel, models that predict different atmospheric compositions can show similar OLR behaviours. Clearly, models' different assumptions and simplifications of climate modelling and radiative transfer have direct implications for their calculated outgoing radiation fluxes, highlighting the impact of these modelling \textit{choices}. These modelling differences exacerbate choices to neglect or incorporate key opacity sources (e.g. collision-induced absorption continua) which lead to calculated OLR values spanning multiple orders of magnitude between the \chili models. Although models disagree and vary \textit{quantitiately} in their calculated OLR fluxes -- whether compared at constant temperature, melt fraction, or composition (Figure~\ref{fig:olr}a-d) -- all codes qualitatively demonstrate that atmospheric blanketing effects explain prolonged magma ocean cooling timescales (Figure~\ref{fig:cooling_time}).

The relative differences in arrangement of models' OLR curves differs between Figure~\ref{fig:olr}'s panels a~and~b because of the complex relationships mapping mantle melt fraction to magma ocean surface temperature, rather than due to different atmospheric physics. For example, PROTEUS has the lowest OLR of all models when compared at 100\% melt fraction, but the second highest OLR of all models when compared at $T_\mathrm{surf}=3000$\,K. This apparent discrepancy is physical, and is discussed in the following section.

\subsection{Geodynamics and mantle phase change}
\label{sec:results_interior}

\begin{figure}
    \centering
    \includegraphics[width=0.92\linewidth]{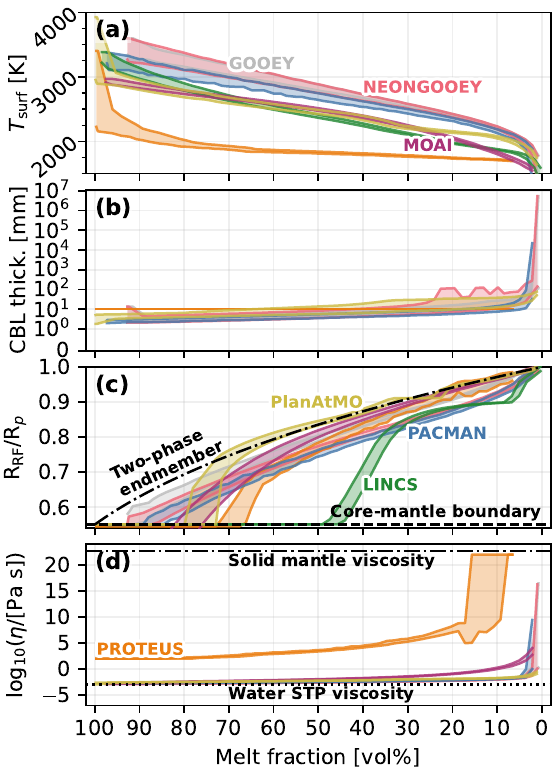}%
    \caption{
    Diagnostics variables of model geodynamics as functions of mantle melt fraction: \textbf{(a)} surface temperature, \textbf{(b)} conductive boundary layer thickness, \textbf{(c)} radius of the solidification front, and \textbf{(d)} effective mantle viscosity. Simulation data dependent variables (y-axes) are flattened over all Earth simulations, for each model, and then re-binned over melt fraction (x-axis) to produce the shaded regions. The radius of the core-mantle boundary is 55\% \citep{lodders_planetary_1998}. The expected analytical relationship $\Phi=(R_p^3-R_\mathrm{RF}^3)/(R_p^3-R_\mathrm{core}^3)$ for a two-phase end-member scenario, in which mantle regions are either fully-molten or fully-solid, is shown by the dashdot line in (c). For reference, the dynamic viscosity material representative of Earth's deep mantle is $\gtrsim$\SI{5e22}{\pascal\second} \citep{peltier_the_1981, mckenzie_the_1967}, and the viscosity of liquid water at $\mathrm{20^\circ C}$ is \SI{e-3}{\pascal\second} \citep{iso_water,huber_new_2009}.
    }
    \label{fig:interior}
\end{figure}

Melt fraction and surface temperature are fundamentally different, but often correlated, physical quantities \citep{abe_thermal_1997}. Mapping these variables depends on their definition \citep{stixrude_melting_2014}, the modelled mantle temperature profile \citep{bower_numerical_2018}, and solidus and liquidus curves of mantle melting \citep{zhang_melting_1994,katz_a_2003,fiquet_melting_2010, Andrault2011}. Figure \ref{fig:interior} presents four diagnostic variables of mantle geodynamics, as functions of volumetric mantle melt fraction, from our Nominal Earth simulations: \textbf{(a)} surface temperature, \textbf{(b)} conductive boundary layer thickness, \textbf{(c)} characteristic radius of solidification, and \textbf{(d)} characteristic mantle viscosity. 

Figure~\ref{fig:interior}a largely explains the differences in simulated Earth solidification timescales presented by Figure~\ref{fig:cooling_time}a. PROTEUS is systematically the `slowest' model to cool, while PlanAtMO and MOAI are slightly more rapid. This line-up is reflected by surface temperature-melt fraction relationships shown in Figure~\ref{fig:interior}a, where PROTEUS has the lowest $T_\mathrm{surf}$ at high melt fractions, followed by PlanAtMO and MOAI with slightly higher temperatures. Atmospheric temperatures and energy fluxes -- which determine the rate of planetary cooling -- are set by chemical composition, incoming stellar radiation, and upwelling surface emission. These quantities are directly tied to surface temperature but not directly to the mantle melt fraction. So, lower surface temperatures at high melt fractions correspond to lower outgoing fluxes (OLR) and these predictions of slower magma ocean cooling. 

All \chili models except LINCS  analytically parametrise a single rheological transition centred around a `critical melt fraction' of 40\% (50\% for PROTEUS), whereas LINCS defines the transition radius where the local temperature equals the solidus temperature (0\% local melt fraction). Figure~\ref{fig:interior}c plots the radial location $R_\mathrm{RF}$ of the rheological transition front versus whole-mantle melt fraction $\Phi$, calculated by each model. This can be interpreted as a measure of radial melt \textit{distribution}, given some bulk amount of molten mantle material. Figure~\ref{fig:interior}c shows that $R_\mathrm{RF}$ and $\Phi$ are predicted to be strongly correlated by each model, as expected, since all scenarios exhibit bottom-up crystallisation modes and adopt qualitatively comparable melting curves (Section~\ref{app:melting_curves}).

LINCS deviates from the other models in $R_\mathrm{RF}-\Phi$ phase space by maintaining a deep magma ocean throughout the mantle, until the \textit{whole} mantle melt fraction ($x$-axis) approaches its rheologically critical value, $\approx50\%$--- a stagnation driven chiefly by a deep adiabatic temperature profile nearly parallel to the solidus, as well as the definition of the transition radius. Its transition radius then increases abruptly to catch up with the other models. 

All models present a bottom-up mode of mantle crystallisation: the rheological transition (Figure~\ref{fig:interior}c) begins at the core-mantle boundary  and progresses radially upwards as crystal cumulates settle \citep{labrosse_fractional_2015, solomatov_nonfraction_1993}. Fractional or batch (equilibrium) crystallisation behaviours depend on the cumulate sizes formed during crystallisation and the magma ocean flow regime \citep{labrosse_fractional_2015, bower_retention_2022}. These models generally consider fractional crystallisation behaviours. None of these simulations correspond to a basal magma ocean scenario. 

Close inter-model correspondence in $R_\mathrm{RF}-\Phi$ space (Figure~\ref{fig:interior}c) contrasts with the substantial differences in $T_\mathrm{surf}-\Phi$ space (Figure~\ref{fig:interior}a). In the case of PROTEUS, a key difference arises from its mixing-length treatment of mantle dynamics, where upward energy transport of heat by convection is partially offset by downward transport of heat by the settling of cumulates and the release of gravitational potential energy \citep{bower_numerical_2018, hirschmann_magma_2012}. Convective suppression causes the mantle temperature profile to deviate from an adiabat, yielding shallower geothermal lapse rates and lower surface temperatures. Additionally, the liquidus adopted by PROTEUS has the lowest temperature at the core-mantle boundary of all models (Figure~\ref{fig:melting_curves}a), so its simulations of Nominal Earth's deep mantle must attain cooler temperatures than other models before solidification begins -- further promoting longer solidification timescales.

Figure~\ref{fig:interior}d shows that PROTEUS (orange region) calculates higher effective mantle viscosities $\eta$ compared to other models, for a given melt fraction. Compared to the other models, the orange PROTEUS curve spans a wider range of viscosity values because its interior geodynamics module \citep[SPIDER;][]{bower_numerical_2018} calculates 1D mantle viscosity profiles, which depend on the melt fraction of each layer, while other models solve for a scalar value representative of the whole mantle \citep{costa_rheology_2009, bottinga_viscosity}. However, even for the initially fully-molten magma ocean conditions, e.g. where $\Phi\approx95\%$, the models predict a range of viscosity values which vary across orders of magnitude (panel d). This reflects their different viscosity parametrisations (Table~\ref{tab:models}) and melting curves (Figure~\ref{fig:melting_curves}). Differences in the models' predicted mantle rheological properties factor directly into the magma ocean cooling timescales through their boundary layer parametrisations (Equation~\ref{eq:fconv}). For example, GOOEY adopts a VFT-type viscosity function evaluated with a single potential temperature to represent the whole mantle, while PROTEUS uses an Arrhenius-type function with a depth-dependent activation volume applied at each mantle layer \citep{bower_numerical_2018}. Higher viscosities make mantle convection less efficient and inhibit heat transport out of the deep interior \citep{bower_retention_2022, turner_convection_1986, solomatov_ToG_2015}, so PROTEUS simulations form sluggish high-viscosity regions which inhibit cooling: correspondent with it finding longer cooling timescales (Figure~\ref{fig:cooling_time}).  

Most of our models predict only late-stage increases in viscosity to values above 1\,Pa\,s, at melt fractions $\lesssim3\%$ (GOOEY, NEONGOOEY, and PACMAN) or $\lesssim11\%$ (MOAI). PROTEUS calculates $\eta\geq10^2$\,Pa\,s under the initially fully-molten conditions (Figure~\ref{fig:interior}d), since its interior geodynamics module adopts a melt-phase viscosity of $10^2$\,Pa\,s. %
These differing behaviours are partially geometric since the volume-averaged bulk melt fraction ($x$-axis) is weighted towards molten upper mantle layers which have larger volumes. However, it is clear from cooling timescales in Figure~\ref{fig:cooling_time} and geodynamical diagnostics in Figure~\ref{fig:interior} that different approaches to parametrising the microphysical processes of silicate crystallisation correlate with predicted cooling timescales differing by orders of magnitude \citep{katz_physics_2022}. Importantly, the modelled Earth and Venus evolutionary pathways spend substantial time within a `mushy' magma ocean regime, defined by mantle conditions between their solidus and liquidus curves; melting curve choices and parametrisations also regulate magma ocean geodynamics and their transition towards solidified conditions (Figure~\ref{fig:melting_curves}).

Unmodelled physics, such as the rheological effects of volatiles dissolved within the magma ocean fluid, would suggest that these calculated magma ocean viscosities are upper-limits on real behaviours. Inter-model agreement of viscosities far below that of a purely solid phase (dash-dot line) across a wide range of melt fractions is therefore consistent with this expectation, although agreement between simplified parametrisations does not by itself establish physical robustness. Real planetary magma oceans are expected to be highly inviscid during largely-molten stages \citep{hirschmann2022magma, Schaefer2018RSPTA}. 

Many of the \chili models incorporate a parametrised, thermally conductive boundary layer (CBL) at the atmosphere-mantle interface (see methods Section~\ref{sec:methods_physics} and Figure~\ref{fig:models}). This CBL is different to the tectonic crust of present day Earth, and is intended to represent a potential floatation `skin' atop the magma ocean induced by  radiative cooling into the lower atmosphere \citep{ElkinsTanton2008, Abe1985}. The CBL cannot be spatially resolved by these models but is suggested to increase magma ocean cooling timescales by decreasing the calculated surface temperature (and thus net thermal emission) relative to the mantle potential temperature \citep{bower_numerical_2018, spaargaren_the_2020}: Equation~\ref{eq:cbl}. In general, the models predict thin CBL thicknesses $d_b\lesssim10\mathrm{\,cm}$ for $\Phi\gtrsim3\%$ (Figure~\ref{fig:interior}b). The models GOOEY, NEONGOOEY, and PlanAtMO predict the most substantial boundary layer thickening at later stages of the Earth simulations, shown by the grey and pink lines in Figure~\ref{fig:interior}b and in Appendix~\ref{app:overview}, as does PACMAN, whose CBL thickens to $\sim$4\,m below $\Phi\approx3\%$. CBL growth exceeds $\sim1$km thickness during the final stages of NEONGOOEY and GOOEY simulations, when mantle melt fractions are less than 0.8\% -- this corresponds with NEONGOOEY and GOOEY predicting the lowest $T_\mathrm{surf}$ at final stages of solidification (Figure~\ref{fig:interior}a). Given its late-stage formation during the final magma ocean phase of these models, this assumed CBL would have only a small opportunity for inhibiting magma ocean cooling. PROTEUS simulations impose a constant small CBL thickness of $1$\,cm while predicting comparable Nominal Earth solidification timescales to GOOEY (Figure~\ref{fig:earth_venus}). However, these models differ in several other components, so similar cooling times could arise from compensating differences and do not by themselves isolate the effect of the CBL. More directly, Figure~\ref{fig:cbl_flux} (shown in Appendix~\ref{app:cbl}) compares each model's interior heat flux against its CBL thickness and the temperature contrast $T_\mathrm{pot}-T_\mathrm{surf}$ at matched melt fractions. Since $F_b=k_b(T_m-T_\mathrm{surf})/d_b$, even a thin surface boundary layer can readily sustain a large temperature contrast when the heat flux is large: up to $\sim$800\,K in PROTEUS while $\Phi>95\%$. Between $\Phi=40\%$ and 5\%, where models spend most of their simulated time, median temperature contrasts are $\lesssim$20\,K in the BLT models. $T_\mathrm{surf}\approx T_\mathrm{pot}$ in this regime, and the models do not depart from linear behaviour in panel (b) of the figure, which together show that planets' cooling rates are more limited by the net atmospheric energy loss than by interior heat transport or by a surface CBL. We conclude that the presence of a CBL atop primordial magma oceans, if it were to form at all \citep{ElkinsTanton2008}, would have a minor effect on their thermal evolution within the magma ocean regime considered here ($\Phi\geq5\%$) --- although it may become important at later near-solidified stages as mantle viscosity increases. The formation of surficial boundary layers could become relevant after a transition to post-magma ocean tectonic regimes \citep{korenaga_initiation_2013, tackley_self_2000, meier_mantle_2026, maurice_onset_2017, gillmann_atmosphere_2014}.

\section{Discussion} 
\label{sec:discuss}

\subsection{Simulated differences arising from model choices}
\label{sec:discuss_difference}

All models' simulations of Nominal Earth and Venus are consistent with empirical constraints on these planets' solidification time-scales derived from in-situ measurements (Figure~\ref{fig:earth_venus}).  The simulated termination ages for an early magma ocean in our Nominal Earth scenario occur long before lunar formation, at ages between $50\sim200$\,Myr \citep{borg_moon_2014, barboni_early_2017}, allowing for multiple distinct magma ocean epochs during the Hadean. This is consistent with indications from measurements of highly fractionated $^3$He/$^{22}$Ne isotope ratios from mid-ocean ridge degassing \citep{tucker_evidence_2014}. Additionally, the solidification time-scales for all Earth cases (Nominal case in Figure~\ref{fig:earth_venus}, and at various H+C inventories in Figure~\ref{fig:cooling_time}) respect the radiometrically estimated ages of Hadean mafic intrusions dated to have solidified within 300\,Myr of Earth's formation \citep{dalrymple_age_2001, sole_hadean_2025}.

We also expect that Venus would take longer to solidify than Earth, all else equal, given its higher insolation \citep{Hamano2013, krissansen2021venus}. Our models reproduce this expectation (Figure~\ref{fig:earth_venus}).

However, there are quantitative differences between the models' simulated evolution of Nominal Earth and Venus. The evolutionary timescales for reaching sequential melt fraction thresholds (Figure~\ref{fig:cooling_time}) vary across orders of magnitude, reflecting our different modelling assumptions and approximations. Part of this spread arises from known limitations of specific models (Section~\ref{sec:results_radtrans}). We have identified the key assumptions which shape these different timescales.

Firstly, atmospheric composition strongly regulates planetary cooling through gas radiative opacities blanketing the interior, inducing a greenhouse effect (Section~\ref{sec:results_chem}). The models' approximations of atmospheric chemistry are thereby reflected in the cooling timescales through climate modelling. In particular, models' approaches to calculating vertical atmospheric temperature structure determines the planet's net outgoing radiation flux (Figure~\ref{fig:olr}). However, our models agree that a planet's post-formation hydrogen budget strongly dictates magma ocean cooling rates \citep{Hamano2015, nicholls_redox_2024}.

Secondly, since these modelled atmospheric compositions are directly calculated from mantle oxidation state, interior redox processes strongly shape atmospheric chemistry. Outgassed \ch{CO2}/\ch{CO} ratios vary as a function of degassing temperature and mantle ferric-ferrous iron partitioning (Section~\ref{sec:results_chem}), so fully-coupled treatments of planetary interior and atmospheric evolution are critical to self-consistent evolution modelling. Atmospheric escape remains a substantial point of model disagreement (Figure~\ref{fig:escape}), although our Venus simulations indicate that volatile element fractionation plays a minor role in regulating magma ocean thermal evolution (Section~\ref{sec:results_escape}).

Thirdly, mantle temperature structure, energy transport processes, and assumed phase-state boundaries (i.e. melting curves) substantially influence when and where planetary interiors are (semi-)molten. Model inter-comparison is best undertaken at states of similar melt fraction, rather than surface temperatures, because model $\Phi-T_\mathrm{surf}$ relations differ substantially. In addition to our models' adoption of different melting curve parametrisations (Appendix~\ref{app:melting_curves}), mantle solidus and liquidus boundaries are also sensitive to multiple parameters and physical effects not considered by these codes \citep{katz_a_2003, ghiorso_the_2002, baumeister_fundamental_2025}. For example, the incorporation of dissolved water (as OH$^-$ ions) into the molten silicate is effective at disrupting crystal polymerisation \citep{katz_a_2003, whittington_water_2000}. These dissolved volatiles have the effect of `solidus depression', which readily reduces the solidus temperatures by $\sim100$s of kelvin, depending on the volatile content, to further sustain molten states \citep{ElkinsTanton2008}. Figures~\ref{fig:cooling_time}~and~\ref{fig:interior} show that differences in these existing thermodynamic modelling choices lead to cooling timescales across multiple orders of magnitude, so careful future consideration of additional physical sensitivities in mantle melting properties is advised \citep{thompson_on_2026}. 

That our models agree upon bottom-up solidification scenarios without basal magma oceans could be viewed as physical robustness. However, it largely arises from all models' parameters being tuned to reproducing a bottom-up crystallisation mode by their choice of melting curves \citep{ElkinsTanton2008, labrosse_fractional_2015, solomatov_nonfraction_1993}. More substantial inter-model deviations would be expected had they adopted different solidii, or considered a diverse range of crystal cumulate sizes that have various proclivities for settling-out of turbulent magma ocean flow \citep{spaargaren_the_2020, bower_numerical_2018, Lichtenberg2023}.

\subsection{Physics of magma ocean environments}
\label{sec:discuss_physics}

The lifetime histories of Earth and Venus raise numerous open questions. For example, there are substantial uncertainties in Earth's current volatile budget \citep{peslier_water_2017, lichtenberg_a_2019, MARTY2025383} and the abundance of light element in its metallic core \citep{labrosse_core_1997, schoenberg_new_2002, Shahar2026}. Synergistic application of planetary formation models and coupled interior-atmosphere simulations can, in principle, constrain the elemental compositions of these reservoirs by tracking volatile budgets. This desire also holds for Venus \citep{Hamano2024arXiv}, Mars \citep{drilleau_constrainin_2026, maurice_onset_2017}, and exoplanet scenarios \citep{nicholls_volatile_2026, barth2021magma}. However, our model intercomparison (Section~\ref{sec:results}) suggests that physical parameters to be inferred (e.g. a planet's initial H content) are largely overprinted by the substantial differences in predicted planetary evolution behaviours which arise from our modelling choices and parametrisations.

We have shown that simulated outcomes depend on multiple physical and chemical processes, which exhibit different levels of sensitivity in their relative importance and modelled parametrisations, and mutually interfere in either constructive or detractive modes. Our controlled intercomparison of planetary evolution models has benchmarked these processes in the context of Venus and Earth. It is critical to identify the physical and chemical processes which (i) must be included in models because they exert leading-order control over planetary composition and evolution, and (ii) must be incorporated into models accurately. Ideally, by focussing on these processes, we can work towards a point where evolutionary models agree on planetary cooling timescales and compositions. Table~\ref{tab:physics} presents the key physical parametrisations adopted by our models and, on the basis of this intercomparison, categorises these processes' relative importance in controlling planetary thermo-compositional evolution as having a `major' or `minor' impact. Processes whose implementation within the \chili codes are typically limited scientific or numerical modelling choices -- rather than by uncertainties in the active physics arising from empirical or experimental gaps -- are indicated with daggers $(\dagger)$ in the leftmost column; these processes are prime targets for future model development. 

\begin{table*}
\centering
\caption{Physical and chemical processes regulating the melting and composition of rocky planets, informed by our analysis of Earth and Venus simulations with different simulation frameworks. Here, processes are qualitatively labelled as having `major' or `minor' impact on thermal and compositional evolution, on the basis of quantitative analysis undertaken in the Results and Discussion sections. Only physical and chemical processes that were explicitly treated and compared in this study are listed. Processes whose implementation within the \chili codes are limited modelling choices, rather than by unknown physics or experimental uncertainty, are indicated with daggers $(\dagger)$. `MO' abbreviates `magma ocean'.}
\begin{tabular}{ l | l l | l l}
Modelled process  & Thermal effect & Impact & Compositional effect & Impact \\
\hline

Hydrodynamic escape & Enables solidification & Major & Enables atmosphere loss & Major \\

Carbon chemistry$^\dagger$ & MO prolonged & Minor & Carbon species form and increases $p_\mathrm{surf}$ & Major \\

Deep storage of trapped melt  & MO foreshortened & Minor & Hydrogen retained in planet & Major \\

Radiative layers and gaseous absorption$^\dagger$  & MO cooling modulated & Major & No explicit effect & Minor \\

Non-convective mantle layers$^\dagger$  & MO at lower $T_\mathrm{surf}$ & Major & Volatile dissolution into melt & Minor \\

Rheology parametrisations  & MO cooling modulated & Major & Volatile dissolution and trapping & Minor \\

Melting curve variations$^\dagger$  & MO prolonged & Major & Volatile dissolution into melt & Minor \\

Iron redox phase partitioning$^\dagger$  & Minimal change & Minor & Late-stage MO oxidation & Major \\

Escape-driven C/O fractionation   & Minimal change & Minor & Atmosphere C-enrichment & Minor \\

Escape-driven H/O fractionation & Oxygen enhancement & Minor & Gradual MO oxidation & Minor \\

Conductive skin layer$^\dagger$  & Minimal change & Minor & None & Minor %

\label{tab:physics}
\end{tabular}
\end{table*}

\subsection{Towards model certainty from empirical constraints}

We can better calibrate our models by adopting additional measurement constraints. In this paper, we have tentatively used estimates for CAI ages \citep{baker_early_2005}, Earth's age at core-differentiation \citep{dalrymple_age_2001}, and the ages of the oldest mineral samples \citep{sole_hadean_2025, Mojzsis2001Natur}, to bracket the valid range of timescales on Earth's early thermal evolution (Figures~\ref{fig:earth_venus}~and~\ref{fig:cooling_time}). All models here respect these age constraints, but these leave orders of magnitude flexibility in cooling timescale, and do not directly constrain compositions or climate states.

The simulation of other governing physics within these codes may enhance model accuracy \citep{Lichtenberg2025Science, nicholls_rocky_2026, Schaefer2018RSPTA}. Meanwhile, further model developments to predict additional historical fingerprints and proxies would be valuable as points for calibration against reality. 

For example, future iterations of these codes could calculate D/H isotopic fractionation due to hydrodynamic escape; comparison of models' predicted D/H ratios against each other \citep{Schaefer2024JGRE}, and against empirical measurements from Earth \citep{peslier_water_2017} and Venus \citep{grinspoon_implication_1993}, would inform us about the accuracy of their escape implementations. Another opportunity for model intercomparison and empirical calibration is through parametrisation of gas-melt and melt-crystal noble gas partitioning into these codes. Earth and Venus measurements of argon \citep{orourke_venus_2015, kaula_venus_1999} and xenon  \citep{rzeplinski_hadean_2022} fractionation are suggestive of particular magma ocean crystallisation scenarios. Yet, the \chili models do not homogeneously quantify metrics for comparison against these measurements \citep{krissansen2021venus, gilbert_a_2026}.

\subsection{Opportunities for the future}
\label{sec:discuss_future}

No model is ever `correct', but we may strive towards sufficient model robustness and complexity in order to understand and explain measurements of the natural world \citep{box_science_1976, Sohl24_cuisines}. In addition to the probed processes indicated by daggers in Table~\ref{tab:physics}, here we discuss the potential for additional, unmodelled physical processes to shape planetary compositions and structures. These additional processes could be considered for incorporation into the \chili codes and similar modelling frameworks with the aim of resolving disagreements presented here (Section~\ref{sec:discuss_difference}) and enabling their valid application to other regimes and planetary environments.

Figure~\ref{fig:interior} highlights that these models' different approaches to simulating mantle temperature structures and mantle rheological properties can partially explain their different predicted magma ocean cooling timescales. Ultimately, these differences in interior-modelling approach reflect the complex microphysics regulating mantle crystallisation, cumulate entrainment, Stokes settling, melt compaction/percolation, and mush dynamics \citep{wordsworth2018redox, lichtenberg_redox_2021, Schaefer2024JGRE,katz_physics_2022}. Future model developments should carefully incorporate robust parametrisations of these effects, which are -- in some scenarios -- also expected to generate basal magma ocean outcomes \citep{boukare_solidificat_2025, boukare_timing_2018, labrosse_fractional_2015, lark_coupled_2026}. A basal magma ocean outcome is an important consideration, since we have shown that deep volatile storage regulates outgassed atmospheric compositions (Section~\ref{sec:results_chem}). Our intercomparison highlights the leading-order role of dynamic melt trapping and the corresponding sequestration of volatiles (Figure~\ref{fig:inv_earth}) -- processes strongly dependent on these microphysics, yet poorly constrained by experiments \citep{katz_physics_2022}. Several \chili models do not resolve radial melt fraction profiles, so our comparisons rely on each model's own method to derive a `bulk' melt fraction $\Phi$ (Section~\ref{sec:methods_physics}). Future developments towards 1D-resolved melt fraction profiles would allow local melting states to be compared directly, removing ambiguity in the averaging domain and potentially permitting enhanced physicality in these simulations.

We have identified important differences between \chili models' predicted Earth and Venus evolution pathways. Although models will necessarily adopt simplifications of reality, given finite computational resources, their parametrisations reflect our epistemological uncertainties in the underlying physics. For example, various atmospheric compositions are generated due to different treatments of solid-phase melt trapping (Section~\ref{sec:results_chem}). Yet, the partitioning coefficients of volatiles into mineral phases remain poorly constrained by experiments \citep{Schaefer2024JGRE}. Sensitivity tests (not plotted) with the LINCS code indicate that -- even for a specific model -- parameter choices relating to the efficiency of melt trapping strongly influence calculated atmospheric compositions. Future studies should carefully consider the microphysics of degassing within magma oceans, beyond the assumption of efficient surficial exchange universally adopted by the \chili codes, to potentially consider the microphysics of bubble formation within subsurface layers and melt trapping within the deep interior \citep{sparks_bubble_1978, bottinga_morb_1990, gaillard_bubble_2021, hier_the_2017, walbecq_the_2025}.

All of these codes treat the planet's metallic core as compositionally inert and leave it spatially unresolved. Yet, core fluid dynamics, cooling,  magnetism, and the core-mantle partitioning of volatiles are rich fields of study \citep{labrosse_core_1997, lay_coremantle_2008, stevenson_core_2001}. The deep core domain of planetary interiors could be incorporated into these interior-atmosphere evolution models \citep{huang_limits_2025, garcia_investigati_2026}. This could, for example, permit model predictions regarding the impact of dynamo-generated magnetism on atmospheric escape \citep{Gulcher2020NatGe}. Magnetic effects are suggested to enhance mass loss from Venus at the present day \citep{lammer_formation_2020, zahnle_mass_1986}.

The \chili protocol specifies fixed orbital parameters to limit intercomparison scope. However, planetary and satellite orbital dynamics are strongly coupled planetary interiors \citep{barnes_tidal_2017, korenaga_tidal_2025, farhat_the_2022}. Lunar tides and resonant states offer one explanation for the in-situ evidence of lunar surface remelting \citep{nimmo_tidally_2024, chen_tidal_2016, farhat_tides_2025}. Theoretical models can resolve the important coupling between tidal heating, orbital evolution, and thermal regulation by overlying atmospheres \citep{zahnle_the_2015, van_onset_2025}. Theoretical modelling of tidal heating within multi-planetary exoplanet systems also suggests that magma oceans may be sustained under temperate irradiation regimes \citep{hay_tides_2019, Herath2024MNRAS, unterborn_inward_2018, pu_low_2019}. Incorporation of tidal heating effects and orbital migration into planetary evolution models would be valuable, but may necessitate a careful intercomparison \citep{sabadini_global_2016}.

The \chili protocol also specifies clear-sky climate modelling. However, clouds strongly determine Earth's present day climate and weather \citep{pierrehumbert_principles_2010, pierrehumbert_the_2002}. Venus' photochemical sulfur clouds  are the source of its high $\approx75\%$ Bond albedo \citep{titov_clouds_2018}, and are crucial to suggested greenhouse feedback effects early in its lifetime \citep{krissansen2021venus, Hamano2024arXiv}. The presence of clouds may have determined whether Venus once hosted surface water \citep{Turbet2021, kitzmann_clouds_2010, pollack_a_1971}. More generally, aerosols can induce a greenhouse or anti-greenhouse effect depending on particle size distributions, particle shapes, compositions, and their vertical distribution \citep{marley_clouds_2014, helling_clouds_2019, lodge_fractal_2026}. Aerosols could extend or foreshorten the magma ocean solidification timescales predicted by the \chili models here \citep{piette_rocky_2023}. The `problem' of aerosols can be approached at various levels of modelling complexity \citep{helling_clouds_2019}, so it remains uncertain how best to study their role in deep planetary histories and resultant impact on present climates. MOAI incorporates Rayleigh scattering but neglects key collisional continua (e.g. \ch{H2O}-\ch{H2O}; \citet{lichtenberg2021}), and predicts Earth and Venus magma ocean solidification timescales intermediate to the other \chili models (Figures~\ref{fig:earth_venus}~and~\ref{fig:cooling_time}). The \chili models also adopt different approaches to calculating atmospheric temperature structure, which, alongside atmospheric composition, sets the planetary cooling rate and potential for cloud formation  \citep{cmiel2025, nicholls_convective_2025, peng2024, janssen_hot_2026}. Differences in models' predicted OLR motivate the adoption of accurate radiative transfer parameterisation, which are especially limited by unconstrained \ch{H2O} continuum absorption coefficients \citep{ABEL2013857, pierrehumbert_principles_2010}.  Overall, understanding the behaviours emergent from the intimate competition between thermodynamic and radiation processes, which we have shown to play a role in setting magma ocean cooling rates (Section~\ref{sec:results_radtrans}), requires a dedicated intercomparison of exoplanet climate models. The ongoing COD-ACCRA and MALBEC intercomparison projects, also part of CUISINES, will directly address these uncertainties \citep{villanueva_modeling_2024, chaverot_codaccra_2026}.

The discovery of sub-Neptune and super-Earth exoplanet populations overturned the picture of planetary formation and evolution established by the Solar System planets \citep{bean_the_2021, Valencia_2006}. Yet, current telescopes (e.g. JWST) and upcoming observational surveys (e.g. PLATO) can only probe exoplanets' bulk compositions and upper-atmosphere chemistry. Understanding the true nature of these new planetary populations requires comprehensive predictive and interpretive modelling approaches  \citep{Lichtenberg2025Science}. However, planetary evolution models have largely targeted Earth-sized planets, which remain the focus for \chili parts I, II, and III. Modelling of sub-Neptune magma oceans and thermal evolution may require a careful handling of additional physics. For example, sub-Neptunes' potential for interior-atmosphere (im)miscibility and high temperature supercritical fluid layers that necessitate the extension of thermodynamic data to extreme pressures and temperatures \citep{rogers_subneptune_2025, young_subneptune_2025, wolf_eos_2018, attia_paleos_2026}. The upper-atmosphere compositions probed by telescopes are also subject to disequilibrium chemical effects, potentially requiring the consideration of complex upper-atmospheric photochemistry within these coupled interior-atmosphere modelling frameworks \citep{nicholls_chemistry_2023, tsai_inferring_2021, yu_how_2021}. 

Ultra-short period exoplanets are highly observable with JWST through secondary eclipse and phase curve measurements \citep{hammond_reliable_2025, Teske2025, demory_a_2016}. The magma oceans on these `lava worlds' may be permanently sustained by intense stellar irradiation \citep{Lichtenberg2025}, so application of planetary evolution and climate modelling to these regimes can enable contemporary insight into the primordial environments of all rocky worlds \citep{janssen_hot_2026, nguyen2024clouds}. \chili Part II and Part III will address the opportunity afforded by lava worlds, to test the validity of coupled interior-atmosphere modelling under intense irradiation and identify the key physics shaping their observations. These planets' exposure to intense irradiation may lead to substantial atmospheric escape, in excess of that found for our Earth and Venus scenarios, and thereby modulate their thermal evolution. 

\section{Conclusions} 
\label{sec:conclude}

We have presented the first controlled intercomparison of planetary evolution models by extending the CUISINES intercomparison framework to Earth and Venus magma ocean evolution. While all seven of these \chili models predict similar thermal and compositional evolutionary behaviours for Earth and Venus, they have also shown substantial quantitative deviations. Our main conclusions are itemised below.

\begin{itemize}
    \item Our nominal Earth and Venus scenarios predict a rapid transition from a global magma ocean to a nearly solidified mantle, consistent with empirical constraints from the geological record. This result allows multiple episodic magma ocean phases during rocky planets' earliest periods. Fully-coupled application of known physical laws governing mantles and atmospheres is a demonstrably successful modelling approach towards understanding planetary lifetime histories.

    \item Venus simulations exhibit large inter-model variance in predicted cooling timescales. LINCS, PROTEUS, and PlanAtMO codes suggest sustained global magma oceans in pseudo steady-states of global radiative equilibrium; solidification is delayed until escape processes sufficiently erode early degassed greenhouse atmospheres. These states are key to understanding the evolutionary divergence between Earth and Venus, and whether a substantial part of the exoplanet population spends a majority of their lifetimes in global magma regimes.

    \item Cooling timescale variations are introduced by models' various physical parametrisations and approximations. Notably, by their different treatments of mantle geodynamics (e.g. mixing length versus boundary layer theory convection), and prescriptions of material melting properties and rheology. These modelling choices induce first-order evolutionary differences, exceeding those induced by variation in planets' post-formation volatile inventories (H and C budgets). 

    \item Gas phase thermochemistry and volatile partitioning behaviours are further points of model disagreement. The codes which incorporate deep-mantle hydrogen storage produce thinner, \ch{CO2}-dominated atmospheres. Other codes yield massive \ch{H2O} greenhouses that prolong the early magma oceans within `mushy' semi-molten states. Atmospheric vertical temperature structure sets outgoing radiation fluxes, which further modulates magma oceans' early thermal evolution.

    \item Multiple processes drive changes in mantle oxidation state during magma ocean cooling and solidification. The \chili models disagree on the processes regulating the transition from early reducing states towards later oxidised conditions. Atmospheric escape is suggested to be important, if sufficiently active, although differential ferric-ferrous partitioning acts simultaneously. Accurate microphysical parametrisations of these effects are needed.    
    
\end{itemize} 

These evolution models' various predictions of volatile thermochemical speciation, redox-dependent magma ocean outgassing, and mantle geodynamics on early Earth and Venus underscore the importance of accurate modelling. Validation of their different treatments of the key physics is a prerequisite for drawing meaningful conclusions from JWST, PLATO, and Ariel observations. Validation can be achieved by appealing to empirical approaches; new laboratory experiments can determine volatile partitioning coefficients under the appropriate conditions \citep{hirschmann_magma_2012, thompson_on_2026}. Additionally, measurements of Venus' composition and structure will allow lifetime comparative planetology against the Earth  (e.g. from the DAVINCI and EnVision missions -- \citet{garvin_revealing_2022, rosenblatt_determinati_2021}).

Most surveyed small exoplanets orbit M-type host, so will experience different lifetime irradiation exposures to Earth and Venus \citep{johnstone2021active, Baraffe2015}. The result of these planets' different stellar contexts are non-trivial to predict, but are thought to have profound implications for their thermal states and potential for habitability  \citep{luger2015extreme, fromont_atmospheric_2024}. \chili Part II will consider three TRAPPIST-1 exoplanet cases as a basis for inter-comparing planetary evolution models, enabling the interpretive tools necessary for exoplanetary science's continued success during the 2030s.


\section{Acknowledgements}

The \chili team thanks our two anonymous reviewers for their thoughtful feedback and constructive suggestions on this manuscript and its model intercomparison. 

\chili belongs to the CUISINES framework, a Nexus for Exoplanet System Science (NExSS) science working group. The \chili team thanks the Netherlands eScience Center (PROTEUS project, NLESC.OEC.2023.017) and the Lorentz Center for funding and support in the organisation of the workshop `Atmospheric and interior evolution of planetary magma oceans' in October 2025 in Leiden.

H.N. acknowledges support from STFC grant UKRI1184. H.N. and T.L. thank the Center for Information Technology of the University of Groningen for their support and for providing access to the H\'abr\'ok high performance computing cluster.

J.K.-T. was supported by a NASA Astrophysics Decadal Survey Precursor Science grant 80NSSC23K1471, the Virtual Planetary Laboratory, a member of the NASA Nexus for Exoplanet System Science (NExSS), funded via the NASA Astrobiology Program grant No.\ 80NSSC23K1398 and the Alfred P.\ Sloan Foundation under grant No.\ G-2025-25204.

T.L. was supported by the Branco Weiss Foundation, the Alfred P.\ Sloan Foundation (AEThER project, G202114194), NASA's Nexus for Exoplanet System Science research coordination network (Alien Earths project, 80NSSC21K0593), and the European Research Council (ERC) under the European Union's Horizon Europe research and innovation programme (101219807, MagmaWorlds).

K.H. was supported by KAKENHI Grant Number JP22H05150 from MEXT and JP25KJ0384 from JSPS.
M.M. acknowledges funding from the European Research Council (ERC) under the European Union's Horizon 2020 research and innovation program (Grant Agreement No.\ 101141606 --- FOREVER).

H.S. and A.d.L acknowledge the support from R\'egion \^Ile de France via the program DIM Origines (project DOMTERRA, under grant number IDF-DIM-ORIGINES-2024-1-03), and from the CNRS/INSU Programme et \'Equipements Prioritaires de Recherche (PEPR-Origins, project DECOMROT). Numerical computations were  performed on the S-CAPAD/DANTE platform, IPGP, France. 

C.O.-Q. acknowledges support from the NSF Graduate Research Fellowship Program (GRFP) Grant No.\ 2035702.

Y.M and L.J.J acknowledge support from the European Research Council (ERC) under the European Union’s Horizon 2020 research and innovation programme (grant agreement no. 101088557, N-GINE). 

P.B. and L.N. acknowledge funding by the European Union (ERC, DIVERSE, 101087755).

R.R. gratefully acknowledges support from the NASA Exoplanets Research Program (XRP) grant No.\ 80NSSC24K0162.

The simulations and analysis presented in this work benefitted from open-source software libraries: NumPy \citep{harris2020array} and Matplotlib \citep{Hunter2007}. Plots are presented using colour schemes developed for Julia by Paul\,Tol \citep{Julia2017}. This research has made use of data obtained from the portal \url{exoplanet.eu} of The Extrasolar Planets Encyclopaedia.

CRediT author statements.
\begin{itemize}

\item Conceptualization: HN, TL, LS, JKT, LJ, HP, MS, AZ, BP, DS, PB.

\item Methodology: HN, TL, LS, JKT, LJ, HP, BP, AZ.

\item Software: HN, TL, LS, JKT, KH, MM, HS, COQ, JK, AdL, EM, LN, EP, RR, MS, AZ, SD.

\item Investigation: HN, LS, KH, JKT, MM, HS, AP, COQ, LN, AZ, SD.

\item Formal analysis: HN, TL, KH, JKT.

\item Data Curation: HN, TL, LS, KH, HS, MS, AZ.

\item Writing -- Original Draft: HN.

\item Writing -- Review \& Editing: HN, TL, JKT, LS, KH, MM, HS, COQ, LJ, AdL, EM, BP, RR, SD.

\item Visualization: HN.

\item Supervision: TL, RR, JKT, LS, YM, LM, RR.

\item Project administration: TL, LS, JKT, DS.

\item Funding acquisition: TL, DS, JKT, LS.

\item Resources: TL, KH, RR, LS, YM, JKT.

\end{itemize}

\appendix
\section{Comparison of melting curves}
\label{app:melting_curves}
\begin{figure}[!h]
    \centering
    \includegraphics[width=0.95\linewidth]{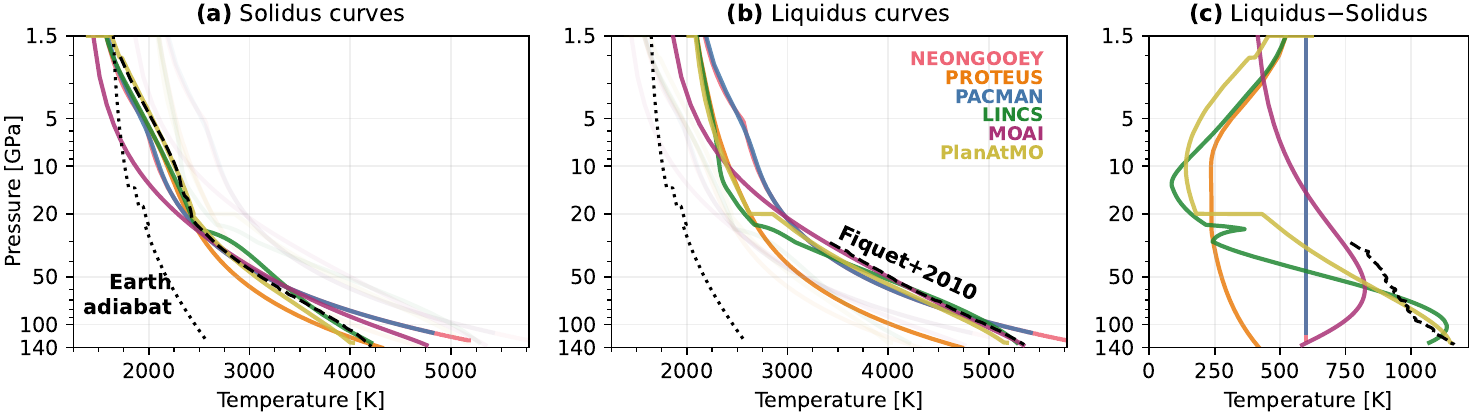}%
    \caption{
    Melting curves used by each model (line colour), which are described in Section~\ref{sec:methods_models}, compared to the literature (dashed black line; \cite{fiquet_melting_2010}) and a recent estimate of Earth's mantle adiabat (dotted black line; \cite{katsura_earth_2022}). The left \textbf{(a)} and middle \textbf{(b)} panels show solidus and liquidus curves, respectively, with their difference plotted on the right \textbf{(c)} panel. 
    }
    \label{fig:melting_curves}
\end{figure}

In this appendix section, we present dry melting curves adopted by each model in their simulated evolutionary calculations of Earth. Figure~\ref{fig:melting_curves} plots the solidus (panel a) and liquidus (panel b) used by each model (line colours),and their difference (panel c). In addition, we include a recent estimate for the interior structure of present-day Earth's mantle adiabat (black line) to contextualise the modelled melting curves. Although mantle melting properties are sensitive to mineralogical composition and dissolved volatile content \citep{ghiorso_the_2002,hirschmann_mantle_2000, katz_a_2003}, these curves\ broadly agree in shape, but differ by up to $\sim$450\,K at upper-mantle pressures ($\leq25$\,GPa) and by up to $\sim$1200\,K (solidus) and $\sim$1300\,K (liquidus) in the lower mantle, while the liquidus-solidus interval ranges from $\sim$90 to $\sim$1150\,K between models (panel c). All curves are consistent with a solidified terrestrial interior at the present day (black line). GOOEY simulations adopt the same melting curves as NEONGOOEY.

\section{\texorpdfstring{Interior heat flux and surface temperature contrast}{Interior heat flux and surface temperature contrast}}
\label{app:cbl}
\begin{figure}[!h]
    \centering
    \includegraphics[width=0.85\linewidth]{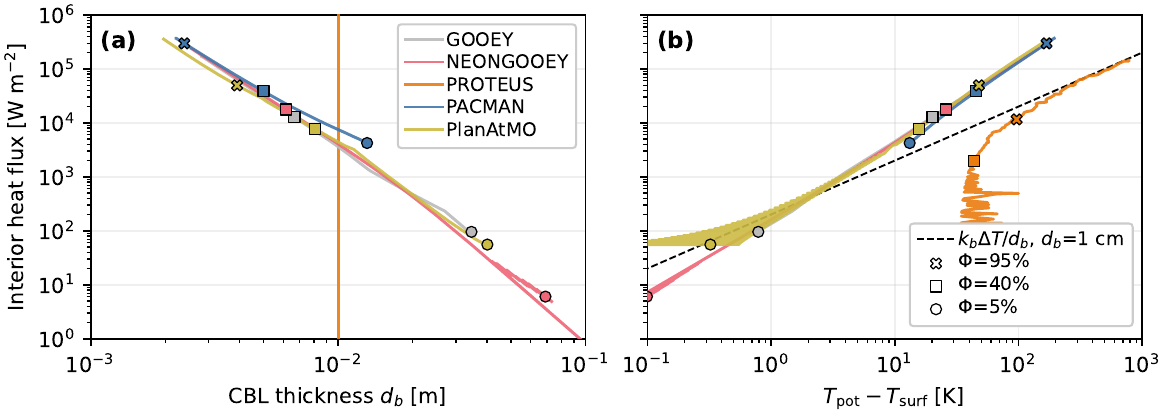}%
    \caption{
    Interior heat flux reported by each model (line colour) for Nominal Earth, versus \textbf{(a)} conductive boundary layer thickness and \textbf{(b)} the temperature contrast between the mantle potential temperature and the surface temperature. Markers indicate melt fractions of 95\% (crosses), 40\% (squares), and 5\% (circles). PROTEUS uses a constant CBL thickness fixed at 1\,cm. The dashed linear black line in (b) shows the conductive flux $k_b\Delta T/d_b$ through an equivalent 1\,cm layer, with $k_b=2$\,W\,m$^{-1}$\,K$^{-1}$. Points on the dashed line correspond to a contrast lying entirely across a 1\,cm conductive layer.}
    \label{fig:cbl_flux}
\end{figure}

To diagnose whether magma ocean cooling is limited by atmospheric radiation or by interior heat transport, Figure~\ref{fig:cbl_flux} plots the interior heat flux reported by each model against \textbf{(a)} the CBL thickness $d_b$ and \textbf{(b)} the temperature contrast $T_\mathrm{pot}-T_\mathrm{surf}$ between the mantle and the surface, for Nominal Earth. Markers indicate states at matched melt fractions of $\Phi=95$, 40, and 5\%. For models that parametrise a CBL, $F_b=k_b(T_m-T_\mathrm{surf})/d_b$ (Equation~\ref{eq:cbl}), so the temperature contrast increases with the heat flux transported out of the planet. During the initial high-flux stage ($\Phi>40\%$, $F_b\gtrsim\SI{e4}{\WPMS}$), far from radiative equilibrium, median contrasts of $\sim$25--200\,K arise in GOOEY, NEONGOOEY, PACMAN, and PlanAtMO (whose CBLs become thinner as the flux increases). In PROTEUS, the prescribed 1\,cm CBL thickness sustains a median contrast of 290\,K (up to $\sim$800\,K) while $\Phi>95\%$. Between $\Phi=40\%$ and 5\%, where all models spend 63--100\% of their simulated time before reaching $\Phi=5\%$, the CBLs remain thinner than $\sim$13\,cm and median temperature contrasts are $\lesssim$20\,K for the BLT models. In PROTEUS, the median contrast of 43\,K in this regime arises mostly within the 1D-resolved mantle profile rather than across the CBL itself. LINCS is omitted since it does not parametrise a CBL.

\section{Atmosphere climate profiles}
\label{app:climate}

Atmospheric climate plays an important role in setting planetary thermal evolution. This appendix section presents temperature-pressure profiles to contextualise the outgoing radiation fluxes plotted in Figure~\ref{fig:olr} and highlight whether water clouds may form within these atmospheres. 

\begin{figure}[!h]
    \centering
    \includegraphics[height=7cm]{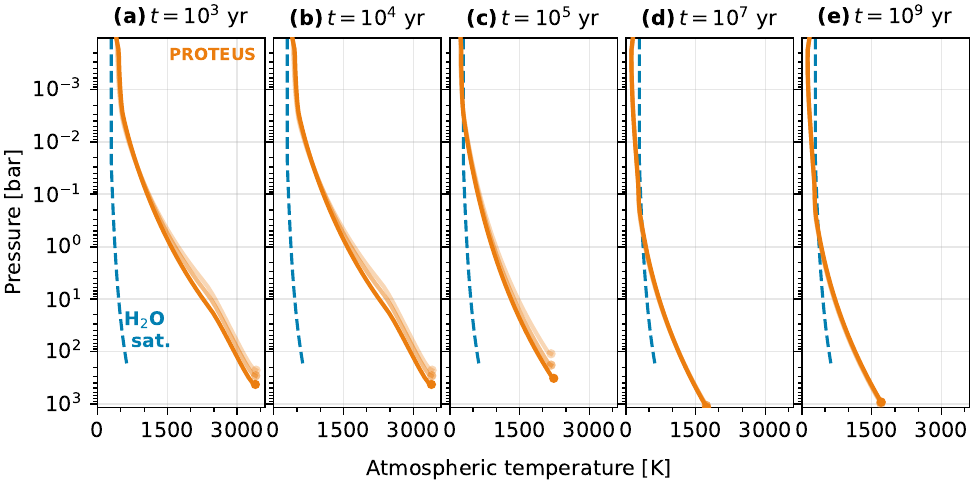}%
    \caption{
    Snapshots of simulated atmospheric $T(p)$ profiles evolving over time (panels \textbf{a}--\textbf{e})  for Earth maximum-hydrogen inventory cases (three carbon inventories per model). Line colour indicates the model used. Each line opacity indicates the carbon inventory. Dashed blue line shows the 100\%-\ch{H2O} saturation curve \citep{wagner_iapws_2002}.
    }
    \label{fig:climate_profiles}
\end{figure}

Figure~\ref{fig:climate_profiles} plots snapshots (panels) of atmosphere temperature profiles for Hhigh maximum-hydrogen cases simulated by PROTEUS, at various carbon inventories (line opacity). These snapshots can be compared against static models in \chili Part III. Furthermore, crossing of the \ch{H2O} saturation curve (dashed blue line) suggests possible water cloud formation during early magma ocean evolution.

\section{Overview plots of Nominal Earth and Venus simulations}
\label{app:overview}

This appendix section presents evolution tracks of model variables for Nominal Earth and Venus scenarios, to help contextualise our main analysis derived from these simulations. Figures~\ref{fig:overview_gooey}--\ref{fig:overview_planatmo} plot volatile partial pressures, reservoir partitioning fractions, surface and mantle temperatures, energy fluxes, mantle rheological properties, and boundary layer depths -- for each \chili evolution model as a function of simulated time.

\foreach \model in {gooey,neongooey,proteus,pacman,lincs,moai,planatmo}{
    \begin{figure}[!htb]
        \centering
        \includegraphics[height=0.52\linewidth]{fig-overview_\model_earth_nominal.pdf}%
        \hspace*{1mm}
        \includegraphics[height=0.52\linewidth]{fig-overview_\model_venus_nominal.pdf}%
        \caption{
        Overview for evolution of Nominal Earth (left) and Nominal Venus (right) simulated by \MakeUppercase{\model}. Rows show diagnostic variables of: total surface pressure, elemental distributions between atmosphere / molten-mantle / solid-mantle reservoirs, temperatures, energy fluxes, mantle dynamic viscosity, and surface boundary layer thickness.
        }
        \label{fig:overview_\model}
    \end{figure}
}

\pagebreak
\clearpage
\bibliography{main}{}

@ARTICLE{Gulcher2020NatGe,
       author = {{G{\"u}lcher}, Anna J.~P. and {Gerya}, Taras V. and {Mont{\'e}si}, Laurent G.~J. and {Munch}, Jessica},
        title = "{Corona structures driven by plume-lithosphere interactions and evidence for ongoing plume activity on Venus}",
      journal = {Nature Geoscience},
         year = 2020,
        month = jul,
       volume = {13},
       number = {8},
        pages = {547-554},
          doi = {10.1038/s41561-020-0606-110.31223/x5jk88},
       adsurl = {https://ui.adsabs.harvard.edu/abs/2020NatGe..13..547G}
}

@inbook{frost_chapter_2018,
    url = {https://doi.org/10.1515/9781501508684-004},
    title = {Chapter 1.INTRODUCTION TO OXYGEN FUGACITY AND ITS PETROLOGIC IMPORTANCE},
    booktitle = {Oxide Minerals},
    author = {B. Ronald Frost},
    publisher = {De Gruyter},
    address = {Berlin, Boston},
    pages = {1--10},
    doi = {doi:10.1515/9781501508684-004},
    isbn = {9781501508684},
    year = {1991},
    lastchecked = {2026-05-16}
}

@book{stamnes2017radiative,
  title={Radiative Transfer in the Atmosphere and Ocean},
  author={Stamnes, K. and Thomas, G.E. and Stamnes, J.J.},
  isbn={9781108210423},
  url={https://books.google.co.uk/books?id=GN0qDwAAQBAJ},
  year={2017},
  publisher={Cambridge University Press}
}

@article{mosenfelder_the_2009,
	author = {Mosenfelder, Jed L. and Asimow, Paul D. and Frost, Daniel J. and Rubie, David C. and Ahrens, Thomas J.},
	title = {{The MgSiO3 system at high pressure: Thermodynamic properties of perovskite, postperovskite, and melt from global inversion of shock and static compression data}},
	journal = {J. Geophys. Res. Solid Earth},
	volume = {114},
	number = {B1},
	year = {2009},
	month = jan,
	issn = {0148-0227},
	publisher = {John Wiley {\&} Sons, Ltd},
	doi = {10.1029/2008JB005900}
}

@article{schiano_primitive_2003,
	author = {Schiano, Pierre},
	title = {{Primitive mantle magmas recorded as silicate melt inclusions in igneous minerals}},
	journal = {Earth-Sci. Rev.},
	volume = {63},
	number = {1},
	pages = {121--144},
	year = {2003},
	month = oct,
	issn = {0012-8252},
	publisher = {Elsevier},
	doi = {10.1016/S0012-8252(03)00034-5}
}

@article{meier_mantle_2026,
	author = {Meier, Tobias G. and Guimond, Claire Marie and Pierrehumbert, Raymond T. and Birkby, Jayne and Chatterjee, Richard D. and Fisher, Chloe E. and Golabek, Gregor J. and others},
	title = {{Mantle convection and nightside volcanism on lava world K2-141{\hspace{0.167em}}b}},
	journal = {Mon. Not. R. Astron. Soc.},
	volume = {547},
	number = {3},
	year = {2026},
	month = mar,
	issn = {0035-8711},
	publisher = {Oxford Academic},
	doi = {10.1093/mnras/stag390}
}

@article{tackley_self_2000,
	author = {Tackley, Paul J.},
	title = {{Self-consistent generation of tectonic plates in time-dependent, three-dimensional mantle convection simulations}},
	journal = {Geochem. Geophys. Geosyst.},
	volume = {1},
	number = {8},
	year = {2000},
	month = aug,
	issn = {1525-2027},
	publisher = {John Wiley {\&} Sons, Ltd},
	doi = {10.1029/2000GC000036}
}

@article{drilleau_constrainin_2026,
	author = {Drilleau, M{\ifmmode\acute{e}\else\'{e}\fi}lanie and Samuel, Henri and Verhoeven, Olivier and Rivoldini, Attilio and Collinet, Max and Garcia, Rapha{\ifmmode\ddot{e}\else\"{e}\fi}l F. and Lognonn{\ifmmode\acute{e}\else\'{e}\fi}, Philippe},
	title = {{Constraining the Thermochemical Structure of Mars Through Joint Inversion of Multidisciplinary Geophysical Data}},
	journal = {J. Geophys. Res. Planets},
	volume = {131},
	number = {2},
	pages = {e2025JE009303},
	year = {2026},
	month = feb,
	issn = {2169-9097},
	publisher = {John Wiley {\&} Sons, Ltd},
	doi = {10.1029/2025JE009303}
}

@article{whittington_water_2000,
	author = {Whittington, Alan and Richet, Pascal and Holtz, Francois},
	title = {{Water and the viscosity of depolymerized aluminosilicate melts}},
	journal = {Geochim. Cosmochim. Acta},
	volume = {64},
	number = {21},
	pages = {3725--3736},
	year = {2000},
	month = nov,
	issn = {0016-7037},
	publisher = {Pergamon},
	doi = {10.1016/S0016-7037(00)00448-8}
}

@book{sabadini_global_2016,
	author = {Sabadini, Roberto and Vermeersen, Bert and Cambiotti, Gabriele},
	title = {{Global Dynamics of the Earth: Applications of Viscoelastic Relaxation Theory to Solid-Earth and Planetary Geophysics}},
    year = 2016,
    month = jun,
	journal = {SpringerLink},
	isbn = {978-94-017-7552-6},
	publisher = {Springer Netherlands},
	address = {Dordrecht, The Netherlands},
	url = {https://link.springer.com/book/10.1007/978-94-017-7552-6}
}

@article{hay_tides_2019,
	author = {Hay, Hamish C. F. C. and Matsuyama, Isamu},
	title = {{Tides Between the TRAPPIST-1 Planets}},
	journal = {Astrophys. J.},
	volume = {875},
	number = {1},
	pages = {22},
	year = {2019},
	month = apr,
	issn = {0004-637X},
	publisher = {The American Astronomical Society},
	doi = {10.3847/1538-4357/ab0c21}
}

@article{farhat_the_2022,
	author = {Farhat, Mohammad and Auclair-Desrotour, Pierre and Bou{\ifmmode\acute{e}\else\'{e}\fi}, Gwena{\ifmmode\ddot{e}\else\"{e}\fi}l and Laskar, Jacques},
	title = {{The resonant tidal evolution of the Earth-Moon distance}},
	journal = {Astron. Astrophys.},
	volume = {665},
	pages = {L1},
	year = {2022},
	month = sep,
	issn = {0004-6361},
	publisher = {EDP Sciences},
	doi = {10.1051/0004-6361/202243445}
}

@article{korenaga_tidal_2025,
	author = {Korenaga, Jun},
	title = {{Tidal dissipation within Earth{'}s solidifying magma ocean: III. Effects of matrix compaction}},
	journal = {Icarus},
	volume = {442},
	pages = {116759},
	year = {2025},
	month = dec,
	issn = {0019-1035},
	publisher = {Academic Press},
	doi = {10.1016/j.icarus.2025.116759}
}

@article{box_science_1976,
    author = {Box, George E. P.},
    title = {Science and Statistics},
    journal = {Journal of the American Statistical Association},
    volume = {71},
    number = {356},
    pages = {791--799},
    year = {1976},
    publisher = {ASA Website},
    doi = {10.1080/01621459.1976.10480949},
    URL = { 
            https://www.tandfonline.com/doi/abs/10.1080/01621459.1976.10480949
    },
    eprint = { 
            https://www.tandfonline.com/doi/pdf/10.1080/01621459.1976.10480949
    }
}

@article{gilbert_a_2026,
	author = {Gilbert-Janizek, Samantha and Barnes, Rory K. and Driscoll, Peter E. and Wogan, Nicholas F. and Mandell, Avi M. and Birky, Jessica L. and Carone, Ludmila and others},
	title = {{A whole-planet model of the Earth without life for terrestrial exoplanet studies}},
	journal = {arXiv},
	year = {2026},
	month = feb,
	eprint = {2602.02267},
	doi = {10.48550/arXiv.2602.02267}
}

@phdthesis{nicholls_rocky_2026,
  edition = {},
  number = {},
  journal = {},
  booktitle = {},
  pages = {},
  publisher = {University of Oxford},
  school = {University of Oxford},
  title = {What happened to rocky planets?},
  volume = {},
  author = {Nicholls, H},
  editor = {},
  year = {2026},
  series = {},
  doi = {10.5287/ora-bmz5xpbrk}
}

@article{baker_early_2005,
	author = {Baker, Joel and Bizzarro, Martin and Wittig, Nadine and Connelly, James and Haack, Henning},
	title = {{Early planetesimal melting from an age of 4.5662{\hspace{0.167em}}Gyr for differentiated meteorites}},
	journal = {Nature},
	volume = {436},
	number = {7054},
	pages = {1127--1131},
	year = {2005},
	month = aug,
	issn = {1476-4687},
	publisher = {Nature Publishing Group},
	doi = {10.1038/nature03882}
}

@article{lark_coupled_2026,
	author = {Lark, Laura H. and Boukar{\ifmmode\acute{e}\else\'{e}\fi}, Charles-{\ifmmode\acute{E}\else\'{E}\fi}douard and Badro, James and Samuel, Henri},
	title = {{Coupled thermochemical evolution of the early Earth{'}s solid mantle and basal magma ocean: The role of melting and melt transport}},
	journal = {Earth Planet. Sci. Lett.},
	volume = {680},
	pages = {119880},
	year = {2026},
	month = apr,
	issn = {0012-821X},
	publisher = {Elsevier},
	doi = {10.1016/j.epsl.2026.119880}
}

@article{baumeister_fundamental_2025,
	author = {Baumeister, Philipp and Miozzi, Francesca and Guimond, Claire Marie and Steinmeyer, Marie-Luise and Dorn, Caroline and Karato, Shun-Ichiro and Bolmont, {\ifmmode\acute{E}\else\'{E}\fi}meline and others},
	title = {{Fundamentals of Interior Modelling and Challenges in the Interpretation of Observed Rocky Exoplanets}},
	journal = {Space Sci. Rev.},
	volume = {221},
	number = {8},
	pages = {123},
	year = {2025},
	month = dec,
	issn = {1572-9672},
	publisher = {Springer Netherlands},
	doi = {10.1007/s11214-025-01248-5}
}

@article{maurice_onset_2017,
	author = {Maurice, Maxime and Tosi, Nicola and Samuel, Henri and Plesa, Ana-Catalina and H{\ifmmode\ddot{u}\else\"{u}\fi}ttig, Christian and Breuer, Doris},
	title = {{Onset of solid-state mantle convection and mixing during magma ocean solidification}},
	journal = {J. Geophys. Res. Planets},
	volume = {122},
	number = {3},
	pages = {577--598},
	year = {2017},
	month = mar,
	issn = {2169-9097},
	publisher = {John Wiley {\&} Sons, Ltd},
	doi = {10.1002/2016JE005250}
}

@ARTICLE{Maurice_MOAI_2024,
       author = {{Maurice}, M. and {Dasgupta}, R. and {Hassanzadeh}, P.},
        title = "{Volatile atmospheres of lava worlds}",
      journal = {\aap},
         year = 2024,
        month = aug,
       volume = {688},
          eid = {A47},
        pages = {A47},
          doi = {10.1051/0004-6361/202347749},
archivePrefix = {arXiv},
       eprint = {2405.09284},
 primaryClass = {astro-ph.EP},
       adsurl = {https://ui.adsabs.harvard.edu/abs/2024A&A...688A..47M}
}

@article{korenaga_initiation_2013,
	author = {Korenaga, Jun},
	title = {{Initiation and Evolution of Plate Tectonics on Earth: Theories and Observations}},
	journal = {Annu. Rev. Earth Planet. Sci.},
	volume = {41},
	pages = {117--151},
	year = {2013},
	month = may,
	publisher = {Annual Reviews},
	doi = {10.1146/annurev-earth-050212-124208}
}

@techreport{iso_water,
type = {Standard},
key = {{ISO/TR\,3666}},
journal = {International Organization for Standardization},
month = dec,
year = {1998},
title = {{Viscosity of water}},
address = {Geneva, CH},
institution = {International Organization for Standardization}
}

@article{huber_new_2009,
	author = {Huber, M. L. and Perkins, R. A. and Laesecke, A. and Friend, D. G. and Sengers, J. V. and Assael, M. J. and Metaxa, I. N. and others},
	title = {{New International Formulation for the Viscosity of H2O}},
	journal = {J. Phys. Chem. Ref. Data},
	volume = {38},
	number = {2},
	pages = {101--125},
	year = {2009},
	month = apr,
	issn = {0047-2689},
	publisher = {AIP Publishing},
	doi = {10.1063/1.3088050}
}

@article{turner_convection_1986,
	author = {Turner, J. S. and Campbell, I. H.},
	title = {{Convection and mixing in magma chambers}},
	journal = {Earth-Sci. Rev.},
	volume = {23},
	number = {4},
	pages = {255--352},
	year = {1986},
	month = aug,
	issn = {0012-8252},
	publisher = {Elsevier},
	doi = {10.1016/0012-8252(86)90015-2}
}

@article{karato_rheology_1993,
	author = {Karato, Shun-ichiro and Wu, Patrick},
	title = {{Rheology of the Upper Mantle: A Synthesis}},
	journal = {Science},
	volume = {260},
	number = {5109},
	pages = {771--778},
	year = {1993},
	month = may,
	issn = {0036-8075},
	publisher = {American Association for the Advancement of Science},
	doi = {10.1126/science.260.5109.771}
}

@article{clough_line_1989,
	author = {Clough, S. A. and Kneizys, F. X. and Davies, R. W.},
	title = {{Line shape and the water vapor continuum}},
	journal = {Atmos. Res.},
	volume = {23},
	number = {3},
	pages = {229--241},
	year = {1989},
	month = oct,
	issn = {0169-8095},
	publisher = {Elsevier},
	doi = {10.1016/0169-8095(89)90020-3}
}

@article{papale_modeling_1997,
	author = {Papale, Paolo},
	title = {{Modeling of the solubility of a one-component H2O or CO2 fluid in silicate liquids}},
	journal = {Contrib. Mineral. Petrol.},
	volume = {126},
	number = {3},
	pages = {237--251},
	year = {1997},
	month = jan,
	issn = {0010-7999},
	doi = {10.1007/s004100050247}
}

@article{russell_models_2022,
	author = {Russell, James K. and Hess, Kai-Uwe and Dingwell, Donald B.},
	title = {{Models for Viscosity of Geological Melts}},
	journal = {Rev. Mineral. Geochem.},
	volume = {87},
	number = {1},
	pages = {841--885},
	year = {2022},
	month = may,
	issn = {1529-6466},
	publisher = {GeoScienceWorld},
	doi = {10.2138/rmg.2022.87.18}
}

@book{turcotte2002geodynamics,
  title={Geodynamics},
  author={Turcotte, Donald L and Schubert, Gerald},
  year={2002},
  publisher={Cambridge university press}
}

@article{kent_melt_2008,
	author = {Kent, Adam J. R.},
	title = {{Melt Inclusions in Basaltic and Related Volcanic Rocks}},
	journal = {Rev. Mineral. Geochem.},
	volume = {69},
	number = {1},
	pages = {273--331},
	year = {2008},
	month = jan,
	issn = {1529-6466},
	publisher = {GeoScienceWorld},
	doi = {10.2138/rmg.2008.69.8}
}

@article{ikeda_understandi_2013,
	author = {Ikeda, Masahiro and Aniya, Masaru},
	title = {{Understanding the Vogel{\textendash}Fulcher{\textendash}Tammann law in terms of the bond strength{\textendash}coordination number fluctuation model}},
	journal = {J. Non-Cryst. Solids},
	volume = {371-372},
	pages = {53--57},
	year = {2013},
	month = jul,
	issn = {0022-3093},
	publisher = {North-Holland},
	doi = {10.1016/j.jnoncrysol.2013.04.034}
}

@article{laurent_earths_2020,
	author = {Laurent, Oscar and Bj{\ifmmode\ddot{o}\else\"{o}\fi}rnsen, Jana and Wotzlaw, J{\ifmmode\ddot{o}\else\"{o}\fi}rn-Frederik and Bretscher, Simone and Pimenta Silva, Manuel and Moyen, Jean-Fran{\ifmmode\mbox{\c{c}}\else\c{c}\fi}ois and Ulmer, Peter and others},
	title = {{Earth{'}s earliest granitoids are crystal-rich magma reservoirs tapped by silicic eruptions}},
	journal = {Nat. Geosci.},
	volume = {13},
	number = {2},
	pages = {163--169},
	year = {2020},
	month = feb,
	issn = {1752-0908},
	publisher = {Nature Publishing Group},
	doi = {10.1038/s41561-019-0520-6}
}

@article{hirschmann_mantle_2000,
	author = {Hirschmann, Marc M.},
	title = {{Mantle solidus: Experimental constraints and the effects of peridotite composition}},
	journal = {Geochem. Geophys. Geosyst.},
	volume = {1},
	number = {10},
	year = {2000},
	month = oct,
	issn = {1525-2027},
	publisher = {John Wiley {\&} Sons, Ltd},
	doi = {10.1029/2000GC000070}
}

@article{katz_a_2003,
	author = {Katz, Richard F. and Spiegelman, Marc and Langmuir, Charles H.},
	title = {{A new parameterization of hydrous mantle melting}},
	journal = {Geochem. Geophys. Geosyst.},
	volume = {4},
	number = {9},
	year = {2003},
	month = sep,
	issn = {1525-2027},
	publisher = {John Wiley {\&} Sons, Ltd},
	doi = {10.1029/2002GC000433}
}

@article{stixrude_thermodynam_2009,
	author = {Stixrude, Lars and de Koker, Nico and Sun, Ni and Mookherjee, Mainak and Karki, Bijaya B.},
	title = {{Thermodynamics of silicate liquids in the deep Earth}},
	journal = {Earth Planet. Sci. Lett.},
	volume = {278},
	number = {3},
	pages = {226--232},
	year = {2009},
	month = feb,
	issn = {0012-821X},
	publisher = {Elsevier},
	doi = {10.1016/j.epsl.2008.12.006}
}

@article{ghiorso_the_2002,
	author = {Ghiorso, Mark S. and Hirschmann, Marc M. and Reiners, Peter W. and Kress, Victor C.},
	title = {{The pMELTS: A revision of MELTS for improved calculation of phase relations and major element partitioning related to partial melting of the mantle to 3 GPa}},
	journal = {Geochem. Geophys. Geosyst.},
	volume = {3},
	number = {5},
	pages = {1--35},
	year = {2002},
	month = may,
	issn = {1525-2027},
	publisher = {John Wiley {\&} Sons, Ltd},
	doi = {10.1029/2001GC000217}
}

@article{zahnle_evolution_1988,
	author = {Zahnle, Kevin J. and Kasting, James F. and Pollack, James B.},
	title = {{Evolution of a steam atmosphere during earth's accretion}},
	journal = {Icarus},
	volume = {74},
	number = {1},
	pages = {62--97},
	year = {1988},
	month = apr,
	issn = {0019-1035},
	publisher = {Academic Press},
	doi = {10.1016/0019-1035(88)90031-0}
}

@article{ingersoll_the_1969,
	author = {Ingersoll, Andrew P.},
	title = {{The Runaway Greenhouse: A History of Water on Venus.}},
	journal = {J. Atmos. Sci.},
	volume = {26},
	number = {6},
	pages = {1191--1198},
	year = {1969},
	month = nov,
	issn = {0022-4928},
	doi = {10.1175/1520-0469(1969)026<1191:TRGAHO>2.0.CO;2}
}

@article{wade_core_2005,
	author = {Wade, J. and Wood, B. J.},
	title = {{Core formation and the oxidation state of the Earth}},
	journal = {Earth Planet. Sci. Lett.},
	volume = {236},
	number = {1},
	pages = {78--95},
	year = {2005},
	month = jul,
	issn = {0012-821X},
	publisher = {Elsevier},
	doi = {10.1016/j.epsl.2005.05.017}
}

@article{kippenhahn2012stellar,
  title={Stellar Structure and Evolution},
  author={Kippenhahn, Rudolf and Weigert, Alfred and Weiss, Achim},
  journal={Astronomy and Astrophysics Library},
  year={2012},
  publisher={Springer Berlin Heidelberg}
}

@article{oneill_conditions_2007,
	author = {O'Neill, C. and Jellinek, A. M. and Lenardic, A.},
	title = {{Conditions for the onset of plate tectonics on terrestrial planets and moons}},
	journal = {Earth Planet. Sci. Lett.},
	volume = {261},
	number = {1},
	pages = {20--32},
	year = {2007},
	month = sep,
	issn = {0012-821X},
	publisher = {Elsevier},
	doi = {10.1016/j.epsl.2007.05.038}
}

@ARTICLE{Rolf2022SSRv,
       author = {{Rolf}, Tobias and {Weller}, Matt and {G{\"u}lcher}, Anna and {Byrne}, Paul and {O'Rourke}, Joseph G. and {Herrick}, Robert and {Bjonnes}, Evan and {Davaille}, Anne and {Ghail}, Richard and {Gillmann}, Cedric and {Plesa}, Ana-Catalina and {Smrekar}, Suzanne},
        title = "{Dynamics and Evolution of Venus' Mantle Through Time}",
      journal = {\ssr},
         year = 2022,
        month = dec,
       volume = {218},
       number = {8},
          eid = {70},
        pages = {70},
          doi = {10.1007/s11214-022-00937-9},
       adsurl = {https://ui.adsabs.harvard.edu/abs/2022SSRv..218...70R}
}

@ARTICLE{Ghail2024SSRv,
       author = {{Ghail}, Richard C. and {Smrekar}, Suzanne E. and {Widemann}, Thomas and {Byrne}, Paul K. and {G{\"u}lcher}, Anna J.~P. and {O'Rourke}, Joseph G. and {Borrelli}, Madison E. and {Gilmore}, Martha S. and {Herrick}, Robert R. and {Ivanov}, Mikhail A. and {Plesa}, Ana-Catalina and {Rolf}, Tobias and {Sabbeth}, Leah and {Schools}, Joe W. and {Gregory Shellnutt}, J.},
        title = "{Volcanic and Tectonic Constraints on the Evolution of Venus}",
      journal = {\ssr},
         year = 2024,
        month = apr,
       volume = {220},
       number = {4},
          eid = {36},
        pages = {36},
          doi = {10.1007/s11214-024-01065-2},
       adsurl = {https://ui.adsabs.harvard.edu/abs/2024SSRv..220...36G}
}

@ARTICLE{Byrne2021PNAS,
       author = {{Byrne}, Paul K. and {Ghail}, Richard C. and {{\c{S}}eng{\"o}r}, A.~M. Cel{\^a}l and {James}, Peter B. and {Klimczak}, Christian and {Solomon}, Sean C.},
        title = "{A globally fragmented and mobile lithosphere on Venus}",
      journal = {Proceedings of the National Academy of Science},
         year = 2021,
        month = jun,
       volume = {118},
       number = {26},
          eid = {e2025919118},
        pages = {e2025919118},
          doi = {10.1073/pnas.2025919118},
       adsurl = {https://ui.adsabs.harvard.edu/abs/2021PNAS..11825919B}
}

@misc{Krijt2023ASPC5341031K,
  title = {Chemical {{Habitability}}: {{Supply}} and {{Retention}} of {{Life}}'s {{Essential Elements During Planet Formation}}},
  shorttitle = {Chemical {{Habitability}}},
  author = {Krijt, Sebastiaan and Kama, Mihkel and McClure, Melissa and Teske, Johanna and Bergin, Edwin A. and Shorttle, Oliver and Walsh, Kevin J. and Raymond, Sean N.},
  year = 2022,
  month = mar,
  number = {arXiv:2203.10056},
  eprint = {2203.10056},
  primaryclass = {astro-ph},
  publisher = {arXiv},
  doi = {10.48550/arXiv.2203.10056},
  urldate = {2026-04-27},
  archiveprefix = {arXiv}
}

@article{Lock2024PSJ528L,
  title = {Atmospheric {{Loss}} in {{Giant Impacts Depends}} on {{Preimpact Surface Conditions}}},
  author = {Lock, Simon J. and Stewart, Sarah T.},
  year = 2024,
  month = feb,
  journal = {The Planetary Science Journal},
  volume = {5},
  number = {2},
  pages = {28},
  doi = {10.3847/PSJ/ad0b16}
}

@article{Hirose2013,
  title = {Composition and {{State}} of the {{Core}}},
  author = {Hirose, Kei and Labrosse, St{\'e}phane and Hernlund, John},
  year = 2013,
  month = may,
  journal = {Annual Review of Earth and Planetary Sciences},
  volume = {41},
  number = {1},
  pages = {657--691},
  issn = {0084-6597, 1545-4495},
  doi = {10.1146/annurev-earth-050212-124007},
  urldate = {2026-05-15},
  langid = {english}
}

@article{Dauphas2017Natur541521D,
  title = {The Isotopic Nature of the {{Earth}}'s Accreting Material through Time},
  author = {Dauphas, Nicolas},
  year = 2017,
  month = jan,
  journal = {Nature},
  volume = {541},
  number = {7638},
  pages = {521--524},
  doi = {10.1038/nature20830}
}

@article{Shahar2026,
	author = {Shahar, Anat and Young, Edward D. and Hirose, Kei and Yokoo, Shunpei},
	title = {{The Compositions of Planetary Cores}},
	journal = {Annu. Rev. Earth Planet. Sci.},
	year = {2026},
	month = feb,
	publisher = {Annual Reviews},
	doi = {10.1146/annurev-earth-040722-094945}
}

@article{Kegerreis2020ApJ901L31K,
  title = {Atmospheric {{Erosion}} by {{Giant Impacts}} onto {{Terrestrial Planets}}: {{A Scaling Law}} for Any {{Speed}}, {{Angle}}, {{Mass}}, and {{Density}}},
  author = {Kegerreis, J. A. and Eke, V. R. and Catling, D. C. and Massey, R. J. and Teodoro, L. F. A. and Zahnle, K. J.},
  year = 2020,
  month = oct,
  journal = {The Astrophysical Journal Letters},
  volume = {901},
  number = {2},
  pages = {L31},
  doi = {10.3847/2041-8213/abb5fb}
}

@misc{Bergin2026arXiv260210308B,
  title = {Carbon from {{Interstellar Clouds}} to {{Habitable Worlds}}},
  author = {Bergin, Edwin A. and Hirschmann, Marc M. and Izidoro, Andre},
  year = 2026,
  month = feb,
  number = {arXiv:2602.10308},
  eprint = {2602.10308},
  primaryclass = {astro-ph},
  publisher = {arXiv},
  doi = {10.48550/arXiv.2602.10308},
  urldate = {2026-02-21},
  archiveprefix = {arXiv},
  langid = {english}
}

@inproceedings{Drazkowska2023ASPC534717D,
  title = {Planet {{Formation Theory}} in the {{Era}} of {{ALMA}} and {{Kepler}}: From {{Pebbles}} to {{Exoplanets}}},
  booktitle = {Protostars and {{Planets VII}}},
  author = {Dr{\k a}{\.z}kowska, J. and Bitsch, B. and Lambrechts, M. and Mulders, G. D. and Harsono, D. and Vazan, A. and Liu, B. and Ormel, C. W. and Kretke, K. and Morbidelli, A.},
  editor = {Inutsuka, S. and Aikawa, Y. and Muto, T. and Tomida, K. and Tamura, M.},
  year = 2023,
  month = jul,
  series = {Astronomical {{Society}} of the {{Pacific Conference Series}}},
  volume = {534},
  pages = {717},
  doi = {10.48550/arXiv.2203.09759}
}

@ARTICLE{Suer2023FrEaS,
       author = {{Suer}, Terry-Ann and {Jackson}, Colin and {Grewal}, Damanveer S. and {Dalou}, Celia and {Lichtenberg}, Tim},
        title = "{The distribution of volatile elements during rocky planet formation}",
      journal = {Frontiers in Earth Science},
         year = 2023,
        month = nov,
       volume = {11},
          eid = {1159412},
        pages = {1159412},
          doi = {10.3389/feart.2023.1159412},
archivePrefix = {arXiv},
       eprint = {2311.18262},
 primaryClass = {astro-ph.EP},
       adsurl = {https://ui.adsabs.harvard.edu/abs/2023FrEaS..1159412S}
}

@article{barboni_early_2017,
	author = {Barboni, Melanie and Boehnke, Patrick and Keller, Brenhin and Kohl, Issaku E. and Schoene, Blair and Young, Edward D. and McKeegan, Kevin D.},
	title = {{Early formation of the Moon 4.51 billion years ago}},
	journal = {Sci. Adv.},
	volume = {3},
	number = {1},
	year = {2017},
	month = jan,
	issn = {2375-2548},
	publisher = {American Association for the Advancement of Science},
	doi = {10.1126/sciadv.1602365}
}

@ARTICLE{Molliere2019AA,
       author = {{Molli{\`e}re}, P. and {Wardenier}, J.~P. and {van Boekel}, R. and {Henning}, Th. and {Molaverdikhani}, K. and {Snellen}, I.~A.~G.},
        title = "{petitRADTRANS. A Python radiative transfer package for exoplanet characterization and retrieval}",
      journal = {\aap},
         year = 2019,
        month = jul,
       volume = {627},
          eid = {A67},
        pages = {A67},
          doi = {10.1051/0004-6361/201935470},
archivePrefix = {arXiv},
       eprint = {1904.11504},
 primaryClass = {astro-ph.EP},
       adsurl = {https://ui.adsabs.harvard.edu/abs/2019A&A...627A..67M}
}

@ARTICLE{Bower2025ApJ,
       author = {{Bower}, Dan J. and {Thompson}, Maggie A. and {Hakim}, Kaustubh and {Tian}, Meng and {Sossi}, Paolo A.},
        title = "{Diversity of Low-mass Planet Atmospheres in the C─H─O─N─S─Cl System with Interior Dissolution, Nonideality, and Condensation: Application to TRAPPIST-1e and Sub-Neptunes}",
      journal = {\apj},
         year = 2025,
        month = dec,
       volume = {995},
       number = {1},
          eid = {59},
        pages = {59},
          doi = {10.3847/1538-4357/ae1479},
archivePrefix = {arXiv},
       eprint = {2507.00499},
 primaryClass = {astro-ph.EP},
       adsurl = {https://ui.adsabs.harvard.edu/abs/2025ApJ...995...59B}
}

@article{Andrault2011,
author = {Andrault, Denis and Bolfan-Casanova, Nathalie and Nigro, Giacomo Lo and Bouhifd, Mohamed A and Garbarino, Gaston and Mezouar, Mohamed},
doi = {10.1016/j.epsl.2011.02.006},
issn = {0012-821X},
journal = {Earth and Planetary Science Letters},
number = {1--2},
pages = {251--259},
title = {{Solidus and liquidus profiles of chondritic mantle: Implication for melting of the Earth across its history}},
volume = {304},
year = {2011}
}

@article{wang_the_2018,
	author = {Wang, Haiyang S. and Lineweaver, Charles H. and Ireland, Trevor R.},
	title = {{The elemental abundances (with uncertainties) of the most Earth-like planet}},
	journal = {Icarus},
	volume = {299},
	pages = {460--474},
	year = {2018},
	month = jan,
	issn = {0019-1035},
	publisher = {Academic Press},
	doi = {10.1016/j.icarus.2017.08.024}
}

@article{nakajima_scaling_2021,
	author = {Nakajima, Miki and Golabek, Gregor J. and W{\ifmmode\ddot{u}\else\"{u}\fi}nnemann, Kai and Rubie, David C. and Burger, Christoph and Melosh, Henry J. and Jacobson, Seth A. and others},
	title = {{Scaling laws for the geometry of an impact-induced magma ocean}},
	journal = {Earth Planet. Sci. Lett.},
	volume = {568},
	pages = {116983},
	year = {2021},
	month = aug,
	issn = {0012-821X},
	publisher = {Elsevier},
	doi = {10.1016/j.epsl.2021.116983}
}

@article{peslier_water_2017,
	author = {Peslier, Anne H. and Sch{\ifmmode\ddot{o}\else\"{o}\fi}nb{\ifmmode\ddot{a}\else\"{a}\fi}chler, Maria and Busemann, Henner and Karato, Shun-Ichiro},
	title = {{Water in the Earth{'}s Interior: Distribution and Origin}},
	journal = {Space Sci. Rev.},
	volume = {212},
	number = {1},
	pages = {743--810},
	year = {2017},
	month = oct,
	issn = {1572-9672},
	publisher = {Springer Netherlands},
	doi = {10.1007/s11214-017-0387-z}
}

@article{cavosie_magmatic_2005,
	author = {Cavosie, A. J. and Valley, J. W. and Wilde, S. A. and E. I. M. F.},
	title = {{Magmatic {$\delta$}18O in 4400{\textendash}3900 Ma detrital zircons: A record of the alteration and recycling of crust in the Early Archean}},
	journal = {Earth Planet. Sci. Lett.},
	volume = {235},
	number = {3},
	pages = {663--681},
	year = {2005},
	month = jul,
	issn = {0012-821X},
	publisher = {Elsevier},
	doi = {10.1016/j.epsl.2005.04.028}
}

@article{dalrymple_age_2001,
author = {G. Brent Dalrymple },
title = {The age of the Earth in the twentieth century: a problem (mostly) solved},
journal = {Geological Society, London, Special Publications},
volume = {190},
number = {1},
pages = {205-221},
year = {2001},
doi = {10.1144/GSL.SP.2001.190.01.14},
URL = {https://www.lyellcollection.org/doi/abs/10.1144/GSL.SP.2001.190.01.14},
eprint = {https://www.lyellcollection.org/doi/pdf/10.1144/GSL.SP.2001.190.01.14},
}

@article{canup_origin_2001,
	author = {Canup, Robin M. and Asphaug, Erik},
	title = {{Origin of the Moon in a giant impact near the end of the Earth's formation}},
	journal = {Nature},
	volume = {412},
	number = {6848},
	pages = {708--712},
	year = {2001},
	month = aug,
	issn = {1476-4687},
	publisher = {Nature Publishing Group},
	doi = {10.1038/35089010}
}

@article{citron_the_2018,
	author = {Citron, Robert I. and Perets, Hagai B. and Aharonson, Oded},
	title = {{The Role of Multiple Giant Impacts in the Formation of the Earth{\textendash}Moon System}},
	journal = {Astrophys. J.},
	volume = {862},
	number = {1},
	pages = {5},
	year = {2018},
	month = jul,
	issn = {0004-637X},
	publisher = {The American Astronomical Society},
	doi = {10.3847/1538-4357/aaca2d}
}

@article{rzeplinski_hadean_2022,
	author = {Rzeplinski, Igor and Sanloup, Chryst{\ifmmode\grave{e}\else\`{e}\fi}le and Gilabert, Eric and Horlait, Denis},
	title = {{Hadean isotopic fractionation of xenon retained in deep silicates}},
	journal = {Nature},
	volume = {606},
	number = {7915},
	pages = {713--717},
	year = {2022},
	month = jun,
	issn = {1476-4687},
	publisher = {Nature Publishing Group},
	doi = {10.1038/s41586-022-04710-4}
}

@article{schoenberg_new_2002,
	author = {Schoenberg, Ronny and Kamber, Balz S. and Collerson, Kenneth D. and Eugster, Otto},
	title = {{New W-isotope evidence for rapid terrestrial accretion and very early core formation}},
	journal = {Geochim. Cosmochim. Acta},
	volume = {66},
	number = {17},
	pages = {3151--3160},
	year = {2002},
	month = sep,
	issn = {0016-7037},
	publisher = {Pergamon},
	doi = {10.1016/S0016-7037(02)00911-0}
}

@article{halliday_the_2023,
	author = {Halliday, Alex N. and Canup, Robin M.},
	title = {{The accretion of planet Earth}},
	journal = {Nat. Rev. Earth Environ.},
	volume = {4},
	number = {1},
	pages = {19--35},
	year = {2023},
	month = jan,
	issn = {2662-138X},
	publisher = {Nature Publishing Group},
	doi = {10.1038/s43017-022-00370-0}
}

@article{rollinson_archaean_2017,
	author = {Rollinson, Hugh and Adetunji, Jacob and Lenaz, Davide and Szilas, Kristoffer},
	title = {{Archaean chromitites show constant Fe3+/{$\Sigma$}Fe in Earth's asthenospheric mantle since 3.8Ga}},
	journal = {Lithos},
	volume = {282-283},
	pages = {316--325},
	year = {2017},
	month = jun,
	issn = {0024-4937},
	publisher = {Elsevier},
	doi = {10.1016/j.lithos.2017.03.020}
}

@article{roche_atmospheric_2025,
	author = {Roche, Matthew J. and Lock, Simon J. and Dou, Jingyao and Carter, Philip J. and Kegerreis, Jacob A. and Leinhardt, Zo{\ifmmode\ddot{e}\else\"{e}\fi} M.},
	title = {{Atmospheric Loss during Giant Impacts: Mechanisms and Scaling of Near- and Far-field Loss}},
	journal = {Planet. Sci. J.},
	volume = {6},
	number = {6},
	pages = {149},
	year = {2025},
	month = jun,
	issn = {2632-3338},
	publisher = {The American Astronomical Society},
	doi = {10.3847/PSJ/add929}
}

@article{sossi_physicochem_2025,
	author = {Sossi, Paolo A. and Hin, Remco C. and Kleine, Thorsten and Morbidelli, Alessandro and Nimmo, Francis},
	title = {{Physicochemical Controls on the Compositions of the Earth and Planets}},
	journal = {Space Sci. Rev.},
	volume = {221},
	number = {8},
	pages = {118},
	year = {2025},
	month = nov,
	issn = {1572-9672},
	publisher = {Springer Netherlands},
	doi = {10.1007/s11214-025-01243-w}
}

@article{fromont_atmospheric_2024,
	author = {Fromont, Emeline F. and Ahlers, John P. and do Amaral, Laura N. R. and Barnes, Rory and Gilbert, Emily A. and Quintana, Elisa V. and Peacock, Sarah and others},
	title = {{Atmospheric Escape From Three Terrestrial Planets in the L 98-59 System}},
	journal = {Astrophys. J.},
	volume = {961},
	number = {1},
	pages = {115},
	year = {2024},
	month = jan,
	issn = {0004-637X},
	publisher = {The American Astronomical Society},
	doi = {10.3847/1538-4357/ad0e0e}
}

@article{owen_atmospheric_2019,
	author = {Owen, James E.},
	title = {{Atmospheric Escape and the Evolution of Close-In Exoplanets}},
	journal = {Annu. Rev. Earth Planet. Sci.},
	volume = {47},
	pages = {67--90},
	year = {2019},
	month = may,
	publisher = {Annual Reviews},
	doi = {10.1146/annurev-earth-053018-060246}
}

@article{hirschmann_magma_2012,
	author = {Hirschmann, Marc M.},
	title = {{Magma ocean influence on early atmosphere mass and composition}},
	journal = {Earth Planet. Sci. Lett.},
	volume = {341-344},
	pages = {48--57},
	year = {2012},
	month = aug,
	issn = {0012-821X},
	publisher = {Elsevier},
	doi = {10.1016/j.epsl.2012.06.015}
}

@article{rosenblatt_determinati_2021,
	author = {Rosenblatt, Pascal and Dumoulin, Caroline and Marty, Jean-Charles and Genova, Antonio},
	title = {{Determination of Venus{'} Interior Structure with EnVision}},
	journal = {Remote Sens.},
	volume = {13},
	number = {9},
	pages = {1624},
	year = {2021},
	month = apr,
	issn = {2072-4292},
	publisher = {Multidisciplinary Digital Publishing Institute},
	doi = {10.3390/rs13091624}
}

@article{garvin_revealing_2022,
	author = {Garvin, James B. and Getty, Stephanie A. and Arney, Giada N. and Johnson, Natasha M. and Kohler, Erika and Schwer, Kenneth O. and Sekerak, Michael and others},
	title = {{Revealing the Mysteries of Venus: The DAVINCI Mission}},
	journal = {Planet. Sci. J.},
	volume = {3},
	number = {5},
	pages = {117},
	year = {2022},
	month = may,
	issn = {2632-3338},
	publisher = {The American Astronomical Society},
	doi = {10.3847/PSJ/ac63c2}
}

@article{thompson_on_2026,
	author = {Thompson, Maggie A.},
	title = {{On the importance of laboratory experiments for interpreting exoplanet observations}},
	journal = {Astrophys. Space Sci.},
	volume = {371},
	number = {3},
	pages = {33},
	year = {2026},
	month = mar,
	issn = {1572-946X},
	publisher = {Springer Netherlands},
	doi = {10.1007/s10509-026-04564-6}
}

@article{hammond_reliable_2025,
	author = {Hammond, Mark and Guimond, Claire Marie and Lichtenberg, Tim and Nicholls, Harrison and Fisher, Chloe and Luque, Rafael and Meier, Tobias G. and others},
	title = {{Reliable Detections of Atmospheres on Rocky Exoplanets with Photometric JWST Phase Curves}},
	journal = {Astrophys. J. Lett.},
	volume = {978},
	number = {2},
	pages = {L40},
	year = {2025},
	month = jan,
	issn = {2041-8205},
	publisher = {The American Astronomical Society},
	doi = {10.3847/2041-8213/ada0bc}
}

@article{attia_paleos_2026,
	author = {Attia, Mara and Lichtenberg, Tim and Jungov{\ifmmode\acute{a}\else\'{a}\fi}, Ema and Sastre, Mariana},
	title = {{PALEOS: Multiphase Equations of State and Mass-Radius Relations for Exoplanet Interiors}},
	journal = {arXiv},
	year = {2026},
	month = may,
	eprint = {2605.03741},
	doi = {10.48550/arXiv.2605.03741}
}

@article{bean_the_2021,
	author = {Bean, Jacob L. and Raymond, Sean N. and Owen, James E.},
	title = {{The Nature and Origins of Sub-Neptune Size Planets}},
	journal = {J. Geophys. Res. Planets},
	volume = {126},
	number = {1},
	pages = {e2020JE006639},
	year = {2021},
	month = jan,
	issn = {2169-9097},
	publisher = {John Wiley {\&} Sons, Ltd},
	doi = {10.1029/2020JE006639}
}

@article{titov_clouds_2018,
	author = {Titov, Dmitrij V. and Ignatiev, Nikolay I. and McGouldrick, Kevin and Wilquet, Val{\ifmmode\acute{e}\else\'{e}\fi}rie and Wilson, Colin F.},
	title = {{Clouds and Hazes of Venus}},
	journal = {Space Sci. Rev.},
	volume = {214},
	number = {8},
	pages = {126},
	year = {2018},
	month = nov,
	issn = {1572-9672},
	publisher = {Springer Netherlands},
	doi = {10.1007/s11214-018-0552-z}
}

@article{villanueva_modeling_2024,
	author = {Villanueva, Geronimo L. and Fauchez, Thomas J. and Kofman, Vincent and Alei, Eleonora and Lee, Elspeth K. H. and Janin, Estelle and Himes, Michael D. and others},
	title = {{Modeling Atmospheric Lines by the Exoplanet Community (MALBEC) Version 1.0: A CUISINES Radiative Transfer Intercomparison Project}},
	journal = {Planet. Sci. J.},
	volume = {5},
	number = {3},
	pages = {64},
	year = {2024},
	month = mar,
	issn = {2632-3338},
	publisher = {The American Astronomical Society},
	doi = {10.3847/PSJ/ad2681}
}

@article{chaverot_codaccra_2026,
	author = {Chaverot, Geronimo L. and others},
	journal = {Planet. Sci. J.},
	year = {In prep.},
	month = mar,
	issn = {2632-3338},
	publisher = {The American Astronomical Society},
}

@article{ikoma_constraints_2006,
	author = {Ikoma, Masahiro and Genda, Hidenori},
	title = {{Constraints on the Mass of a Habitable Planet with Water of Nebular Origin}},
	journal = {Astrophys. J.},
	volume = {648},
	number = {1},
	pages = {696},
	year = {2006},
	month = sep,
	issn = {0004-637X},
	publisher = {IOP Publishing},
	doi = {10.1086/505780}
}

@article{lichtenberg_a_2019,
	author = {Lichtenberg, Tim and Golabek, Gregor J. and Burn, Remo and Meyer, Michael R. and Alibert, Yann and Gerya, Taras V. and Mordasini, Christoph},
	title = {{A water budget dichotomy of rocky protoplanets from 26Al-heating}},
	journal = {Nat. Astron.},
	volume = {3},
	number = {4},
	pages = {307--313},
	year = {2019},
	month = apr,
	issn = {2397-3366},
	publisher = {Nature Publishing Group},
	doi = {10.1038/s41550-018-0688-5}
}

@article{pluriel_modeling_2019,
	author = {Pluriel, W. and Marcq, E. and Turbet, M.},
	title = {{Modeling the albedo of Earth-like magma ocean planets with H2O-CO2 atmospheres}},
	journal = {Icarus},
	volume = {317},
	pages = {583--590},
	year = {2019},
	month = jan,
	issn = {0019-1035},
	publisher = {Academic Press},
	doi = {10.1016/j.icarus.2018.08.023}
}

@article{constantinou_venus_2026,
	author = {Constantinou, Tereza and Shorttle, Oliver and Nicholls, Harrison},
	title = {{Equifinality of Venus-like CO2 atmospheres}},
	journal = {Mon. Not. R. Astron. Soc.},
	volume = {548},
	number = {4},
	pages = {stag823},
	year = {2026},
	month = jun,
	issn = {0035-8711},
	publisher = {Oxford Academic},
	doi = {10.1093/mnras/stag823}
}

@article{sim_volatile_2024,
	author = {Sim, Shi J. and Hirschmann, Marc M. and Hier-Majumder, Saswata},
	title = {{Volatile and Trace Element Storage in a Crystallizing Martian Magma Ocean}},
	journal = {J. Geophys. Res. Planets},
	volume = {129},
	number = {8},
	pages = {e2024JE008346},
	year = {2024},
	month = aug,
	issn = {2169-9097},
	publisher = {John Wiley {\&} Sons, Ltd},
	doi = {10.1029/2024JE008346}
}

@article{petford_rheology_2003,
	author = {Petford, Nick},
	title = {{Rheology of Granitic Magmas During Ascent and Emplacement}},
	journal = {Annu. Rev. Earth Planet. Sci.},
	volume = {31},
	pages = {399--427},
	year = {2003},
	month = may,
	publisher = {Annual Reviews},
	doi = {10.1146/annurev.earth.31.100901.141352}
}

@article{mckenzie_the_1967,
	author = {McKenzie, D. P.},
	title = {{The Viscosity of the Mantle}},
	journal = {Geophys. J. Int.},
	volume = {14},
	number = {1-4},
	pages = {297--305},
	year = {1967},
	month = nov,
	issn = {0956-540X},
	publisher = {Oxford Academic},
	doi = {10.1111/j.1365-246X.1967.tb06246.x}
}

@article{ahlers_heat_2009,
	author = {Ahlers, Guenter and Grossmann, Siegfried and Lohse, Detlef},
	title = {{Heat transfer and large scale dynamics in turbulent Rayleigh-Benard convection}},
	journal = {Rev. Mod. Phys.},
	volume = {81},
	number = {2},
	pages = {503--537},
	year = {2009},
	month = apr,
	publisher = {American Physical Society},
	doi = {10.1103/RevModPhys.81.503}
}

@article{siggia_high_1994,
	author = {Siggia, E. D.},
	title = {{High Rayleigh Number Convection}},
	journal = {Annu. Rev. Fluid Mech.},
	number = {Volume 26, 1994},
	pages = {137--168},
	year = {1994},
	month = jan,
	publisher = {Annual Reviews},
	doi = {10.1146/annurev.fl.26.010194.001033}
}

@article{bottinga_viscosity,
	author = {Bottinga, Yan and Weill, Daniel F.},
	journal = {American Journal of Science},
	doi = {10.2475/ajs.272.5.438},
	number = {5},
	year = {1972},
	month = {may 1},
	pages = {438--475},
	publisher = {American Journal of Science},
	title = {The viscosity of magmatic silicate liquids; a model calculation},
	volume = {272},
}

@article{schulik_aiolos_2023,
	author = {Schulik, Matth{\ifmmode\ddot{a}\else\"{a}\fi}us and Booth, Richard A.},
	title = {{aiolos {\textendash} A multipurpose 1D hydrodynamics code for planetary atmospheres}},
	journal = {Mon. Not. R. Astron. Soc.},
	volume = {523},
	number = {1},
	pages = {286--304},
	year = {2023},
	month = may,
	issn = {0035-8711},
	publisher = {Oxford Academic},
	doi = {10.1093/mnras/stad1251}
}

@article{brain_atmospheric_2016,
	author = {Brain, D. A. and Bagenal, F. and Ma, Y.-J. and Nilsson, H. and Wieser, G. Stenberg},
	title = {{Atmospheric escape from unmagnetized bodies}},
	journal = {J. Geophys. Res. Planets},
	volume = {121},
	number = {12},
	pages = {2364--2385},
	year = {2016},
	month = dec,
	issn = {2169-9097},
	publisher = {John Wiley {\&} Sons, Ltd},
	doi = {10.1002/2016JE005162}
}

@article{sossi_redox_2020,
	author = {Sossi, Paolo A. and Burnham, Antony D. and Badro, James and Lanzirotti, Antonio and Newville, Matt and O{'}Neill, Hugh {\relax St}. C.},
	title = {{Redox state of Earth{'}s magma ocean and its Venus-like early atmosphere}},
	journal = {Sci. Adv.},
	volume = {6},
	number = {48},
	year = {2020},
	month = nov,
	issn = {2375-2548},
	publisher = {American Association for the Advancement of Science},
	doi = {10.1126/sciadv.abd1387}
}

@article{javoy_where_2005,
	author = {Javoy, Marc},
	title = {{Where do the oceans come from?}},
	journal = {C. R. Geosci.},
	volume = {337},
	number = {1},
	pages = {139--158},
	year = {2005},
	month = jan,
	issn = {1631-0713},
	publisher = {No longerElsevier},
	doi = {10.1016/j.crte.2004.10.005}
}

@article{gueymard_proposed_2002,
	author = {Gueymard, C. A. and Myers, D. and Emery, K.},
	title = {{Proposed reference irradiance spectra for solar energy systems testing}},
	journal = {Sol. Energy},
	volume = {73},
	number = {6},
	pages = {443--467},
	year = {2002},
	month = dec,
	issn = {0038-092X},
	publisher = {Pergamon},
	doi = {10.1016/S0038-092X(03)00005-7}
}

@article{frost_the_2008,
	author = {Frost, Daniel J. and McCammon, Catherine A.},
	title = {{The Redox State of Earth's Mantle}},
	journal = {Annu. Rev. Earth Planet. Sci.},
	volume = {36},
	pages = {389--420},
	year = {2008},
	month = may,
	publisher = {Annual Reviews},
	doi = {10.1146/annurev.earth.36.031207.124322}
}

@article{genda_origin_2016,
	author = {Genda, Hidenori},
	title = {{Origin of Earth{'}s oceans: An assessment of the total amount, history and supply of water}},
	journal = {Geochem. J.},
	volume = {50},
	number = {1},
	pages = {27--42},
	year = {2016},
	issn = {0016-7002},
	publisher = {GEOCHEMICAL SOCIETY OF JAPAN},
	doi = {10.2343/geochemj.2.0398}
}

@article{scott_the_2006,
	author = {Scott, T. and Kohlstedt, D. L.},
	title = {{The effect of large melt fraction on the deformation behavior of peridotite}},
	journal = {Earth Planet. Sci. Lett.},
	volume = {246},
	number = {3},
	pages = {177--187},
	year = {2006},
	month = jun,
	issn = {0012-821X},
	publisher = {Elsevier},
	doi = {10.1016/j.epsl.2006.04.027}
}

@incollection{peltier_the_1981,
	author = {Peltier, W. R. and Wu, P. and Yuen, D. A.},
	title = {{The Viscosities of the Earth's Mantle}},
	booktitle = {{Anelasticity in the Earth}},
	journal = {AGU Journals},
	pages = {59--77},
	year = {1981},
	month = jan,
	isbn = {978-1-11866994-5},
	publisher = {John Wiley {\&} Sons, Ltd.},
	address = {Chichester, England, UK},
	doi = {10.1029/GD004p0059}
}

@article{ABEL2013857,
title = {Collision-induced spectra and current astronomical research1},
journal = {Canadian Journal of Physics},
volume = {91},
number = {11},
pages = {857-869},
year = {2013},
issn = {0008-4204},
doi = {https://doi.org/10.1139/cjp-2012-0532},
url = {https://www.sciencedirect.com/science/article/pii/S0008420413002328},
author = {Martin Abel and Lothar Frommhold}
}

@article{garcia_investigati_2026,
	author = {Garcia, Rodolfo and Barnes, Rory and Driscoll, Peter E. and Meadows, Victoria S. and Gialluca, Megan},
	title = {{Investigation of Venus{'} Thermal History, Crustal Evolution, and Core Dynamics with a Coupled Interior-lithosphere-atmosphere Model}},
	journal = {Planet. Sci. J.},
	volume = {7},
	number = {5},
	pages = {120},
	year = {2026},
	month = may,
	issn = {2632-3338},
	publisher = {The American Astronomical Society},
	doi = {10.3847/PSJ/ae5248}
}

@article{pollack_a_1971,
	author = {Pollack, James B.},
	title = {{A nongrey calculation of the runaway greenhouse: Implications for Venus' past and present}},
	journal = {Icarus},
	volume = {14},
	number = {3},
	pages = {295--306},
	year = {1971},
	month = jun,
	issn = {0019-1035},
	publisher = {Academic Press},
	doi = {10.1016/0019-1035(71)90001-7}
}

@article{gaillard_bubble_2021,
	author = {Gaillard, F. and Bouhifd, M. A. and F{\ifmmode\ddot{u}\else\"{u}\fi}ri, E. and Malavergne, V. and Marrocchi, Y. and Noack, L. and Ortenzi, G. and others},
	title = {{The Diverse Planetary Ingassing/Outgassing Paths Produced over Billions of Years of Magmatic Activity}},
	journal = {Space Sci. Rev.},
	volume = {217},
	number = {1},
	pages = {22},
	year = {2021},
	month = feb,
	issn = {1572-9672},
	publisher = {Springer Netherlands},
	doi = {10.1007/s11214-021-00802-1}
}

@article{sparks_bubble_1978,
	author = {Sparks, R. S. J.},
	title = {{The dynamics of bubble formation and growth in magmas: A review and analysis}},
	journal = {J. Volcanol. Geotherm. Res.},
	volume = {3},
	number = {1},
	pages = {1--37},
	year = {1978},
	month = mar,
	issn = {0377-0273},
	publisher = {Elsevier},
	doi = {10.1016/0377-0273(78)90002-1}
}

@article{katz_physics_2022,
	author = {Katz, Richard F. and Jones, David W. Rees and Rudge, John F. and Keller, Tobias},
	title = {{Physics of Melt Extraction from the Mantle: Speed and Style}},
	journal = {Annu. Rev. Earth Planet. Sci.},
	number = {Volume 50, 2022},
	pages = {507--540},
	year = {2022},
	month = may,
	publisher = {Annual Reviews},
	doi = {10.1146/annurev-earth-032320-083704}
}

@article{hirschmann_solubility_2012,
	author = {Hirschmann, M. M. and Withers, A. C. and Ardia, P. and Foley, N. T.},
	title = {{Solubility of molecular hydrogen in silicate melts and consequences for volatile evolution of terrestrial planets}},
	journal = {Earth Planet. Sci. Lett.},
	volume = {345-348},
	pages = {38--48},
	year = {2012},
	month = sep,
	issn = {0012-821X},
	publisher = {Elsevier},
	doi = {10.1016/j.epsl.2012.06.031}
}

@article{bottinga_morb_1990,
	author = {Bottinga, Y. and Javoy, M.},
	title = {{MORB degassing: Bubble growth and ascent}},
	journal = {Chem. Geol.},
	volume = {81},
	number = {4},
	pages = {255--270},
	year = {1990},
	month = feb,
	issn = {0009-2541},
	publisher = {Elsevier},
	doi = {10.1016/0009-2541(90)90050-H}
}

@article{rozel_continental_2017,
	author = {Rozel, A. B. and Golabek, G. J. and Jain, C. and Tackley, P. J. and Gerya, T.},
	title = {{Continental crust formation on early Earth controlled by intrusive magmatism}},
	journal = {Nature},
	volume = {545},
	number = {7654},
	pages = {332--335},
	year = {2017},
	month = may,
	issn = {1476-4687},
	publisher = {Nature Publishing Group},
	doi = {10.1038/nature22042}
}

@article{grimm_helios_2021,
	author = {Grimm, Simon L. and Malik, Matej and Kitzmann, Daniel and Guzm{\ifmmode\acute{a}\else\'{a}\fi}n-Mesa, Andrea and Hoeijmakers, H. Jens and Fisher, Chloe and Mendon{\ifmmode\mbox{\c{c}}\else\c{c}\fi}a, Jo{\ifmmode\tilde{a}\else\~{a}\fi}o M. and others},
	title = {{HELIOS-K 2.0 Opacity Calculator and Open-source Opacity Database for Exoplanetary Atmospheres}},
	journal = {Astrophys. J. Suppl. Ser.},
	volume = {253},
	number = {1},
	pages = {30},
	year = {2021},
	month = mar,
	issn = {0067-0049},
	eprint = {arXiv:2101.02005},
	doi = {10.3847/1538-4365/abd773}
}

@article{tennyson_exomol_2012,
	author = {Tennyson, Jonathan and Yurchenko, Sergei N.},
	title = {{ExoMol: molecular line lists for exoplanet and other atmospheres}},
	journal = {Mon. Not. R. Astron. Soc.},
	volume = {425},
	number = {1},
	pages = {21--33},
	year = {2012},
	month = sep,
	issn = {0035-8711},
	publisher = {Oxford Academic},
	doi = {10.1111/j.1365-2966.2012.21440.x}
}

@article{di_rheology_2026,
	author = {Di Fiore, Fabrizio and Cassetta, Michele},
	title = {{Rheology and Structure of Fe-Mg-Ca Enriched Silicate Melt: Benchmarking Viscosity Models for an Exotic Planetary Composition}},
	journal = {J. Geophys. Res. Planets},
	volume = {131},
	number = {8},
	pages = {e2026JE009904},
	year = {2026},
	month = aug,
	issn = {2169-9097},
	publisher = {John Wiley {\&} Sons, Ltd},
	doi = {10.1029/2026JE009904}
}

@article{iacono_new_2012,
	author = {Iacono-Marziano, Giada and Morizet, Yann and Le Trong, Emmanuel and Gaillard, Fabrice},
	title = {{New experimental data and semi-empirical parameterization of H2O-CO2 solubility in mafic melts}},
	journal = {Geochim. Cosmochim. Acta},
	volume = {97},
	pages = {1--23},
	year = {2012},
	month = nov,
	issn = {0016-7037},
	doi = {10.1016/j.gca.2012.08.035}
}

@article{hier_the_2017,
	author = {Hier-Majumder, Saswata and Hirschmann, Marc M.},
	title = {{The origin of volatiles in the Earth's mantle}},
	journal = {Geochem. Geophys. Geosyst.},
	volume = {18},
	number = {8},
	pages = {3078--3092},
	year = {2017},
	month = aug,
	issn = {1525-2027},
	publisher = {John Wiley {\&} Sons, Ltd},
	doi = {10.1002/2017GC006937}
}

@article{rubie_tungsten_2025,
	author = {Rubie, D. C. and Dale, K. I. and Nathan, G. and Nakajima, M. and Jennings, E. S. and Golabek, G. J. and Jacobson, S. A. and others},
	title = {{Tungsten isotope evolution during Earth's formation and new constraints on the viability of accretion simulations}},
	journal = {Earth Planet. Sci. Lett.},
	volume = {651},
	pages = {119139},
	year = {2025},
	month = feb,
	issn = {0012-821X},
	eprint = {arXiv:2411.17681},
	doi = {10.1016/j.epsl.2024.119139}
}

@article{gilmore_venus_2017,
	author = {Gilmore, Martha and Treiman, Allan and Helbert, J{\ifmmode\ddot{o}\else\"{o}\fi}rn and Smrekar, Suzanne},
	title = {{Venus Surface Composition Constrained by Observation and Experiment}},
	journal = {Space Sci. Rev.},
	volume = {212},
	number = {3},
	pages = {1511--1540},
	year = {2017},
	month = nov,
	issn = {1572-9672},
	publisher = {Springer Netherlands},
	doi = {10.1007/s11214-017-0370-8}
}

@article{gillmann_the_2022,
	author = {Gillmann, Cedric and Way, M. J. and Avice, Guillaume and Breuer, Doris and Golabek, Gregor J. and H{\ifmmode\ddot{o}\else\"{o}\fi}ning, Dennis and Krissansen-Totton, Joshua and others},
	title = {{The Long-Term Evolution of the Atmosphere of Venus: Processes and Feedback Mechanisms}},
	journal = {Space Sci. Rev.},
	volume = {218},
	number = {7},
	pages = {56},
	year = {2022},
	month = oct,
	issn = {1572-9672},
	publisher = {Springer Netherlands},
	doi = {10.1007/s11214-022-00924-0}
}

@article{cameron_the_1976,
	author = {Cameron, A. G. W. and Ward, W. R.},
	title = {{The Origin of the Moon}},
	journal = {Lunar and Planetary Science Conference},
	volume = {7},
	pages = {120},
	year = {1976},
	month = mar,
	url = {https://ui.adsabs.harvard.edu/abs/1976LPI.....7..120C/abstract}
}

@book{lodders_planetary_1998,
	author = {Lodders, Katharina and Fegley, Bruce},
	title = {{The Planetary Scientist{'}s Companion}},
	journal = {OUP Academic},
	year = {1998},
	month = dec,
	isbn = {978-0-19770972-6},
	publisher = {Oxford University Press},
	address = {Oxford, England, UK},
	doi = {10.1093/oso/9780195116946.001.0001}
}

@article{cronin_on_2014,
	author = {Cronin, Timothy W.},
	title = {{On the Choice of Average Solar Zenith Angle}},
	journal = {J. Atmos. Sci.},
	volume = {71},
	number = {8},
	pages = {2994--3003},
	year = {2014},
	month = aug,
	issn = {0022-4928},
	publisher = {American Meteorological Society},
	doi = {10.1175/JAS-D-13-0392.1}
}

@article{costa_rheology_2009,
	author = {Costa, Antonio and Caricchi, Luca and Bagdassarov, Nickolai},
	title = {{A model for the rheology of particle-bearing suspensions and partially molten rocks}},
	journal = {Geochem. Geophys. Geosyst.},
	volume = {10},
	number = {3},
	year = {2009},
	month = mar,
	issn = {1525-2027},
	publisher = {John Wiley {\&} Sons, Ltd},
	doi = {10.1029/2008GC002138}
}

@article{fischer_equation_2011,
	author = {Fischer, Rebecca A. and Campbell, Andrew J. and Shofner, Gregory A. and Lord, Oliver T. and Dera, Przemyslaw and Prakapenka, Vitali B.},
	title = {{Equation of state and phase diagram of FeO}},
	journal = {Earth Planet. Sci. Lett.},
	volume = {304},
	number = {3},
	pages = {496--502},
	year = {2011},
	month = apr,
	issn = {0012-821X},
	publisher = {Elsevier},
	doi = {10.1016/j.epsl.2011.02.025}
}

@article{oneill_the_2002,
	author = {O'Neill, Hugh {\relax St}. C. and Eggins, Stephen M.},
	title = {{The effect of melt composition on trace element partitioning: an experimental investigation of the activity coefficients of FeO, NiO, CoO, MoO2 and MoO3 in silicate melts}},
	journal = {Chem. Geol.},
	volume = {186},
	number = {1},
	pages = {151--181},
	year = {2002},
	month = jun,
	issn = {0009-2541},
	publisher = {Elsevier},
	doi = {10.1016/S0009-2541(01)00414-4}
}

@article{spaargaren_the_2020,
	author = {Spaargaren, Rob J. and Ballmer, Maxim D. and Bower, Dan J. and Dorn, Caroline and Tackley, Paul J.},
	title = {{The influence of bulk composition on the long-term interior-atmosphere evolution of terrestrial exoplanets}},
	journal = {Astron. Astrophys.},
	volume = {643},
	pages = {A44},
	year = {2020},
	month = nov,
	issn = {0004-6361},
	publisher = {EDP Sciences},
	doi = {10.1051/0004-6361/202037632}
}

@incollection{solomatov_ToG_2015,
title = {9.04 - Magma Oceans and Primordial Mantle Differentiation},
editor = {Gerald Schubert},
booktitle = {Treatise on Geophysics (Second Edition)},
publisher = {Elsevier},
edition = {Second Edition},
address = {Oxford},
pages = {81-104},
year = {2015},
isbn = {978-0-444-53803-1},
doi = {https://doi.org/10.1016/B978-0-444-53802-4.00155-X},
url = {https://www.sciencedirect.com/science/article/pii/B978044453802400155X},
author = {V. Solomatov}
}

@article{solomatov_nonfraction_1993,
	author = {Solomatov, Viatcheslav S. and Stevenson, David J.},
	title = {{Nonfractional crystallization of a terrestrial magma ocean}},
	journal = {J. Geophys. Res. Planets},
	volume = {98},
	number = {E3},
	pages = {5391--5406},
	year = {1993},
	month = mar,
	issn = {0148-0227},
	publisher = {John Wiley {\&} Sons, Ltd},
	doi = {10.1029/92JE02579}
}

@incollection{labrosse_fractional_2015,
	author = {Labrosse, St{\ifmmode\acute{e}\else\'{e}\fi}phane and Hernlund, John W. and Hirose, Kei},
	title = {{Fractional Melting and Freezing in the Deep Mantle and Implications for the Formation of a Basal Magma Ocean}},
	booktitle = {{The Early Earth}},
	journal = {AGU Journals},
	pages = {123--142},
	year = {2015},
	month = oct,
	isbn = {978-1-11886035-9},
	publisher = {John Wiley {\&} Sons, Ltd.},
	address = {Chichester, England, UK},
	doi = {10.1002/9781118860359.ch7}
}

@article{nicholls_volatile_2026,
	author = {Nicholls, Harrison and Lichtenberg, Tim and Chatterjee, Richard D. and Guimond, Claire Marie and Postolec, Emma and Pierrehumbert, Raymond T.},
	title = {{Volatile-rich evolution of molten super-Earth L 98-59 d}},
	journal = {Nat. Astron.},
	pages = {1--9},
	year = {2026},
	month = mar,
	issn = {2397-3366},
	publisher = {Nature Publishing Group},
	doi = {10.1038/s41550-026-02815-8}
}

@article{tucker_evidence_2014,
	author = {Tucker, Jonathan M. and Mukhopadhyay, Sujoy},
	title = {{Evidence for multiple magma ocean outgassing and atmospheric loss episodes from mantle noble gases}},
	journal = {Earth Planet. Sci. Lett.},
	volume = {393},
	pages = {254--265},
	year = {2014},
	month = may,
	issn = {0012-821X},
	publisher = {Elsevier},
	doi = {10.1016/j.epsl.2014.02.050}
}

@article{bower_numerical_2018,
	title = {Numerical solution of a non-linear conservation law applicable to the interior dynamics of partially molten planets},
	volume = {274},
	issn = {00319201},
	url = {https://linkinghub.elsevier.com/retrieve/pii/S0031920117302042},
	doi = {10.1016/j.pepi.2017.11.004},
	language = {en},
	urldate = {2023-01-10},
	journal = {Physics of the Earth and Planetary Interiors},
	author = {Bower, Dan J. and Sanan, Patrick and Wolf, Aaron S.},
	month = jan,
	year = {2018},
	pages = {49--62},
}

@article{bower_retention_2022,
	author = {Bower, Dan J. and Hakim, Kaustubh and Sossi, Paolo A. and Sanan, Patrick},
	title = {{Retention of Water in Terrestrial Magma Oceans and Carbon-rich Early Atmospheres}},
	journal = {Planet. Sci. J.},
	volume = {3},
	number = {4},
	pages = {93},
	year = {2022},
	month = apr,
	issn = {2632-3338},
	publisher = {The American Astronomical Society},
	doi = {10.3847/PSJ/ac5fb1}
}

@article{sossi_solubility_2023,
	author = {Sossi, Paolo A. and Tollan, Peter M. E. and Badro, James and Bower, Dan J.},
	title = {{Solubility of water in peridotite liquids and the prevalence of steam atmospheres on rocky planets}},
	journal = {Earth Planet. Sci. Lett.},
	volume = {601},
	pages = {117894},
	year = {2023},
	month = jan,
	issn = {0012-821X},
	publisher = {Elsevier},
	doi = {10.1016/j.epsl.2022.117894}
}

@article{way_venus_2016,
	author = {Way, M. J. and Del Genio, Anthony D. and Kiang, Nancy Y. and Sohl, Linda E. and Grinspoon, David H. and Aleinov, Igor and Kelley, Maxwell and others},
	title = {{Was Venus the first habitable world of our solar system?}},
	journal = {Geophys. Res. Lett.},
	volume = {43},
	number = {16},
	pages = {8376--8383},
	year = {2016},
	month = aug,
	issn = {0094-8276},
	publisher = {John Wiley {\&} Sons, Ltd},
	doi = {10.1002/2016GL069790}
}

@article{warren_narrow_2023,
	author = {Warren, Alexandra O. and Kite, Edwin S.},
	title = {{Narrow range of early habitable Venus scenarios permitted by modeling of oxygen loss and radiogenic argon degassing}},
	journal = {Proc. Natl. Acad. Sci. U.S.A.},
	volume = {120},
	number = {11},
	pages = {e2209751120},
	year = {2023},
	month = mar,
	publisher = {Proceedings of the National Academy of Sciences},
	doi = {10.1073/pnas.2209751120}
}

@article{orourke_venus_2015,
	author = {O{'}Rourke, Joseph G. and Korenaga, Jun},
	title = {{Thermal evolution of Venus with argon degassing}},
	journal = {Icarus},
	volume = {260},
	pages = {128--140},
	year = {2015},
	month = nov,
	issn = {0019-1035},
	publisher = {Academic Press},
	doi = {10.1016/j.icarus.2015.07.009}
}

@article{kaula_venus_1999,
	author = {Kaula, William M.},
	title = {{Constraints on Venus Evolution from Radiogenic Argon}},
	journal = {Icarus},
	volume = {139},
	number = {1},
	pages = {32--39},
	year = {1999},
	month = may,
	issn = {0019-1035},
	publisher = {Academic Press},
	doi = {10.1006/icar.1999.6082}
}

@article{constantinou_venus_2025,
	author = {Constantinou, Tereza and Shorttle, Oliver and Rimmer, Paul B.},
	title = {{A dry Venusian interior constrained by atmospheric chemistry}},
	journal = {Nat. Astron.},
	volume = {9},
	number = {2},
	pages = {189--198},
	year = {2025},
	month = feb,
	issn = {2397-3366},
	publisher = {Nature Publishing Group},
	doi = {10.1038/s41550-024-02414-5}
}

@article{strom_the_1994,
	author = {Strom, Robert G. and Schaber, Gerald G. and Dawson, Douglas D.},
	title = {{The global resurfacing of Venus}},
	journal = {J. Geophys. Res. Planets},
	volume = {99},
	number = {E5},
	pages = {10899--10926},
	year = {1994},
	month = may,
	issn = {0148-0227},
	publisher = {John Wiley {\&} Sons, Ltd},
	doi = {10.1029/94JE00388}
}

@article{wagner_iapws_2002,
	author = {Wagner, W. and Pru{\ss}, A.},
	title = {{The IAPWS Formulation 1995 for the Thermodynamic Properties of Ordinary Water Substance for General and Scientific Use}},
	journal = {J. Phys. Chem. Ref. Data},
	volume = {31},
	number = {2},
	pages = {387--535},
	year = {2002},
	month = jun,
	issn = {0047-2689},
	publisher = {AIP Publishing},
	doi = {10.1063/1.1461829}
}

@article{nakajima_olr_1992,
	author = {Nakajima, Shinichi and Hayashi, Yoshi-Yuki and Abe, Yutaka},
	title = {{A study on the 'runaway greenhouse effect' with a one-dimensional radiative-convective equilibrium model}},
	journal = {J. Atmos. Sci.},
	volume = {49},
	number = {23},
	pages = {2256--2266},
	year = {1992},
	month = dec,
	issn = {0022-4928},
	doi = {10.1175/1520-0469(1992)049<2256:ASOTGE>2.0.CO;2}
}

@article{sole_hadean_2025,
	author = {Sole, C. and O{'}Neil, J. and Rizo, H. and Paquette, J.-L. and Benn, D. and Plakholm, J.},
	title = {{Evidence for Hadean mafic intrusions in the Nuvvuagittuq Greenstone Belt, Canada}},
	journal = {Science},
	volume = {388},
	number = {6754},
	pages = {1431--1435},
	year = {2025},
	month = jun,
	issn = {0036-8075},
	publisher = {American Association for the Advancement of Science},
	doi = {10.1126/science.ads8461}
}

@article{borg_moon_2014,
	author = {Borg, Lars E. and Gaffney, Amy M. and Shearer, Charles K.},
	title = {{A review of lunar chronology revealing a preponderance of 4.34{\textendash}4.37 Ga ages}},
	journal = {Meteorit. Planet. Sci.},
	volume = {50},
	number = {4},
	pages = {715--732},
	year = {2014},
	month = nov,
	issn = {1945-5100},
	publisher = {John Wiley {\&} Sons, Ltd.},
	doi = {10.1111/maps.12373}
}

@article{katsura_earth_2022,
	author = {Katsura, Tomoo},
	title = {{A Revised Adiabatic Temperature Profile for the Mantle}},
	journal = {J. Geophys. Res. Solid Earth},
	volume = {127},
	number = {2},
	pages = {e2021JB023562},
	year = {2022},
	month = feb,
	issn = {2169-9313},
	publisher = {John Wiley {\&} Sons, Ltd},
	doi = {10.1029/2021JB023562}
}

@article{johnstone2021active,
  title={The active lives of stars: A complete description of the rotation and XUV evolution of F, G, K, and M dwarfs},
  author={Johnstone, Colin P and Bartel, Michael and G{\"u}del, Manuel},
  journal={Astronomy \& Astrophysics},
  volume={649},
  pages={A96},
  year={2021},
  publisher={EDP Sciences}
}

@article{yu_how_2021,
	author = {Yu, Xinting and Moses, Julianne I. and Fortney, Jonathan J. and Zhang, Xi},
	title = {{How to Identify Exoplanet Surfaces Using Atmospheric Trace Species in Hydrogen-dominated Atmospheres}},
	journal = {Astrophys. J.},
	volume = {914},
	number = {1},
	pages = {38},
	year = {2021},
	month = jun,
	issn = {0004-637X},
	eprint = {arXiv:2104.09843},
	doi = {10.3847/1538-4357/abfdc7}
}

@article{tsai_inferring_2021,
	author = {Tsai, Shang-Min and Innes, Hamish and Lichtenberg, Tim and Taylor, Jake and Malik, Matej and Chubb, Katy and Pierrehumbert, Raymond},
	title = {{Inferring Shallow Surfaces on Sub-Neptune Exoplanets with JWST}},
	journal = {Astrophys. J. Lett.},
	volume = {922},
	number = {2},
	pages = {L27},
	year = {2021},
	month = nov,
	issn = {2041-8205},
	publisher = {The American Astronomical Society},
	doi = {10.3847/2041-8213/ac399a}
}

@article{demory_a_2016,
	author = {Demory, Brice-Olivier and Gillon, Michael and de Wit, Julien and Madhusudhan, Nikku and Bolmont, Emeline and Heng, Kevin and Kataria, Tiffany and others},
	title = {{A map of the large day{\textendash}night temperature gradient of a super-Earth exoplanet}},
	journal = {Nature},
	volume = {532},
	number = {7598},
	pages = {207--209},
	year = {2016},
	month = apr,
	issn = {1476-4687},
	publisher = {Nature Publishing Group},
	doi = {10.1038/nature17169}
}

@article{ElkinsTanton2008,
  author = {L. T. Elkins-Tanton},
  title = {Linked magma ocean solidification and atmospheric growth for Earth and Mars},
  journal = {Earth Planet. Sci. Lett.},
  year = {2008},
  volume = {271},
  pages = {181--191},
  doi = {10.1016/j.epsl.2008.03.062}
}

@incollection{Hamano2024arXiv,
title = {The evolutionary divergence of Mars, Venus, and Earth},
editor = {Ariel Anbar and Dominique Weis},
booktitle = {Treatise on Geochemistry (Third edition)},
publisher = {Elsevier},
edition = {Third edition},
address = {Oxford},
pages = {541-574},
year = {2025},
isbn = {978-0-323-99763-8},
doi = {https://doi.org/10.1016/B978-0-323-99762-1.00104-2},
url = {https://www.sciencedirect.com/science/article/pii/B9780323997621001042},
author = {Keiko Hamano and Cedric Gillmann and Gregor J. Golabek and Diogo Lourenco and Frances Westall},
}

@ARTICLE{Schaefer2018RSPTA,
       author = {{Schaefer}, Laura and {Elkins-Tanton}, Linda T.},
        title = "{Magma oceans as a critical stage in the tectonic development of rocky planets}",
      journal = {Philosophical Transactions of the Royal Society of London Series A},
         year = 2018,
        month = nov,
       volume = {376},
       number = {2132},
          eid = {20180109},
        pages = {20180109},
          doi = {10.1098/rsta.2018.0109},
archivePrefix = {arXiv},
       eprint = {1809.01629},
 primaryClass = {astro-ph.EP},
       adsurl = {https://ui.adsabs.harvard.edu/abs/2018RSPTA.37680109S}
}

@article{Elkins-Tanton2012,
author = {Elkins-Tanton, Linda T},
doi = {10.1146/annurev-earth-042711-105503},
editor = {Jeanloz, R},
isbn = {978-0-8243-2040-9},
issn = {0084-6597},
journal = {Annual review of Earth and Planetary Sciences},
pages = {113--139},
title = {{Magma Oceans in the Inner Solar System}},
volume = {40},
year = {2012}
}

@ARTICLE{Herath2024MNRAS,
       author = {{Herath}, Mahesh and {Boukar{\'e}}, Charles-{\'E}douard and {Cowan}, Nicolas B.},
        title = "{Thermal evolution of lava planets}",
      journal = {\mnras},
         year = 2024,
        month = dec,
       volume = {535},
       number = {3},
        pages = {2404-2414},
          doi = {10.1093/mnras/stae2431},
       adsurl = {https://ui.adsabs.harvard.edu/abs/2024MNRAS.535.2404H}
}

@article{Hamano2013,
  author = {K. Hamano and Y. Abe and H. Genda},
  title = {Emergence of two types of terrestrial planet on hot and cold migration tracks},
  journal = {Nature},
  year = {2013},
  volume = {497},
  pages = {607--610},
  doi = {10.1038/nature12163}
}

@ARTICLE{Lupu2014ApJ,
       author = {{Lupu}, R.~E. and {Zahnle}, Kevin and {Marley}, Mark S. and {Schaefer}, Laura and {Fegley}, Bruce and {Morley}, Caroline and {Cahoy}, Kerri and {Freedman}, Richard and {Fortney}, Jonathan J.},
        title = "{The Atmospheres of Earthlike Planets after Giant Impact Events}",
      journal = {\apj},
         year = 2014,
        month = mar,
       volume = {784},
       number = {1},
          eid = {27},
        pages = {27},
          doi = {10.1088/0004-637X/784/1/27},
archivePrefix = {arXiv},
       eprint = {1401.1499},
 primaryClass = {astro-ph.EP},
       adsurl = {https://ui.adsabs.harvard.edu/abs/2014ApJ...784...27L}
}

@ARTICLE{Lichtenberg2025,
       author = {{Lichtenberg}, Tim and {Miguel}, Yamila},
        title = "{Super-Earths and Earth-like Exoplanets}",
      journal = {Treatise on Geochemistry},
         year = 2025,
        month = jan,
       volume = {7},
        pages = {51-112},
          doi = {10.1016/B978-0-323-99762-1.00122-4},
archivePrefix = {arXiv},
       eprint = {2405.04057},
 primaryClass = {astro-ph.EP},
       adsurl = {https://ui.adsabs.harvard.edu/abs/2025TrGeo...7...51L}
}

@INPROCEEDINGS{Lichtenberg2023,
       author = {{Lichtenberg}, Tim and {Schaefer}, Laura K. and {Nakajima}, Miki and {Fischer}, Rebecca A.},
        title = "{Geophysical Evolution During Rocky Planet Formation}",
    booktitle = {Protostars and Planets VII},
         year = 2023,
       editor = {{Inutsuka}, S. and {Aikawa}, Y. and {Muto}, T. and {Tomida}, K. and {Tamura}, M.},
       series = {Astronomical Society of the Pacific Conference Series},
       volume = {534},
        month = jul,
        pages = {907},
          doi = {10.48550/arXiv.2203.10023},
archivePrefix = {arXiv},
       eprint = {2203.10023},
 primaryClass = {astro-ph.EP},
       adsurl = {https://ui.adsabs.harvard.edu/abs/2023ASPC..534..907L}
}

@ARTICLE{Lichtenberg2025Science,
       author = {{Lichtenberg}, Tim and {Shorttle}, Oliver and {Teske}, Johanna and {Kempton}, Eliza M. -R.},
        title = "{Constraining exoplanet interiors using observations of their atmospheres}",
      journal = {Science},
         year = 2025,
       volume = {390},
        month = oct,
          eid = {eads3660},
        pages = {eads3660},
          doi = {10.1126/science.ads3360},
archivePrefix = {arXiv},
       eprint = {2510.08844}
}

@ARTICLE{Mojzsis2001Natur,
       author = {{Mojzsis}, Stephen J. and {Harrison}, T. Mark and {Pidgeon}, Robert T.},
        title = "{Oxygen-isotope evidence from ancient zircons for liquid water at the Earth's surface 4,300Myr ago}",
      journal = {\nat},
         year = 2001,
        month = jan,
       volume = {409},
       number = {6817},
        pages = {178-181},
          doi = {10.1038/35051557},
       adsurl = {https://ui.adsabs.harvard.edu/abs/2001Natur.409..178M}
}

@article{Marcq2017,
author = {Marcq, E and Salvador, A and Massol, H and Davaille, A},
doi = {10.1002/2016JE005224},
issn = {2169-9100},
journal = {Journal of Geophysical Research: Planets},
number = {7},
pages = {1539--1553},
title = {{Thermal radiation of magma ocean planets using a 1-D radiative-convective model of H$_{2}$O-CO$_{2}$ atmospheres}},
url = {http://dx.doi.org/10.1002/2016JE005224},
volume = {122},
year = {2017}
}

@article{Nikolaou2019,
author = {Nikolaou, Athanasia and Katyal, Nisha and Tosi, Nicola and Godolt, Mareike and {Lee Grenfell}, John and Rauer, Heike},
doi = {10.3847/1538-4357/ab08ed},
eprint = {1903.07436v1},
journal = {The Astrophysical Journal},
number = {11},
pages = {24},
title = {{What factors affect the duration and outgassing of the terrestrial magma ocean?}},
url = {https://doi.org/10.3847/1538-4357/ab08ed},
volume = {875},
year = {2019}
}

@article{seidler_impact_2024,
	author = {Seidler, Fabian L. and Sossi, Paolo A. and Grimm, Simon L.},
	title = {{Impact of oxygen fugacity on the atmospheric structure and emission spectra of ultra-hot rocky exoplanets}},
	journal = {Astron. Astrophys.},
	volume = {691},
	pages = {A159},
	year = {2024},
	month = nov,
	issn = {0004-6361},
	publisher = {EDP Sciences},
	doi = {10.1051/0004-6361/202450546}
}

@ARTICLE{rogers_subneptune_2025,
	author = {Rogers, James G. and Young, Edward D. and Schlichting, Hilke E.},
	title = {{Redefining interiors and envelopes: hydrogen{\textendash}silicate miscibility and its consequences for the structure and evolution of sub-Neptunes}},
	journal = {Mon. Not. R. Astron. Soc.},
	volume = {544},
	number = {4},
	pages = {3496--3511},
	year = {2025},
	month = dec,
	issn = {0035-8711},
	publisher = {Oxford Academic},
	doi = {10.1093/mnras/staf1940}
}

@article{Salvador2017,
  author = {A. Salvador and H. Massol and A. Davaille and E. Marcq and P. Sarda and E. Chassefi\`ere},
  title = {The relative influence of H$_2$O and CO$_2$ on the primitive surface conditions and evolution of rocky planets},
  journal = {J. Geophys. Res. Planets},
  year = {2017},
  volume = {122},
  pages = {1458--1486},
  doi = {10.1002/2017JE005286}
}

@article{Salvador2023_SSR,
author = {Salvador, Arnaud and Avice, Guillaume and Breuer, Doris and Gillmann, C{\'{e}}dric and Lammer, Helmut and Marcq, Emmanuel and Raymond, Sean N and Sakuraba, Haruka and Scherf, Manuel and Way, M J},
doi = {10.1007/s11214-023-00995-7},
issn = {1572-9672},
journal = {Space Science Reviews},
number = {7},
pages = {51},
title = {{Magma Ocean, Water, and the Early Atmosphere of Venus}},
url = {https://doi.org/10.1007/s11214-023-00995-7},
volume = {219},
year = {2023}
}

@ARTICLE{Schaefer2017ApJ,
       author = {{Schaefer}, Laura and {Fegley}, Jr., Bruce},
        title = "{Redox States of Initial Atmospheres Outgassed on Rocky Planets and Planetesimals}",
      journal = {\apj},
         year = 2017,
        month = jul,
       volume = {843},
       number = {2},
          eid = {120},
        pages = {120},
          doi = {10.3847/1538-4357/aa784f},
       adsurl = {https://ui.adsabs.harvard.edu/abs/2017ApJ...843..120S}
}

@ARTICLE{schaefer2016,
       author = {{Schaefer}, Laura and {Wordsworth}, Robin D. and {Berta-Thompson}, Zachory and {Sasselov}, Dimitar},
        title = "{Predictions of the Atmospheric Composition of GJ 1132b}",
      journal = {\apj},
         year = 2016,
        month = oct,
       volume = {829},
       number = {2},
          eid = {63},
        pages = {63},
          doi = {10.3847/0004-637X/829/2/63},
archivePrefix = {arXiv},
       eprint = {1607.03906},
 primaryClass = {astro-ph.EP},
       adsurl = {https://ui.adsabs.harvard.edu/abs/2016ApJ...829...63S}
}

@ARTICLE{Schaefer2024JGRE,
       author = {{Schaefer}, Laura and {Pahlevan}, Kaveh and {Elkins-Tanton}, Linda T.},
        title = "{Ferric Iron Evolution During Crystallization of the Earth and Mars}",
      journal = {Journal of Geophysical Research (Planets)},
         year = 2024,
        month = sep,
       volume = {129},
       number = {9},
          eid = {e2023JE008262},
        pages = {e2023JE008262},
          doi = {10.1029/2023JE008262},
       adsurl = {https://ui.adsabs.harvard.edu/abs/2024JGRE..12908262S}
}

@ARTICLE{Sim2024JGRE,
       author = {{Sim}, Shi J. and {Hirschmann}, Marc M. and {Hier-Majumder}, Saswata},
        title = "{Volatile and Trace Element Storage in a Crystallizing Martian Magma Ocean}",
      journal = {Journal of Geophysical Research (Planets)},
         year = 2024,
        month = aug,
       volume = {129},
       number = {8},
          eid = {e2024JE008346},
        pages = {e2024JE008346},
          doi = {10.1029/2024JE008346},
       adsurl = {https://ui.adsabs.harvard.edu/abs/2024JGRE..12908346S}
}

@ARTICLE{Teske2025,
       author = {{Teske}, Johanna K. and {Wallack}, Nicole L. and {Piette}, Anjali A.~A. and {Dang}, Lisa and {Lichtenberg}, Tim and {Plotnykov}, Mykhaylo and {Pierrehumbert}, Raymond T. and {Postolec}, Emma and {Boucher}, Samuel and {McGinty}, Alex and {Peng}, Bo and {Valencia}, Diana and {Hammond}, Mark},
        title = "{A Thick Volatile Atmosphere on the Ultra-Hot Super-Earth TOI-561 b}",
      journal = {arXiv e-prints},
         year = 2025,
        month = sep,
          eid = {arXiv:2509.17231},
        pages = {arXiv:2509.17231},
          doi = {10.48550/arXiv.2509.17231},
archivePrefix = {arXiv},
       eprint = {2509.17231},
 primaryClass = {astro-ph.EP},
       adsurl = {https://ui.adsabs.harvard.edu/abs/2025arXiv250917231T}
}

@ARTICLE{Valencia_2006,
       author = {{Valencia}, Diana and {O'Connell}, Richard J. and {Sasselov}, Dimitar},
        title = "{Internal structure of massive terrestrial planets}",
      journal = {\icarus},
         year = 2006,
        month = apr,
       volume = {181},
       number = {2},
        pages = {545-554},
          doi = {10.1016/j.icarus.2005.11.021},
archivePrefix = {arXiv},
       eprint = {astro-ph/0511150},
 primaryClass = {astro-ph},
       adsurl = {https://ui.adsabs.harvard.edu/abs/2006Icar..181..545V}
}

@article{wolf_eos_2018,
title = {An equation of state for high pressure-temperature liquids (RTpress) with application to MgSiO3 melt},
journal = {Physics of the Earth and Planetary Interiors},
volume = {278},
pages = {59-74},
year = {2018},
issn = {0031-9201},
doi = {https://doi.org/10.1016/j.pepi.2018.02.004},
url = {https://www.sciencedirect.com/science/article/pii/S0031920117301449},
author = {Aaron S. Wolf and Dan J. Bower},
}

@article{wordsworth_ABIOTICO_2014,
	author = {Wordsworth, Robin and Pierrehumbert, Raymond},
	title = {{ABIOTIC OXYGEN-DOMINATED ATMOSPHERES ON TERRESTRIAL HABITABLE ZONE PLANETS}},
	journal = {Astrophys. J. Lett.},
	volume = {785},
	number = {2},
	pages = {L20},
	year = {2014},
	month = apr,
	issn = {2041-8205},
	publisher = {The American Astronomical Society},
	doi = {10.1088/2041-8205/785/2/L20}
}

@ARTICLE{young_subneptune_2025,
       author = {{Young}, Edward D. and {Werlen}, Aaron and {Marcum}, Sarah P. and {Stixrude}, Lars and {Dullemond}, Cornelis P.},
        title = "{Differentiation, the exception not the rule -- Evidence for full miscibility in sub-Neptune interiors}",
      journal = {arXiv e-prints},
         year = 2025,
        month = jul,
          eid = {arXiv:2507.00947},
        pages = {arXiv:2507.00947},
          doi = {10.48550/arXiv.2507.00947},
archivePrefix = {arXiv},
       eprint = {2507.00947},
 primaryClass = {astro-ph.EP},
       adsurl = {https://ui.adsabs.harvard.edu/abs/2025arXiv250700947Y}
}

@ARTICLE{zilinskas2025,
       author = {{Zilinskas}, M. and {van Buchem}, C.~P.~A. and {Zieba}, S. and {Miguel}, Y. and {Sandford}, E. and {Hu}, R. and {Patel}, J.~A. and {Bello-Arufe}, A. and {Janssen}, L.~J. and {Tsai}, S. -M. and {Dragomir}, D. and {Zhang}, M.},
        title = "{Characterising the atmosphere of 55 Cancri e: 1D forward model grid for current and future JWST observations}",
      journal = {\aap},
         year = 2025,
        month = may,
       volume = {697},
          eid = {A34},
        pages = {A34},
          doi = {10.1051/0004-6361/202554062},
archivePrefix = {arXiv},
       eprint = {2503.15844},
 primaryClass = {astro-ph.EP},
       adsurl = {https://ui.adsabs.harvard.edu/abs/2025A&A...697A..34Z}
}

@article{kite_water_2021,
	author = {Kite, Edwin S. and Schaefer, Laura},
	title = {{Water on Hot Rocky Exoplanets}},
	journal = {Astrophys. J. Lett.},
	volume = {909},
	number = {2},
	pages = {L22},
	year = {2021},
	month = mar,
	issn = {2041-8205},
	publisher = {The American Astronomical Society},
	doi = {10.3847/2041-8213/abe7dc}
}

@article{miozzi_experiments_2025,
	author = {Miozzi, F. and Shahar, A. and Young, E. D. and Wang, J. and Steele, A. and Borensztajn, S. and Vitale, S. M. and others},
	title = {{Experiments reveal extreme water generation during planet formation}},
	journal = {Nature},
	volume = {648},
	number = {8094},
	pages = {551--555},
	year = {2025},
	month = dec,
	issn = {1476-4687},
	publisher = {Nature Publishing Group},
	doi = {10.1038/s41586-025-09816-z}
}

@article{zhang_melting_1994,
	author = {Zhang, Jianzhong and Herzberg, Claude},
	title = {{Melting experiments on anhydrous peridotite KLB-1 from 5.0 to 22.5 GPa}},
	journal = {J. Geophys. Res. Solid Earth},
	volume = {99},
	number = {B9},
	pages = {17729--17742},
	year = {1994},
	month = sep,
	issn = {0148-0227},
	publisher = {John Wiley {\&} Sons, Ltd},
	doi = {10.1029/94JB01406}
}

@article{fiquet_melting_2010,
	author = {Fiquet, G. and Auzende, A. L. and Siebert, J. and Corgne, A. and Bureau, H. and Ozawa, H. and Garbarino, G.},
	title = {{Melting of Peridotite to 140 Gigapascals}},
	journal = {Science},
	volume = {329},
	number = {5998},
	pages = {1516--1518},
	year = {2010},
	month = sep,
	issn = {0036-8075},
	publisher = {American Association for the Advancement of Science},
	doi = {10.1126/science.1192448}
}

@article{hauri_partitionin_2006,
	author = {Hauri, Erik H. and Gaetani, Glenn A. and Green, Trevor H.},
	title = {{Partitioning of water during melting of the Earth's upper mantle at H2O-undersaturated conditions}},
	journal = {Earth Planet. Sci. Lett.},
	volume = {248},
	number = {3},
	pages = {715--734},
	year = {2006},
	month = aug,
	issn = {0012-821X},
	publisher = {Elsevier},
	doi = {10.1016/j.epsl.2006.06.014}
}

@article{bell_water_1992,
	author = {Bell, David R. and Rossman, George R.},
	title = {{Water in Earth's Mantle: The Role of Nominally Anhydrous Minerals}},
	journal = {Science},
	volume = {255},
	number = {5050},
	pages = {1391--1397},
	year = {1992},
	month = mar,
	issn = {0036-8075},
	publisher = {American Association for the Advancement of Science},
	doi = {10.1126/science.255.5050.1391}
}

@article{chao_lava_2021,
	author = {Chao, Keng-Hsien and DeGraffenried, Rebecca and Lach, Mackenzie and Nelson, William and Truax, Kelly and Gaidos, Eric},
	title = {{Lava worlds: From early earth to exoplanets}},
	journal = {Geochemistry},
	volume = {81},
	number = {2},
	pages = {125735},
	year = {2021},
	month = may,
	issn = {0009-2819},
	publisher = {Urban {\&} Fischer},
	doi = {10.1016/j.chemer.2020.125735}
}

@article{nicholls_convective_2025,
    author = {Nicholls, Harrison and Pierrehumbert, Raymond T and Lichtenberg, Tim and Soucasse, Laurent and Smeets, Stef},
    title = {Convective shutdown in the atmospheres of lava worlds},
    journal = {Monthly Notices of the Royal Astronomical Society},
    volume = {536},
    number = {3},
    pages = {2957-2971},
    year = {2025},
    month = 1,
    issn = {0035-8711},
    doi = {10.1093/mnras/stae2772},
    url = {https://doi.org/10.1093/mnras/stae2772},
    eprint = {https://academic.oup.com/mnras/article-pdf/536/3/2957/61218697/stae2772.pdf},
}

@article{nicholls_redox_2024,
author = {Nicholls, Harrison and Lichtenberg, Tim and Bower, Dan J. and Pierrehumbert, Raymond},
title = {Magma Ocean Evolution at Arbitrary Redox State},
journal = {Journal of Geophysical Research: Planets},
volume = {129},
number = {12},
pages = {e2024JE008576},
doi = {https://doi.org/10.1029/2024JE008576},
url = {https://agupubs.onlinelibrary.wiley.com/doi/abs/10.1029/2024JE008576},
eprint = {https://agupubs.onlinelibrary.wiley.com/doi/pdf/10.1029/2024JE008576},
note = {e2024JE008576 2024JE008576},
year = {2024}
}

@ARTICLE{lichtenberg2021,
       author = {{Lichtenberg}, Tim and {Bower}, Dan J. and {Hammond}, Mark and {Boukrouche}, Ryan and {Sanan}, Patrick and {Tsai}, Shang-Min and {Pierrehumbert}, Raymond T.},
        title = "{Vertically Resolved Magma Ocean-Protoatmosphere Evolution: H$_{2}$, H$_{2}$O, CO$_{2}$, CH$_{4}$, CO, O$_{2}$, and N$_{2}$ as Primary Absorbers}",
      journal = {Journal of Geophysical Research (Planets)},
         year = 2021,
        month = feb,
       volume = {126},
       number = {2},
          eid = {e06711},
        pages = {e06711},
          doi = {10.1029/2020JE006711},
archivePrefix = {arXiv},
       eprint = {2101.10991},
 primaryClass = {astro-ph.EP},
       adsurl = {https://ui.adsabs.harvard.edu/abs/2021JGRE..12606711L}
}

@ARTICLE{peng2024,
       author = {{Peng}, Bo and {Valencia}, Diana},
        title = "{Puffy Venuses: The Mass{\textendash}Radius Impact of Carbon-rich Atmospheres on Lava Worlds}",
      journal = {\apj},
         year = 2024,
        month = dec,
       volume = {976},
       number = {2},
          eid = {202},
        pages = {202},
          doi = {10.3847/1538-4357/ad6f03},
archivePrefix = {arXiv},
       eprint = {2405.08998},
 primaryClass = {astro-ph.EP},
       adsurl = {https://ui.adsabs.harvard.edu/abs/2024ApJ...976..202P}
}

@ARTICLE{krissansen-totton2021,
       author = {{Krissansen-Totton}, Joshua and {Fortney}, Jonathan J. and {Nimmo}, Francis and {Wogan}, Nicholas},
        title = "{Oxygen False Positives on Habitable Zone Planets Around Sun-Like Stars}",
      journal = {AGU Advances},
         year = {2021a},
        month = jun,
       volume = {2},
       number = {2},
          eid = {e00294},
        pages = {e00294},
          doi = {10.1029/2020AV000294},
archivePrefix = {arXiv},
       eprint = {2104.06463},
 primaryClass = {astro-ph.EP},
       adsurl = {https://ui.adsabs.harvard.edu/abs/2021AGUA....200294K}
}

@article{Krissansen-Totton_2024, 
title={The erosion of large primary atmospheres typically leaves behind substantial secondary atmospheres on temperate rocky planets}, volume={15}, 
ISSN={2041-1723}, 
DOI={10.1038/s41467-024-52642-6},
number={1}, 
journal={Nature Communications}, 
author={Krissansen-Totton, Joshua and Wogan, Nicholas and Thompson, Maggie and Fortney, Jonathan J.}, 
year={2024}, month=sep, pages={8374}, language={en} }

@article{Abe1985,
author = {Abe, Yutaka and Matsui, Takafumi},
doi = {10.1029/jb090is02p0c545},
issn = {0148-0227},
journal = {Journal of Geophysical Research},
pages = {C545--C559},
title = {{The formation of an impact-generated H$_2$O atmosphere and its implications for the early thermal history of the Earth}},
volume = {90},
year = {1985}
}

@ARTICLE{cmiel2025,
       author = {{Cmiel}, Jessica and {Wordsworth}, Robin and {Seeley}, Jacob T.},
        title = "{Characterizing the Radiative{\textendash}Convective Structure of Dense Rocky Planet Atmospheres}",
      journal = {\psj},
         year = 2025,
        month = may,
       volume = {6},
       number = {5},
          eid = {123},
        pages = {123},
          doi = {10.3847/PSJ/adcd5f},
archivePrefix = {arXiv},
       eprint = {2505.00775},
 primaryClass = {astro-ph.EP},
       adsurl = {https://ui.adsabs.harvard.edu/abs/2025PSJ.....6..123C}
}

@article{nguyen2024clouds,
  title={Clouds in Partial Atmospheres of Lava Planets and Where to Find Them},
  author={Nguyen, T Giang and Cowan, Nicolas B and Dang, Lisa},
  journal={The Astronomical Journal},
  volume={168},
  number={6},
  pages={287},
  year={2024},
  publisher={IOP Publishing}
}

@article{Sohl24_cuisines,
  title = {The {{CUISINES Framework}} for {{Conducting Exoplanet Model Intercomparison Projects}}, {{Version}} 1.0},
  author = {Sohl, Linda E. and Fauchez, Thomas J. and {Domagal-Goldman}, Shawn and Christie, Duncan A. and Deitrick, Russell and {Haqq-Misra}, Jacob and Harman, C. E. and Iro, Nicolas and Mayne, Nathan J. and Tsigaridis, Kostas and Villanueva, Geronimo L. and Young, Amber V. and Chaverot, Guillaume},
  year = {2024},
  month = aug,
  journal = {The Planetary Science Journal},
  volume = {5},
  number = {8},
  pages = {175},
  publisher = {IOP Publishing},
  issn = {2632-3338},
  doi = {10.3847/PSJ/ad5830},
  urldate = {2024-08-25},
  langid = {english}
}

@ARTICLE{Agol2021,
       author = {{Agol}, Eric and {Dorn}, Caroline and {Grimm}, Simon L. and {Turbet}, Martin and {Ducrot}, Elsa and {Delrez}, Laetitia and {Gillon}, Micha{\"e}l and {Demory}, Brice-Olivier and {Burdanov}, Artem and {Barkaoui}, Khalid and {Benkhaldoun}, Zouhair and {Bolmont}, Emeline and {Burgasser}, Adam and {Carey}, Sean and {de Wit}, Julien and {Fabrycky}, Daniel and {Foreman-Mackey}, Daniel and {Haldemann}, Jonas and {Hernandez}, David M. and {Ingalls}, James and {Jehin}, Emmanuel and {Langford}, Zachary and {Leconte}, J{\'e}r{\'e}my and {Lederer}, Susan M. and {Luger}, Rodrigo and {Malhotra}, Renu and {Meadows}, Victoria S. and {Morris}, Brett M. and {Pozuelos}, Francisco J. and {Queloz}, Didier and {Raymond}, Sean N. and {Selsis}, Franck and {Sestovic}, Marko and {Triaud}, Amaury H.~M.~J. and {Van Grootel}, Valerie},
        title = "{Refining the Transit-timing and Photometric Analysis of TRAPPIST-1: Masses, Radii, Densities, Dynamics, and Ephemerides}",
      journal = {\psj},
         year = 2021,
        month = feb,
       volume = {2},
       number = {1},
          eid = {1},
        pages = {1},
          doi = {10.3847/PSJ/abd022},
archivePrefix = {arXiv},
       eprint = {2010.01074},
 primaryClass = {astro-ph.EP},
       adsurl = {https://ui.adsabs.harvard.edu/abs/2021PSJ.....2....1A}
}

@article{luger2015extreme,
  title={Extreme water loss and abiotic O2 buildup on planets throughout the habitable zones of M dwarfs},
  author={Luger, Rodrigo and Barnes, Rory},
  journal={Astrobiology},
  volume={15},
  number={2},
  pages={119--143},
  year={2015},
  publisher={Mary Ann Liebert, Inc. 140 Huguenot Street, 3rd Floor New Rochelle, NY 10801 USA}
}

@article{Baraffe2015,
  author       = {Baraffe, Isabelle and Homeier, Derek and Allard, France and Chabrier, Gilles},
  title        = {New evolutionary models for pre-main sequence and main sequence low-mass stars down to the hydrogen-burning limit},
  journal      = {Astronomy \& Astrophysics},
  year         = {2015},
  volume       = {577},
  pages        = {A42},
  doi          = {10.1051/0004-6361/201425481}
}

@ARTICLE{Hamano2015,
       author = {{Hamano}, Keiko and {Kawahara}, Hajime and {Abe}, Yutaka and {Onishi}, Masanori and {Hashimoto}, George L.},
        title = "{Lifetime and Spectral Evolution of a Magma Ocean with a Steam Atmosphere: Its Detectability by Future Direct Imaging}",
      journal = {\apj},
         year = 2015,
        month = jun,
       volume = {806},
       number = {2},
          eid = {216},
        pages = {216},
          doi = {10.1088/0004-637X/806/2/216},
archivePrefix = {arXiv},
       eprint = {1505.03552},
 primaryClass = {astro-ph.EP},
       adsurl = {https://ui.adsabs.harvard.edu/abs/2015ApJ...806..216H}
}

@article{Rothman2010,
title = {HITEMP, the high-temperature molecular spectroscopic database},
journal = {Journal of Quantitative Spectroscopy and Radiative Transfer},
volume = {111},
number = {15},
pages = {2139-2150},
year = {2010},
note = {XVIth Symposium on High Resolution Molecular Spectroscopy (HighRus-2009)},
issn = {0022-4073},
doi = {https://doi.org/10.1016/j.jqsrt.2010.05.001},
author = {L.S. Rothman and I.E. Gordon and R.J. Barber and H. Dothe and R.R. Gamache and A. Goldman and V.I. Perevalov and S.A. Tashkun and J. Tennyson}
}

@article{Mlawer2012,
  title={Development and recent evaluation of the MT\_CKD model of continuum absorption},
  author={Mlawer, Eli J and Payne, Vivienne H and Moncet, Jean-Luc and Delamere, Jennifer S and Alvarado, Matthew J and Tobin, David C},
  journal={Philosophical Transactions of the Royal Society A: Mathematical, Physical and Engineering Sciences},
  volume={370},
  number={1968},
  pages={2520--2556},
  year={2012},
  publisher={The Royal Society Publishing}
}

@incollection{Presnall&Hoover1987,
  author     = {Presnall, D. C. and Hoover, J. D.},
  title     = {High pressure phase equilibrium constraints on the origin of mid-ocean ridge basalts},
  booktitle = {Magmatic Processes: Physicochemical Principles},
  editor    = {Mysen, B. O.},
  year      = {1987},
  publisher = {The Geochemical Society},
  series    = {Special Publication},
  volume    = {1},
  pages     = {75--89}
}

@article{Herzberg2000,
  author    = {Herzberg, C. and Raterron, P. and Zhang, J.},
  title     = {New experimental observations on the anhydrous solidus for peridotite {KLB-1}},
  journal   = {Geochemistry, Geophysics, Geosystems},
  volume    = {1},
  number    = {11},
  pages     = {1051},
  year      = {2000},
  publisher = {Wiley-Blackwell},
}

@article{Gordon2022,
  author       = {Gordon, Iouli E. and Rothman, Laurence S. and Hargreaves, Robert J. and Hashemi, Rozenn and Karlovets, Elena V. and Skinner, Frances M. and Conway, Emma K. and Hill, Calum and Kochanov, Roman V. and Tan, Yan and Wcisło, Paweł and Finenko, Andrew A. and Nelson, Kevin and Bernath, Peter F. and Boudon, Vincent and Campargue, Alain and Chance, Kelly and Coustenis, Athena and Drouin, Brian J. and Flaud, Jean-Marie and Gamache, Robert R. and Hodges, Joseph T. and Jacquemart, David and Mlawer, Eli J. and Moazzen-Ahmadi, Nasser and Nikitin, Andrei V. and Perevalov, Victor I. and Perrin, Agnes and Shine, Keith P. and Smith, Mary A. H. and Sung, Keeyoon and Tennyson, Jonathan and Toon, Geoff C. and Tran, Ha and Tyuterev, Vladimir G. and Adkins, Erika M. and Baker, Anne and Barbe, Alain and Boone, Chris and Cich, Matthew J. and Clarisse, Lieven and Clerbaux, Cathy and Coheur, Pierre-François and Dana, Victor and Devi, Vidhya M. and Fayt, Alan and Fittschen, Christa and Flaud, J.-M. and Gamache, R. R. and Harrison, Jeremy J. and Hatton, Chloë and Hodnebrog, Øivind and Hu, Sun and Jacquinet-Husson, Nicole and Jeong, Sang-Hun and Li, Guanhua and Long, David A. and Lyulin, Oleg M. and Mikhailenko, Sergey N. and Müller, Holger S. P. and Nemtchinov, Ivan and Orphal, Johannes and Pickett, Herbert M. and Polyansky, Oleg L. and Rey, Michael and Rotger, Marc and Sharpe, Steven W. and Smith, Kevin M. and Starikova, Elena and Sung, Keeyoon and Tashkun, Sergey A. and Wagner, Georg and Wilzewski, Jonas S. and Wong, Alan and Zak, Ewelina J.},
  title        = {The HITRAN2020 molecular spectroscopic database},
  journal      = {Journal of Quantitative Spectroscopy and Radiative Transfer},
  volume       = {277},
  pages        = {107949},
  year         = {2022},
  doi          = {10.1016/j.jqsrt.2021.107949}
}

@article{hirschmann2022magma,
  title={Magma oceans, iron and chromium redox, and the origin of comparatively oxidized planetary mantles},
  author={Hirschmann, MM},
  journal={Geochimica et Cosmochimica Acta},
  volume={328},
  pages={221--241},
  year={2022},
  publisher={Elsevier}
}

@article{Lebrun2013,
  author = {T. Lebrun and H. Massol and E. Chassefi\`ere and A. Davaille and E. Marcq and P. Sarda and F. Leblanc and G. Brandeis},
  title = {Thermal evolution of an early magma ocean in interaction with the atmosphere},
  journal = {J. Geophys. Res. Planets},
  year = {2013},
  volume = {118},
  pages = {1155--1176},
  doi = {10.1002/jgre.20068}
}

@article{Wordsworth2013,
  author = {R. Wordsworth and R. Pierrehumbert},
  title = {Hydrogen-nitrogen greenhouse warming in Earth's early atmosphere},
  journal = {Science},
  year = {2013},
  volume = {339},
  pages = {64--67},
  doi = {10.1126/science.1225759}
}

@article{Turbet2021,
  author = {M. Turbet and G. Chaverot and J. Leconte and F. Selsis and E. Bolmont and F. Forget},
  title = {Day--night cloud asymmetry prevents early oceans on Venus but not Earth},
  journal = {Nature},
  year = {2021},
  volume = {598},
  pages = {276--280},
  doi = {10.1038/s41586-021-03873-w}
}

@article{selsis_cool_2023,
	title = {A cool runaway greenhouse without surface magma ocean},
	volume = {620},
	copyright = {2023 The Author(s), under exclusive licence to Springer Nature Limited},
	issn = {1476-4687},
	url = {https://www.nature.com/articles/s41586-023-06258-3},
	doi = {10.1038/s41586-023-06258-3},
	language = {en},
	number = {7973},
	urldate = {2025-04-08},
	journal = {Nature},
	author = {Selsis, Franck and Leconte, Jeremy and Turbet, Martin and Chaverot, Guillaume and Bolmont, Emeline},
	month = aug,
	year = {2023},
	note = {Publisher: Nature Publishing Group},
	pages = {287--291},
}

@article{leconte_spectral_2021,
	title = {Spectral binning of precomputed correlated-k coefficients},
	volume = {645},
	copyright = {© J. Leconte 2020},
	issn = {0004-6361, 1432-0746},
	url = {https://www.aanda.org/articles/aa/abs/2021/01/aa39040-20/aa39040-20.html},
	doi = {10.1051/0004-6361/202039040},
	language = {en},
	urldate = {2025-04-08},
	journal = {Astronomy \& Astrophysics},
	author = {Leconte, Jérémy},
	month = jan,
	year = {2021},
	note = {Publisher: EDP Sciences},
	pages = {A20},
}

@article{kress_the_1991,
	author = {Kress, Victor C. and Carmichael, Ian S. E.},
	title = {{The compressibility of silicate liquids containing Fe2O3 and the effect of composition, temperature, oxygen fugacity and pressure on their redox states}},
	journal = {Contrib. Mineral. Petrol.},
	volume = {108},
	number = {1},
	pages = {82--92},
	year = {1991},
	month = jul,
	issn = {1432-0967},
	publisher = {Springer-Verlag},
	doi = {10.1007/BF00307328}
}

@article{katyal_effect_2020,
	author = {Katyal, Nisha and Ortenzi, Gianluigi and Grenfell, John Lee and Noack, Lena and Sohl, Frank and Godolt, Mareike and Mu{\ifmmode\tilde{n}\else\~{n}\fi}oz, Antonio Garc{\ifmmode\acute{\imath}\else\'{\i}\fi}a and others},
	title = {{Effect of mantle oxidation state and escape upon the evolution of Earth{'}s magma ocean atmosphere}},
	journal = {Astron. Astrophys.},
	volume = {643},
	pages = {A81},
	year = {2020},
	month = nov,
	issn = {0004-6361},
	publisher = {EDP Sciences},
	doi = {10.1051/0004-6361/202038779}
}

@article{kasting_mantle_1993,
	author = {Kasting, James F. and Eggler, David H. and Raeburn, Stuart P.},
	title = {{Mantle Redox Evolution and the Oxidation State of the Archean Atmosphere}},
	journal = {J. Geol.},
	year = {1993},
	month = mar,
	publisher = {University of Chicago Press},
	doi = {10.1086/648219}
}

@article{sorbadere_the_2018,
	author = {Sorbadere, Fanny and Laurenz, Vera and Frost, Daniel J. and Wenz, Michelle and Rosenthal, Anja and McCammon, Catherine and Rivard, Camille},
	title = {{The behaviour of ferric iron during partial melting of peridotite}},
	journal = {Geochim. Cosmochim. Acta},
	volume = {239},
	pages = {235--254},
	year = {2018},
	month = oct,
	issn = {0016-7037},
	publisher = {Pergamon},
	doi = {10.1016/j.gca.2018.07.019}
}

@article{Miyazaki2019,
  title={On the timescale of magma ocean solidification and its chemical consequences: 1. Thermodynamic database for liquid at high pressures},
  author={Miyazaki, Yoshinori and Korenaga, Jun},
  journal={Journal of Geophysical Research: Solid Earth},
  volume={124},
  number={4},
  pages={3382--3398},
  year={2019},
  publisher={Wiley Online Library}
}

@article{wordsworth2018redox,
  title={Redox evolution via gravitational differentiation on low-mass planets: implications for abiotic oxygen, water loss, and habitability},
  author={Wordsworth, RD and Schaefer, LK and Fischer, RA},
  journal={The Astronomical Journal},
  volume={155},
  number={5},
  pages={195},
  year={2018},
  publisher={IOP Publishing}
}

@article{barth2021magma,
  title={Magma ocean evolution of the TRAPPIST-1 planets},
  author={Barth, Patrick and Carone, Ludmila and Barnes, Rory and Noack, Lena and Molli{\`e}re, Paul and Henning, Thomas},
  journal={Astrobiology},
  volume={21},
  number={11},
  pages={1325--1349},
  year={2021},
  publisher={Mary Ann Liebert, Inc., publishers 140 Huguenot Street, 3rd Floor New~…}
}

@article{krissansen2022predictions,
  title={Predictions for observable atmospheres of trappist-1 planets from a fully coupled atmosphere--interior evolution model},
  author={Krissansen-Totton, Joshua and Fortney, Jonathan J},
  journal={The Astrophysical Journal},
  volume={933},
  number={1},
  pages={115},
  year={2022},
  publisher={IOP Publishing}
}

@article{stixrude_melting_2014,
	author = {Stixrude, Lars},
	title = {{Melting in super-earths}},
	journal = {Philos. Trans. Royal Soc. A},
	volume = {372},
	number = {2014},
	year = {2014},
	month = apr,
	issn = {1364-503X},
	publisher = {The Royal Society},
	doi = {10.1098/rsta.2013.0076}
}

@article{manners_a_2024,
	author = {Manners, James},
	title = {{A fast and flexible scheme for photolysis and radiative heating of the whole atmosphere}},
	journal = {AIP Conf. Proc.},
	volume = {2988},
	number = {1},
	year = {2024},
	month = jan,
	issn = {0094-243X},
	publisher = {AIP Publishing},
	doi = {10.1063/5.0185476}
}

@article{joyce_a_2023,
	author = {Joyce, Meridith and Tayar, Jamie},
	title = {{A Review of the Mixing Length Theory of Convection in 1D Stellar Modeling}},
	journal = {Galaxies},
	volume = {11},
	number = {3},
	pages = {75},
	year = {2023},
	month = jun,
	issn = {2075-4434},
	publisher = {Multidisciplinary Digital Publishing Institute},
	doi = {10.3390/galaxies11030075}
}

@article{marchi_long_2023,
	author = {Marchi, Simone and Rufu, Raluca and Korenaga, Jun},
	title = {{Long-lived volcanic resurfacing of Venus driven by early collisions}},
	journal = {Nat. Astron.},
	volume = {7},
	number = {10},
	pages = {1180--1187},
	year = {2023},
	month = oct,
	issn = {2397-3366},
	publisher = {Nature Publishing Group},
	doi = {10.1038/s41550-023-02037-2}
}

@article{ormel_how_2021,
	author = {Ormel, Chris W. and Vazan, Allona and Brouwers, Marc G.},
	title = {{How planets grow by pebble accretion. III. Emergence of an interior composition gradient}},
	journal = {Astron. Astrophys.},
	volume = {647},
	pages = {A175},
	year = {2021},
	month = mar,
	issn = {0004-6361},
	eprint = {arXiv:2010.14213},
	doi = {10.1051/0004-6361/202039706}
}

@article{chili_protocol_2025,
	author = {Lichtenberg, Tim and Schaefer, Laura and Krissansen-Totton, Joshua and Miguel, Yamila and Sergeev, Denis E. and Baumeister, Philipp and Cmiel, Jessica and others},
	title = {{Coupled atmospHere Interior modeL Intercomparison (CHILI){\ifmmode---\else\textemdash\fi}Protocol Version 1.0: A CUISINES Intercomparison Project of Magma Ocean Models}},
	journal = {Planet. Sci. J.},
	volume = {7},
	number = {5},
	pages = {108},
	year = {2026},
	month = may,
	issn = {2632-3338},
	publisher = {The American Astronomical Society},
	doi = {10.3847/PSJ/ae593b}
}

@article{walbecq_the_2025,
	author = {Walbecq, Arthur and Samuel, Henri and Limare, Angela},
	title = {{The effect of out-of-equilibrium outgassing on the cooling of planetary magma oceans}},
	journal = {Icarus},
	volume = {434},
	pages = {116513},
	year = {2025},
	month = jul,
	issn = {0019-1035},
	publisher = {Academic Press},
	doi = {10.1016/j.icarus.2025.116513}
}

@article{maurice_redox_2023,
	author = {Maurice, Maxime and Dasgupta, Rajdeep and Hassanzadeh, Pedram},
	title = {{RETRACTED: Redox Evolution of the Crystallizing Terrestrial Magma Ocean and Its Influence on the Outgassed Atmosphere}},
	journal = {Planet. Sci. J.},
	volume = {4},
	number = {2},
	pages = {31},
	year = {2023},
	month = feb,
	issn = {2632-3338},
	publisher = {The American Astronomical Society},
	doi = {10.3847/PSJ/acb2ca}
}

@article{cherubim2025oxidation,
  title={An Oxidation Gradient Straddling the Small Planet Radius Valley},
  author={Cherubim, Collin and Wordsworth, Robin and Bower, Dan J and Sossi, Paolo A and Adams, Danica and Hu, Renyu},
  journal={The Astrophysical Journal},
  volume={983},
  number={2},
  pages={97},
  year={2025},
  publisher={IOP Publishing}
}

@article{abe_thermal_1997,
	author = {Abe, Yutaka},
	title = {{Thermal and chemical evolution of the terrestrial magma ocean}},
	journal = {Phys. Earth Planet. Inter.},
	volume = {100},
	number = {1},
	pages = {27--39},
	year = {1997},
	month = mar,
	issn = {0031-9201},
	publisher = {Elsevier},
	doi = {10.1016/S0031-9201(96)03229-3}
}

@article{solomatov_stagnant_1996,
	author = {Solomatov, V. S. and Moresi, L.-N.},
	title = {{Stagnant lid convection on Venus}},
	journal = {J. Geophys. Res. Planets},
	volume = {101},
	number = {E2},
	pages = {4737--4753},
	year = {1996},
	month = feb,
	issn = {2156-2202},
	publisher = {John Wiley {\&} Sons, Ltd.},
	doi = {10.1029/95JE03361}
}

@article{stevenson_core_2001,
	author = {Stevenson, David J.},
	title = {{Mars' core and magnetism}},
	journal = {Nature},
	volume = {412},
	number = {6843},
	pages = {214--219},
	year = {2001},
	month = jul,
	issn = {1476-4687},
	publisher = {Nature Publishing Group},
	doi = {10.1038/35084155}
}

@article{lay_coremantle_2008,
	author = {Lay, Thorne and Hernlund, John and Buffett, Bruce A.},
	title = {{Core{\textendash}mantle boundary heat flow}},
	journal = {Nat. Geosci.},
	volume = {1},
	number = {1},
	pages = {25--32},
	year = {2008},
	month = jan,
	issn = {1752-0908},
	publisher = {Nature Publishing Group},
	doi = {10.1038/ngeo.2007.44}
}

@article{zahnle_the_2015,
	author = {Zahnle, Kevin J. and Lupu, Roxana and Dobrovolskis, Anthony and Sleep, Norman H.},
	title = {{The tethered Moon}},
	journal = {Earth Planet. Sci. Lett.},
	volume = {427},
	pages = {74--82},
	year = {2015},
	month = oct,
	issn = {0012-821X},
	publisher = {Elsevier},
	doi = {10.1016/j.epsl.2015.06.058}
}

@article{pu_low_2019,
	author = {Pu, Bonan and Lai, Dong},
	title = {{Low-eccentricity migration of ultra-short-period planets in multiplanet systems}},
	journal = {Mon. Not. R. Astron. Soc.},
	volume = {488},
	number = {3},
	pages = {3568--3587},
	year = {2019},
	month = sep,
	issn = {0035-8711},
	publisher = {Oxford Academic},
	doi = {10.1093/mnras/stz1817}
}

@article{unterborn_inward_2018,
	author = {Unterborn, Cayman T. and Desch, Steven J. and Hinkel, Natalie R. and Lorenzo, Alejandro},
	title = {{Inward migration of the TRAPPIST-1 planets as inferred from their water-rich compositions}},
	journal = {Nat. Astron.},
	volume = {2},
	number = {4},
	pages = {297--302},
	year = {2018},
	month = apr,
	issn = {2397-3366},
	publisher = {Nature Publishing Group},
	doi = {10.1038/s41550-018-0411-6}
}

@article{labrosse_core_1997,
	author = {Labrosse, St{\ifmmode\acute{e}\else\'{e}\fi}phane and Poirier, Jean-Paul and Le Mou{\ifmmode\ddot{e}\else\"{e}\fi}l, Jean-Louis},
	title = {{On cooling of the Earth's core}},
	journal = {Phys. Earth Planet. Inter.},
	volume = {99},
	number = {1},
	pages = {1--17},
	year = {1997},
	month = jan,
	issn = {0031-9201},
	publisher = {Elsevier},
	doi = {10.1016/S0031-9201(96)03207-4}
}

@article{boukare_timing_2018,
	author = {Boukar{\ifmmode\acute{e}\else\'{e}\fi}, C.-E. and Parmentier, E. M. and Parman, S. W.},
	title = {{Timing of mantle overturn during magma ocean solidification}},
	journal = {Earth Planet. Sci. Lett.},
	volume = {491},
	pages = {216--225},
	year = {2018},
	month = jun,
	issn = {0012-821X},
	publisher = {Elsevier},
	doi = {10.1016/j.epsl.2018.03.037}
}

@article{boukare_solidificat_2025,
	author = {Boukar{\ifmmode\acute{e}\else\'{e}\fi}, Charles-{\ifmmode\acute{E}\else\'{E}\fi}douard and Badro, James and Samuel, Henri},
	title = {{Solidification of Earth{'}s mantle led inevitably to a basal magma ocean}},
	journal = {Nature},
	volume = {640},
	number = {8057},
	pages = {114--119},
	year = {2025},
	month = apr,
	issn = {1476-4687},
	publisher = {Nature Publishing Group},
	doi = {10.1038/s41586-025-08701-z}
}

@article{lichtenberg_redox_2021,
	author = {Lichtenberg, Tim},
	title = {{Redox Hysteresis of Super-Earth Exoplanets from Magma Ocean Circulation}},
	journal = {Astrophys. J. Lett.},
	volume = {914},
	number = {1},
	pages = {L4},
	year = {2021},
	month = jun,
	issn = {2041-8205},
	publisher = {The American Astronomical Society},
	doi = {10.3847/2041-8213/ac0146}
}

@book{clark_carbon_1982,
    author = {{Clark}, W. C.},
    booktitle = "Carbon Dioxide Review",
    year = 1982,
    pages = 469,
    title = {Carbon Dioxide Review},
    address = {New York},
    publisher = {Oxford University Press},
}

@article{nicholls_chemistry_2023,
    author = {Nicholls, Harrison and Hébrard, Eric and Venot, Olivia and Drummond, Benjamin and Evans, Elise},
    title = {Temperature–chemistry coupling in the evolution of gas giant atmospheres driven by stellar flares},
    journal = {Monthly Notices of the Royal Astronomical Society},
    volume = {523},
    number = {4},
    pages = {5681-5702},
    year = {2023},
    month = {08},
    issn = {0035-8711},
    doi = {10.1093/mnras/stad1734},
    url = {https://doi.org/10.1093/mnras/stad1734},
    eprint = {https://academic.oup.com/mnras/article-pdf/523/4/5681/50740394/stad1734.pdf},
}

@article{hunten_mass_1987,
	author = {Hunten, Donald M. and Pepin, Robert O. and Walker, James C. G.},
	title = {{Mass fractionation in hydrodynamic escape}},
	journal = {Icarus},
	volume = {69},
	number = {3},
	pages = {532--549},
	year = {1987},
	month = mar,
	issn = {0019-1035},
	publisher = {Academic Press},
	doi = {10.1016/0019-1035(87)90022-4}
}

@article{zahnle_mass_1986,
	author = {Zahnle, Kevin J. and Kasting, James F.},
	title = {{Mass fractionation during transonic escape and implications for loss of water from Mars and Venus}},
	journal = {Icarus},
	volume = {68},
	number = {3},
	pages = {462--480},
	year = {1986},
	month = dec,
	issn = {0019-1035},
	publisher = {Academic Press},
	doi = {10.1016/0019-1035(86)90051-5}
}

@article{huang_limits_2025,
	author = {Huang, Dongyang and Dorn, Caroline},
	title = {{Limits on forming coreless terrestrial worlds in the TRAPPIST-1 system}},
	journal = {arXiv},
	year = {2025},
	month = nov,
	eprint = {2511.01231},
	doi = {10.48550/arXiv.2511.01231}
}

@incollection{MARTY2025383,
title = {Geochemical and dynamical views on the origin of Earth's atmosphere and oceans},
editor = {Ariel Anbar and Dominique Weis},
booktitle = {Treatise on Geochemistry (Third edition)},
publisher = {Elsevier},
edition = {Third edition},
address = {Oxford},
pages = {383-416},
year = {2025},
isbn = {978-0-323-99763-8},
doi = {https://doi.org/10.1016/B978-0-323-99762-1.00106-6},
url = {https://www.sciencedirect.com/science/article/pii/B9780323997621001066},
author = {Bernard Marty and Hidenori Genda},
}

@article{ivanov_the_2013,
	author = {Ivanov, Mikhail A. and Head, James W.},
	title = {{The history of volcanism on Venus}},
	journal = {Planet. Space Sci.},
	volume = {84},
	pages = {66--92},
	year = {2013},
	month = aug,
	issn = {0032-0633},
	publisher = {Pergamon},
	doi = {10.1016/j.pss.2013.04.018}
}

@article{dasgupta_ingassing_2013,
	author = {Dasgupta, Rajdeep},
	title = {{Ingassing, Storage, and Outgassing of Terrestrial Carbon through Geologic Time}},
	journal = {Rev. Mineral. Geochem.},
	volume = {75},
	number = {1},
	pages = {183--229},
	year = {2013},
	month = jan,
	issn = {1529-6466},
	publisher = {GeoScienceWorld},
	doi = {10.2138/rmg.2013.75.7}
}

@article{lodge_fractal_2026,
	author = {Lodge, Matt G. and Moran, Sarah E. and Wakeford, Hannah R. and Leinhardt, Zo{\ifmmode\ddot{e}\else\"{e}\fi} M. and Marley, Mark S.},
	title = {{Fractal Aggregate Aerosols in the Virga Cloud Code. II. Exploring the Effects of Key Cloud Parameters in Warm Neptune, Hot Jupiter and Brown Dwarf Atmospheres}},
	journal = {Astrophys. J.},
	volume = {997},
	number = {2},
	pages = {317},
	year = {2026},
	month = jan,
	issn = {0004-637X},
	publisher = {The American Astronomical Society},
	doi = {10.3847/1538-4357/ae2752}
}

@article{janssen_hot_2026,
	author = {Janssen, Leoni J. and Miguel, Yamila and Min, Michiel and Huang, Helong and Zilinskas, Mantas and van Buchem, Christiaan P. A.},
	title = {{Hot and cloudy: High temperature clouds in super-Earths and sub-Neptunes}},
	journal = {arXiv},
	year = {2026},
	month = jan,
	eprint = {2601.15927},
	doi = {10.48550/arXiv.2601.15927}
}

@article{kitzmann_clouds_2010,
	author = {Kitzmann, D. and Patzer, A. B. C. and von Paris, P. and Godolt, M. and Stracke, B. and Gebauer, S. and Grenfell, J. L. and others},
	title = {{Clouds in the atmospheres of extrasolar planets - I. Climatic effects of multi-layered clouds for Earth-like planets and implications for habitable zones}},
	journal = {Astron. Astrophys.},
	volume = {511},
	pages = {A66},
	year = {2010},
	month = feb,
	issn = {0004-6361},
	publisher = {EDP Sciences},
	doi = {10.1051/0004-6361/200913491}
}

@incollection{marley_clouds_2014,
	author = {{M. S. Marley} and {A. S. Ackerman}},
	title = {{Clouds and Hazes in Exoplanet Atmospheres}},
	booktitle = {{Comparative Climatology of Terrestrial Planets}},
	pages = {367--392},
	year = {2014},
	isbn = {978-0-81659975-2},
	publisher = {University of Arizona Press},
	address = {Tucson, AZ, USA},
	url = {https://muse.jhu.edu/chapter/1207495}
}

@article{helling_clouds_2019,
	author = {Helling, Christiane},
	title = {{Exoplanet Clouds}},
	journal = {Annu. Rev. Earth Planet. Sci.},
	volume = {47},
	pages = {583--606},
	year = {2019},
	month = may,
	publisher = {Annual Reviews},
	doi = {10.1146/annurev-earth-053018-060401}
}

@article{farhat_tides_2025,
	author = {Farhat, Mohammad and Auclair-Desrotour, Pierre and Bou{\ifmmode\acute{e}\else\'{e}\fi}, Gwena{\ifmmode\ddot{e}\else\"{e}\fi}l and Lichtenberg, Tim and Laskar, Jacques},
	title = {{Tides on Lava Worlds: Application to Close-in Exoplanets and the Early Earth{\textendash}Moon System}},
	journal = {Astrophys. J.},
	volume = {979},
	number = {2},
	pages = {133},
	year = {2025},
	month = jan,
	issn = {0004-637X},
	publisher = {The American Astronomical Society},
	doi = {10.3847/1538-4357/ad9b93}
}

@article{chen_tidal_2016,
	author = {Chen, Erinna M. A. and Nimmo, Francis},
	title = {{Tidal dissipation in the lunar magma ocean and its effect on the early evolution of the Earth{\textendash}Moon system}},
	journal = {Icarus},
	volume = {275},
	pages = {132--142},
	year = {2016},
	month = sep,
	issn = {0019-1035},
	publisher = {Academic Press},
	doi = {10.1016/j.icarus.2016.04.012}
}

@article{nimmo_tidally_2024,
	author = {Nimmo, Francis and Kleine, Thorsten and Morbidelli, Alessandro},
	title = {{Tidally driven remelting around 4.35 billion years ago indicates the Moon is old}},
	journal = {Nature},
	volume = {636},
	number = {8043},
	pages = {598--602},
	year = {2024},
	month = dec,
	issn = {1476-4687},
	publisher = {Nature Publishing Group},
	doi = {10.1038/s41586-024-08231-0}
}

@article{grinspoon_implication_1993,
	author = {Grinspoon, David Harry},
	title = {{Implications of the high D/H ratio for the sources of water in Venus' atmosphere}},
	journal = {Nature},
	volume = {363},
	number = {6428},
	pages = {428--431},
	year = {1993},
	month = jun,
	issn = {1476-4687},
	publisher = {Nature Publishing Group},
	doi = {10.1038/363428a0}
}

@article{piette_rocky_2023,
	author = {Piette, Anjali A. A. and Gao, Peter and Brugman, Kara and Shahar, Anat and Lichtenberg, Tim and Miozzi, Francesca and Driscoll, Peter},
	title = {{Rocky Planet or Water World? Observability of Low-density Lava World Atmospheres}},
	journal = {Astrophys. J.},
	volume = {954},
	number = {1},
	pages = {29},
	year = {2023},
	month = aug,
	issn = {0004-637X},
	publisher = {The American Astronomical Society},
	doi = {10.3847/1538-4357/acdef2}
}

@article{barnes_tidal_2017,
	author = {Barnes, Rory},
	title = {{Tidal locking of habitable exoplanets}},
	journal = {Celest. Mech. Dyn. Astr.},
	volume = {129},
	number = {4},
	pages = {509--536},
	year = {2017},
	month = dec,
	issn = {1572-9478},
	publisher = {Springer Netherlands},
	doi = {10.1007/s10569-017-9783-7}
}

@article{gillmann_atmosphere_2014,
	author = {Gillmann, Cedric and Tackley, Paul},
	title = {{Atmosphere/mantle coupling and feedbacks on Venus}},
	journal = {J. Geophys. Res. Planets},
	volume = {119},
	number = {6},
	pages = {1189--1217},
	year = {2014},
	month = jun,
	issn = {2169-9097},
	publisher = {John Wiley {\&} Sons, Ltd},
	doi = {10.1002/2013JE004505}
}

@article{van_onset_2025,
	author = {van Dijk, Marijn R. and Nicholls, Harrison and Lichtenberg, Tim},
	title = {{Onset of Habitable Conditions on the Hadean Earth Set by Feedback between Tides and Greenhouse Forcing}},
	journal = {Planet. Sci. J.},
	volume = {7},
	number = {4},
	pages = {94},
	year = {2026},
	month = apr,
	issn = {2632-3338},
	publisher = {The American Astronomical Society},
	doi = {10.3847/PSJ/ae5928}
}

@book{pierrehumbert_principles_2010,
	author = {Pierrehumbert, Raymond T.},
	title = {{Principles of Planetary Climate}},
	journal = {Cambridge Aspire website},
	year = {2010},
	month = dec,
	isbn = {978-0-51178078-3},
	publisher = {Cambridge University Press},
	address = {Cambridge, England, UK},
	doi = {10.1017/CBO9780511780783}
}

@article{nicholls_massradius_2026,
	author = {Nicholls, Harrison and Shorttle, Oliver and Lichtenberg, Tim and Pascal, Flavia},
	title = {{Beyond the mass{\textendash}radius plane: integrated radiative{\textendash}convective and interior structure simulations of the exoplanet continuum}},
	journal = {Mon. Not. R. Astron. Soc.},
	volume = {551},
	number = {2},
	pages = {stag1489},
	year = {2026},
	month = sep,
	issn = {0035-8711},
	publisher = {Oxford Academic},
	doi = {10.1093/mnras/stag1489}
}

@article{nicholls_agni_2025,
	author = {Nicholls, Harrison and Pierrehumbert, Raymond and Lichtenberg, Tim},
	title = {{AGNI: A radiative-convective model for lava planet atmospheres}},
	journal = {Journal of Open Source Software},
	volume = {10},
	number = {109},
	pages = {7726},
	year = {2025},
	month = may,
	issn = {2475-9066},
	doi = {10.21105/joss.07726}
}

@article{shorttle_distinguish_2024,
	author = {Shorttle, Oliver and Jordan, Sean and Nicholls, Harrison and Lichtenberg, Tim and Bower, Dan J.},
	title = {{Distinguishing Oceans of Water from Magma on Mini-Neptune K2-18b}},
	journal = {Astrophys. J. Lett.},
	volume = {962},
	number = {1},
	pages = {L8},
	year = {2024},
	month = feb,
	issn = {2041-8205},
	publisher = {The American Astronomical Society},
	doi = {10.3847/2041-8213/ad206e}
}

@article{lammer_formation_2020,
	author = {Lammer, Helmut and Brasser, Ramon and Johansen, Anders and Scherf, Manuel and Leitzinger, Martin},
	title = {{Formation of Venus, Earth and Mars: Constrained by Isotopes}},
	journal = {Space Sci. Rev.},
	volume = {217},
	number = {1},
	pages = {7},
	year = {2020},
	month = dec,
	issn = {1572-9672},
	publisher = {Springer Netherlands},
	doi = {10.1007/s11214-020-00778-4}
}

@article{pierrehumbert_the_2002,
	author = {Pierrehumbert, Raymond T.},
	title = {{The hydrologic cycle in deep-time climate problems}},
	journal = {Nature},
	volume = {419},
	number = {6903},
	pages = {191--198},
	year = {2002},
	month = sep,
	issn = {1476-4687},
	publisher = {Nature Publishing Group},
	doi = {10.1038/nature01088}
}

@article{foley_the_2015,
	author = {Foley, Bradford J.},
	title = {{THE ROLE OF PLATE TECTONIC{\textendash}CLIMATE COUPLING AND EXPOSED LAND AREA IN THE DEVELOPMENT OF HABITABLE CLIMATES ON ROCKY PLANETS}},
	journal = {Astrophys. J.},
	volume = {812},
	number = {1},
	pages = {36},
	year = {2015},
	month = oct,
	issn = {0004-637X},
	publisher = {The American Astronomical Society},
	doi = {10.1088/0004-637X/812/1/36}
}

@article{zahnle_earths_2010,
	author = {Zahnle, Kevin and Schaefer, Laura and Fegley, Bruce},
	title = {{Earth{'}s Earliest Atmospheres}},
	journal = {Cold Spring Harbor Perspect. Biol.},
	volume = {2},
	number = {10},
	pages = {a004895},
	year = {2010},
	month = jun,
	publisher = {Cold Spring Harbor Lab},
	doi = {10.1101/cshperspect.a004895}
}

@article{krissansen2022understanding,
  title={Understanding planetary context to enable life detection on exoplanets and test the Copernican principle},
  author={Krissansen-Totton, Joshua and Thompson, Maggie and Galloway, Max L and Fortney, Jonathan J},
  journal={Nature Astronomy},
  volume={6},
  number={2},
  pages={189--198},
  year={2022},
  publisher={Nature Publishing Group UK London}
}

@article{krissansen2021venus,
  title={Was Venus ever habitable? Constraints from a coupled interior--atmosphere--redox evolution model},
  author={Krissansen-Totton, Joshua and Fortney, Jonathan J and Nimmo, Francis},
  journal={The Planetary Science Journal},
  volume={2},
  number={5},
  pages={216},
  year={2021b},
  publisher={IOP Publishing}
}

@dataset{github_zenodo,
  author       = {Nicholls, Harrison},
  title        = {CHILI (CUISINES) Part I - Evolution of Earth and
                   Venus, simulation data and plotting utilities
                  },
  month        = jun,
  year         = 2026,
  publisher    = {Zenodo},
  doi          = {10.5281/zenodo.20680020},
  url          = {https://doi.org/10.5281/zenodo.20680020},
}

@article{ramirez_warming_2014,
	author = {Ramirez, Ramses M. and Kopparapu, Ravi and Zugger, Michael E. and Robinson, Tyler D. and Freedman, Richard and Kasting, James F.},
	title = {{Warming early Mars with CO2 and H2}},
	journal = {Nat. Geosci.},
	volume = {7},
	number = {1},
	pages = {59--63},
	year = {2014},
	month = jan,
	issn = {1752-0908},
	publisher = {Nature Publishing Group},
	doi = {10.1038/ngeo2000}
}

@article{Julia2017,
    title={Julia: A fresh approach to numerical computing},
    author={Bezanson, Jeff and Edelman, Alan and Karpinski, Stefan and Shah, Viral B},
    journal={SIAM {R}eview},
    volume={59},
    number={1},
    pages={65--98},
    year={2017},
    publisher={SIAM},
    doi={10.1137/141000671},
    url={https://epubs.siam.org/doi/10.1137/141000671}
}

@Article{Hunter2007,
  Author    = {Hunter, J. D.},
  Title     = {Matplotlib: A 2D graphics environment},
  Journal   = {Computing in Science \& Engineering},
  Volume    = {9},
  Number    = {3},
  Pages     = {90--95},
  publisher = {IEEE COMPUTER SOC},
  doi       = {10.1109/MCSE.2007.55},
  year      = 2007
}

@Article{harris2020array,
 title         = {Array programming with {NumPy}},
 author        = {Charles R. Harris and K. Jarrod Millman and St{\'{e}}fan J.
                 van der Walt and Ralf Gommers and Pauli Virtanen and David
                 Cournapeau and Eric Wieser and Julian Taylor and Sebastian
                 Berg and Nathaniel J. Smith and Robert Kern and Matti Picus
                 and Stephan Hoyer and Marten H. van Kerkwijk and Matthew
                 Brett and Allan Haldane and Jaime Fern{\'{a}}ndez del
                 R{\'{i}}o and Mark Wiebe and Pearu Peterson and Pierre
                 G{\'{e}}rard-Marchant and Kevin Sheppard and Tyler Reddy and
                 Warren Weckesser and Hameer Abbasi and Christoph Gohlke and
                 Travis E. Oliphant},
 year          = {2020},
 month         = sep,
 journal       = {Nature},
 volume        = {585},
 number        = {7825},
 pages         = {357--362},
 doi           = {10.1038/s41586-020-2649-2},
 publisher     = {Springer Science and Business Media {LLC}},
 url           = {https://doi.org/10.1038/s41586-020-2649-2}
}

@ARTICLE{odert2016,
       author = {{Odert}, P. and {Lammer}, H. and {Erkaev}, N.~V. and {Nikolaou}, A. and {Lichtenegger}, H.~I.~M. and {Johnstone}, C.~P. and {Kislyakova}, K.~G. and {Leitzinger}, M. and {Tosi}, N.},
        title = "{Escape and fractionation of volatiles and noble gases from Mars-sized planetary embryos and growing protoplanets}",
      journal = {\icarus},
         year = 2018,
        month = jun,
       volume = {307},
        pages = {327-346},
          doi = {10.1016/j.icarus.2017.10.031},
archivePrefix = {arXiv},
       eprint = {1706.06988},
 primaryClass = {astro-ph.EP},
       adsurl = {https://ui.adsabs.harvard.edu/abs/2018Icar..307..327O}
}

@ARTICLE{zahnle1990,
       author = {{Zahnle}, Kevin and {Kasting}, James F. and {Pollack}, James B.},
        title = "{Mass fractionation of noble gases in diffusion-limited hydrodynamic hydrogen escape}",
      journal = {\icarus},
         year = 1990,
        month = apr,
       volume = {84},
       number = {2},
        pages = {502-527},
          doi = {10.1016/0019-1035(90)90050-J},
       adsurl = {https://ui.adsabs.harvard.edu/abs/1990Icar...84..502Z}
}

@article{kobayashi_viscosity_2000,
	author = {Kobayashi, Hiroshi and Takahashi, Haruyuki and Hiki, Yosio},
	title = {{Viscosity of glasses near and below the glass transition temperature}},
	journal = {J. Appl. Phys.},
	volume = {88},
	number = {6},
	pages = {3776--3778},
	year = {2000},
	month = sep,
	issn = {0021-8979},
	doi = {10.1063/1.1286107}
}

@article{marsh_on_1981,
	author = {Marsh, B. D.},
	title = {{On the crystallinity, probability of occurrence, and rheology of lava and magma}},
	journal = {Contrib. Mineral. Petrol.},
	volume = {78},
	number = {1},
	pages = {85--98},
	year = {1981},
	month = oct,
	issn = {0010-7999},
	doi = {10.1007/BF00371146}
}

@article{karki_viscosity_2010,
	author = {Karki, Bijaya B. and Stixrude, Lars P.},
	title = {{Viscosity of MgSiO3 Liquid at Earth{'}s Mantle Conditions: Implications for an Early Magma Ocean}},
	journal = {Science},
	volume = {328},
	number = {5979},
	pages = {740},
	year = {2010},
	month = may,
	issn = {0036-8075},
	doi = {10.1126/science.1188327}
}

@ARTICLE{carone2025,
       author = {{Carone}, L. and {Barnes}, R. and {Noack}, L. and {Chubb}, K. and {Barth}, P. and {Bitsch}, B. and {Thamm}, A. and {Balduin}, A. and {Garcia}, R. and {Helling}, Ch.},
        title = "{From CO$_{2}$- to H$_{2}$O-dominated atmospheres and back: How mixed outgassing changes the volatile distribution in magma oceans around M dwarf stars}",
      journal = {\aap},
         year = 2025,
        month = jan,
       volume = {693},
          eid = {A303},
        pages = {A303},
          doi = {10.1051/0004-6361/202450307},
archivePrefix = {arXiv},
       eprint = {2412.10192},
 primaryClass = {astro-ph.EP},
       adsurl = {https://ui.adsabs.harvard.edu/abs/2025A&A...693A.303C}
}

@article{holloway_volatiles_1994,
  title={Volatiles in magmas},
  author={Holloway, JR and Blank, JG and Carroll, MR},
  journal={Rev. Mineral},
  volume={30},
  year={1994}
}

@article{pan_the_1991,
	author = {Pan, Vivian and Holloway, John R. and Hervig, Richard L.},
	title = {{The pressure and temperature dependence of carbon dioxide solubility in tholeiitic basalt melts}},
	journal = {Geochim. Cosmochim. Acta},
	volume = {55},
	number = {6},
	pages = {1587--1595},
	year = {1991},
	month = jun,
	issn = {0016-7037},
	publisher = {Pergamon},
	doi = {10.1016/0016-7037(91)90130-W}
}

@article{roscoe1952viscosity,
  title={The viscosity of suspensions of rigid spheres},
  author={Roscoe, R},
  journal={British journal of applied physics},
  volume={3},
  number={8},
  pages={267},
  year={1952},
  publisher={IOP Publishing}
}
\bibliographystyle{aasjournal}

\end{document}